\documentclass{aa}  

\usepackage{multirow}
\usepackage{natbib}
\usepackage{mathtools}
\usepackage{hyperref}
\usepackage{graphicx}
\usepackage{siunitx}                                    
\DeclareSIUnit \parsec {pc}
\usepackage{txfonts}
\begin{document} 

   \title{A sample of dust attenuation laws for Dark Energy Survey supernova host galaxies\thanks{DES-2022-069. FERMILAB-PUB-22-760-PPD.}\thanks{The DES-SN host galaxy photometric data and corresponding SN light-curve parameters are available as part of the DES3YR data release, accessible at \url{https://www.darkenergysurvey.org/des-year-3-supernova-cosmology-results/}. Cornerplots and SED fit plots for the host galaxies can be found at \url{https://github.com/SN-CRISP/DES-SN_Host-Galaxies}.}}


\author{
J.~Duarte \inst{1},
S. Gonz\'alez-Gait\'an \inst{1},
A.~Mour\~ao \inst{1},
A.~Paulino-Afonso \inst{1},
P.~Guilherme-Garcia \inst{2},
J.~\'Aguas \inst{1},
L.~Galbany \inst{3,4},
L.~Kelsey \inst{5,6},
D.~Scolnic \inst{7},
M.~Sullivan \inst{6},
D.~Brout \inst{8},
A.~Palmese \inst{9},
P.~Wiseman \inst{6},
M.~Aguena \inst{10},
O.~Alves \inst{11},
D.~Bacon \inst{5},
E.~Bertin \inst{12,13},
S.~Bocquet \inst{14},
D.~Brooks \inst{15},
D.~L.~Burke \inst{16,17},
A.~Carnero~Rosell \inst{18,10,19},
M.~Carrasco~Kind \inst{20,21},
J.~Carretero \inst{22},
M.~Costanzi \inst{23,24,25},
M.~E.~S.~Pereira \inst{26},
T.~M.~Davis \inst{27},
J.~De~Vicente \inst{28},
S.~Desai \inst{29},
H.~T.~Diehl \inst{30},
P.~Doel \inst{15},
S.~Everett \inst{31},
I.~Ferrero \inst{32},
D.~Friedel \inst{20},
J.~Frieman \inst{30,33},
J.~Garc\'ia-Bellido \inst{34},
M.~Gatti \inst{35},
D.~W.~Gerdes \inst{36,11},
D.~Gruen \inst{14},
R.~A.~Gruendl \inst{20,21},
G.~Gutierrez \inst{30},
S.~R.~Hinton \inst{27},
D.~L.~Hollowood \inst{37},
K.~Honscheid \inst{38,39},
D.~J.~James \inst{8},
K.~Kuehn \inst{40,41},
N.~Kuropatkin \inst{30},
P.~Melchior \inst{42},
R.~Miquel \inst{43,22},
F.~Paz-Chinch\'{o}n \inst{20,44},
A.~Pieres \inst{10,45},
A.~A.~Plazas~Malag\'on \inst{42},
M.~Raveri \inst{35},
M.~Rodriguez-Monroy \inst{28},
E.~Sanchez \inst{28},
V.~Scarpine \inst{30},
I.~Sevilla-Noarbe \inst{28},
M.~Smith \inst{6},
E.~Suchyta \inst{46},
G.~Tarle \inst{11},
C.~To \inst{38},
and N.~Weaverdyck \inst{11,47}\\
\protect\begin{center} (DES Collaboration) \protect\end{center}
\protect\begin{center}
\protect\parbox{3in}{\protect\centering\textit{Affiliations are listed at the end of the paper}}
\protect\end{center}
}

\authorrunning{J.~Duarte}
\institute{}

   \date{Received 29 March 2023 / Accepted 05 October 2023
}

 
  \abstract
   {
   Type Ia supernovae (SNe Ia) are useful distance indicators in cosmology, provided their luminosity is standardized by applying empirical corrections based on light-curve properties. One factor behind these corrections is dust extinction, which is accounted for in the color-luminosity relation of the standardization. This relation is usually assumed to be universal, which can potentially introduce systematics into the standardization. The ``mass step'' observed for SN Ia Hubble residuals has been suggested as one such systematic.} 
   {We seek to obtain a more complete view of dust attenuation properties for a sample of 162 SN Ia host galaxies and to probe their link to the mass step.} 
   { We inferred attenuation laws toward hosts from both global and local (4 kpc) Dark Energy Survey photometry and composite stellar population model fits.} 
   {We recovered a relation between the optical depth and the attenuation slope, best explained by differing star-to-dust geometry for different galaxy orientations, which is significantly different from the optical depth and extinction slope relation observed directly for SNe. We obtain a large variation of attenuation slopes and confirm these change with host properties, such as the stellar mass and age, meaning a universal SN Ia correction should ideally not be assumed. Analyzing the cosmological standardization, we find evidence for a  mass step and a two-dimensional ``dust step'', both more pronounced for red SNe. Although comparable, the two steps are not found to be completely analogous.}
   %
   {We conclude that host galaxy dust data cannot fully account for the mass step, using either an alternative SN standardization with extinction proxied by host attenuation or a dust-step approach.}


   \maketitle
%

\section{Introduction}

\label{sec:intro}
Type Ia supernovae (SNe Ia) are a set of astrophysical transients that make for excellent distance indicators, having famously contributed to the discovery of the accelerated expansion of the Universe \citep{Riess98,Perlmutter_99}. SN Ia data have proven extremely important in constraining the various cosmological parameters, such as the equation of state parameter for the dark energy $w$ \citep{Scolnic_2018,abbott_2019}, as well as the Hubble constant $H_0$, which has recently garnered further interest due to the large discrepancies between estimates from early and late-time probes of the Universe. This discrepancy is usually called Hubble tension \citep[e.g.,][]{Dhawan_18,feeney_2018,Riess_2019}, with current measurements indicating it lies at a confidence level of $\sim5\sigma$ \citep[e.g.,][]{Riess_2020}. 
\par
Despite not being standard candles, SN Ia luminosity can be standardized by applying some empirical corrections. The first two corrections are based on the shape-luminosity \citep{Phillips_1993} and light-curve color-luminosity relations \citep{Tripp98}. The color-luminosity correction, in particular, is introduced to account for the fact that bluer SNe Ia tend to be more luminous. Even so, after the application of these two corrections, SNe Ia originating in higher-mass galaxies appear consistently over-luminous \citep[e.g.,][]{Sullivan10,Lampeitl_2010,kelly_2010}, such that a mass step is evident in the difference between their observed and predicted luminosities. A third correction is often introduced to account for this environmental dependency.
\par
Because of the empirical nature of these corrections, the full impact of the standardization process on cosmology is still not fully understood. Although heavily debated, two main effects seem to be at the origin of the color-luminosity relation: intrinsic color variations among different SNe Ia and the effects of dust extinction, namely reddening \citep{Conley_07}.
\par
Extinction refers to the absorption of light as well as the scattering of that light out of the line of sight, which, as a result of the overall composition, size, and orientation of the dust grains, tend to be stronger for bluer wavelengths \citep{Calzetti_1994,Calzetti_2001,Salim_2020}. This leads both to a dimming and a reddening of the observed spectrum. The reddening law is most commonly parameterized by the total-to-selective extinction parameter $R_V=A_V/E(B-V)$, where $A_V$ is the extinction for the $V$ band and $E(B-V)=A_B-A_V$ is the difference between the object's observed and intrinsic $B-V$ color index, referred to as the color excess. 
\par
For the purposes of SN Ia standardization, the color-luminosity relation is generally assumed to be the same for all SNe Ia in the population under study \citep{Tripp98}. By the same token, when correcting for galactic dust reddening, it is common practice to assume a constant $R_V$ for all galaxies, such as the Milky Way average value of $R_V=3.1$ \citep[e.g.,][]{Schlafly_2016,Cardelli_89}.
Other values exist, such as the \cite{Calzetti_2000} value of $R_V=4.05$ for star-forming galaxies and the peculiarly low values for the Magellanic Clouds \citep[e.g.,][]{Lequeux_82,Nandy_81}. However, it has been shown that dust properties, particularly $R_V$, can vary between galaxies \citep[e.g.,][]{Salim_2020,Salim_2018,Zahid_2013}, and so assuming a common dust law for all galaxies and even for different regions within the same galaxy might introduce systematic errors into the recovered results \citep[e.g.,][]{Draine_2003,Nataf_2015}.
\par
In the case of SN Ia standardization, it has been suggested that the previously mentioned mass step is one such systematic. \cite{Brout_2021} find that the correlation between the corrected luminosity and the host galaxy mass can be explained by differing $R_V$ values between low- and high-mass galaxy bins, while \cite{wiseman_2022} find that the variation of $R_V$ with galaxy age can explain almost the entire observed mass step. Additionally, \cite{johansson2021nearir} {and \cite{Meldorf}} show that the mass step can be eliminated using individual low-redshift SN Ia light-curve fits in the optical/NIR range to directly determine extinction properties and thus constrain the color-luminosity relation for each SN.
\par
However, there is also evidence to suggest that the mass step cannot entirely be explained as a dust systematic. \cite{Thorp_2021} find that a mass step is still observable even when allowing for different $R_V$ values between high- and low-mass host galaxies. \cite{gaitan} obtain similar results when considering separate color-luminosity relations for different SN populations. \cite{uddin_2020} and \cite{ponder2021type} interpret the consistency of the mass steps recovered from both optical and NIR light curves as evidence that dust does not play a large role in the mass-step correlation, as the effects of dust extinction in NIR are greatly reduced. \cite{Jones_2022} also find evidence for a NIR mass step when looking at combined high and low redshift populations. 
\par
More thorough studies on the nature of the SN Ia color-luminosity relation are needed to better grasp the contributions of interstellar dust to the mass-step correction. As shown by \cite{Meldorf} and \cite{popovic2021}, it can prove advantageous to individually determine dust properties for each individual SN Ia or for each individual SN Ia population in order to better constrain the color-luminosity relation in the standardization. This allows one to account for varying dust properties across different lines of sight and environments. However, it can be challenging to directly determine the extinction for SNe Ia, especially at higher redshifts, which require deep infrared photometry. The study of SN Ia host galaxies offers an interesting opportunity, not only to tackle the problem of dust-induced systematics in SN standardization, but also to extend our knowledge about general dust properties affecting extra-galactic observations. 
\par

When dealing with extended objects, such as SN host galaxies, the description of dust effects by extinction becomes too simplistic, as we can no longer presume to be dealing with a single line of sight. In these cases, we refer to the analogous concept of dust attenuation \citep{Calzetti_1994,Calzetti_2001,Salim_2020}. This phenomenon includes not only the abovementioned effects, but also effects arising from the spatial distribution of stars and dust in a galaxy or stellar population, such as scattering of light back into the line of sight, multiple dust cloud densities, and light emitted by unobscured stars. These phenomena are mostly negligible when considering only a point source.

\par 
In this work, we tackle two main questions. First, we study dust attenuation laws for the host galaxies of a set of SNe Ia, examining how they differ from each other and how they compare to point source literature extinctions measured directly from SNe Ia. To this effect, we present an alternative approach to infer dust attenuation laws toward the host galaxies for a cosmological sample of 162 SNe Ia collected as part of the Dark Energy Survey (DES) \citep{DES,Kelsey_2020}. We fit composite stellar population (CSP) models to the ``deep'' host photometric data in the \textit{griz} filter bands\footnote{\url{http://www.ctio.noirlab.edu/noao/content/DECam-filter-information}} \citep{Smith_2020,wiseman_2020}, as detailed in Section \ref{sec:methods}. To validate our approach, in Section \ref{sec:sim} we present the recovered fit results for a series of simulated host galaxies, comparing the best-fit values and the ``true'' parameters used in the simulations. In Section \ref{sec:des_fits} we analyze and characterize the recovered attenuation laws, comparing them to known SN extinction laws. Namely, we analyze the viability of using our host galaxy fit results to approximate the extinction for the respective SNe. Second, we seek to ascertain whether the mass step for SNe Ia can be explained by host dust properties. In Section \ref{sec:hubble} we explore the possibility of a step in luminosity linked to the dust environment surrounding the SNe and whether it can be related to the more usual mass step. We also explore the relation between these Hubble residuals steps and SN color. We assume a spatially flat $\Lambda$CDM model \citep{Condon_2018}, with a matter density $\Omega_m=0.3$, a dark energy density of $\Omega_\Lambda=0.7$, a dark energy equation of state parameter $w=-1$, and a Hubble constant $H_{0}=70\SI{}{km\,s^{-1}\,Mpc^{-1}}$. 

\section{Data and methods}
\label{sec:methods}

\subsection{Dark Energy Survey photometric data}

The Dark Energy Survey (DES) is an imaging survey covering $\sim$ 5100 square degrees of the Southern Hemisphere using the 4-m Blanco telescope at the Cerro Tololo Inter-American Observatory (CTIO), equipped with the 520 megapixel wide-field Dark Energy Camera (DECam) \citep{Flaugher_15} with a 0.263 arcsecond/pixel resolution.
\par
In this work we look at data collected as part of the first three-year cosmological sample of the SN survey (DES3YR)\footnote{\url{https://www.darkenergysurvey.org/des-year-3-supernova-cosmology-results/}} \citep{Brout_2019}, consisting of 206 spectroscopically confirmed SNe, as well as their respective host galaxies \citep{Smith_2020,wiseman_2020}. These data include both global and local broadband photometry for the hosts in the DECam \textit{griz} filter bands, originally computed by \cite{Kelsey_2020}. Only the 162 hosts meeting the selection cuts of \cite{Kelsey_2020} were considered, covering a redshift range of $0.077<z<0.58$. In addition, we use the light-curve parameters recovered for each of the corresponding SNe Ia by \cite{Brout_2019}. 
\par
As detailed in \cite{Kelsey_2020}, the global photometry for each host galaxy was measured from stacked images using {\sc SExtractor} \citep{bertin_96}, employing Kron FLUX\_AUTO measurements \citep{wiseman_2020}. A detection image was used to set the aperture, ensuring it was the same for the measurements in each filter. The images were also corrected for Milky Way dust extinction using Schlegel dust maps \citep{Schlegel_1998} and a Fitzpatrick extinction law \citep{Fitzpatrick_1999} with standard $R_V=3.1$ and fiducial coefficients $R_g=3.186$, $R_r = 2.140$, $R_i = 1.569$ and $R_z = 1.196$ \citep{Kelsey_2020,DES_data1}.
\par
In the case of local photometry, the aperture photometry tool from the {\sc Photutils} Python module \citep{bradley} was used to take measurements in a 4 kpc radius aperture around the SN location. This value is chosen to account for the $1\sigma$ seeing of the DES seeing-optimized stacks, which is $0.55''$ \citep{Kelsey_2020}. Thus, it is ensured that the aperture is larger than a $0.55''$ radius in our redshift range. The resulting fluxes were corrected for Milky Way extinction in the same way as described above. As the local aperture size is constant, the local stellar mass obtained in this case can be thought of as a measure of the local density.

\subsection{Host galaxy fits}
\label{subsec:fits}
We are interested in fitting CSP models to the photometric data for the SN Ia host galaxies. We do this by computing the Spectral Energy Distributions (SEDs) for each CSP and comparing them with the observed data. We rely on Bayesian inference methods to do so, using the {\sc Prospector}\footnote{\url{https://github.com/bd-j/prospector}} \citep{Johnson_2021} and {\sc emcee}\footnote{\url{https://github.com/dfm/emcee}} \citep{Foreman_2013} Python packages. The models themselves are built using the {\sc FSPS}\footnote{\url{https://github.com/cconroy20/fsps}} \citep{Conroy_2009,Conroy_2010} FORTRAN code and the {\sc python-FSPS}\footnote{\url{https://github.com/dfm/python-fsps}} package \citep{Foreman-Mackey}.
 \par
 Our fitting model is built using the following assumptions: a \cite{Kroupa_2001} initial mass function (IMF); a delayed $\tau$-model star formation history (SFH), with an exponentially decreasing star formation rate $\textrm{SFR}(t)=te^{-t/\tau}$, with e-folding time $\tau=1$Gyr and time $t$ measured up to the stellar population age $t_{\textrm{age}}$ \footnote{$t_{\textrm{age}}$ refers to the time between the start of the last burst of star formation and the observing time of these galaxies. Thus, it acts as an upper limit}; a modified Calzetti attenuation law \citep{Noll_09,Kriek_2013}, with an expression given by Eq. \ref{eq:Calzetii_mod}:
     \begin{equation}
    \tau(\lambda)=\frac{\tau_V}{R_{V,0}}\left[k(\lambda)+D(\lambda,n)\right]\left(\frac{\lambda}{\lambda_V}\right)^n.
    \label{eq:Calzetii_mod}
\end{equation}
\par
The metallicity $\log(Z_\star/Z_\odot)$, the age $t_{\textrm{age}}$, the optical depth $\tau_V$, and the dust index $n$ are taken as model free parameters, with overall flat priors. The stellar mass is fixed, referring to the fit values obtained by \cite{Kelsey_2020}.

In Eq. \ref{eq:Calzetii_mod}, $k(\lambda)$ describes the original \cite{Calzetti_2000} attenuation law with an original $R_{V,0}=4.05$ and $D(\lambda,n)$ describes a UV-bump centered at $\lambda_0=2175\si{\angstrom}$ \citep{Noll_09}. We adopt this attenuation law both because of its ability to describe the attenuation for a wide range of star-forming galaxies \citep{Noll_09} and because it allows for a finer probing of the reddening levels by offering a two parameter description. $\tau_V$ describes the optical depth of the medium, such that $A_V\simeq1.086\tau_V$, while $n$ is introduced as a dust index parameter in order to produce different slopes without the need to alter $\tau_V$; $n$ is constrained between the values of $-2.2$ and $0.4$, with more negative values corresponding to a larger level of reddening, for the same value of $\tau_V$. The more commonly used $R_V$ parameter can be obtained using $R_V=\frac{\tau_V}{\tau_B-\tau_V}$, such that $\ln(R_V)$ is roughly proportional to $n$ ($\ln{(R_V)}\sim n$). The average value for the Milky Way of $R_V=3.1$ corresponds approximately to $n=-0.3$.
\par

With regard to the model parameters, the stellar metallicity does not vary with age in our model, as no age-metallicity relation is implemented \citep{Leja_2017}. In addition, the age parameter $t_{\textrm{age}}$ for a given population is constrained such that no time-steps older than the universe at the corresponding redshift are allowed.
\par
The real SFH for the galaxies may be much more intricate than the simplistic delayed $\tau$-model we assume. However, given how few filter bands we have to work with, more complex models are not feasible in this case. The effect of the assumptions made for the SFH and dust models on the recovered parameters will be further explored in the next section.
\par
The choice to fix the stellar mass in the fits is made because, although Bayesian inference does not limit the number of fit parameters one can use, increasing their number would increase the number of iterations necessary for convergence, making the computation time much longer. While other parameters might exhibit some degeneracy, the stellar mass is fairly easy to constrain, as it represents a normalization of the SED. As such, the mass values obtained from a five-parameter sampling do not significantly differ from those previously obtained for these galaxies by \cite{Kelsey_2020}, with test fits exhibiting a median relative difference of $\sim4\%$ for both global and local measurements. A deeper look at these results can be found in Appendix \ref{app:mass}. The differences between a four- and five-parameter sampling on the recovered global dust parameters are also not very noticeable, with test fit  median relative differences of $\sim1\%$ and $\sim4\%$ for $\tau_V$ and $n$, respectively. These differences are more pronounced when using local measurements, with median relative differences of $\sim8\%$ and $\sim60\%$ for $\tau_V$ and $n$, respectively. These relatively high values are mainly a product of convergence problems in the four-parameter sampling process. 
\par
As a final note, although the posterior distributions obtained from Bayesian inference are very informative, a best-fit value is often useful for data analysis. For a given galaxy and a given parameter, we take this value to be the median of the distribution, with the uncertainty range given by the
16th and 84th percentiles, which define a 68\% credible region around the median.

\section{Host galaxy simulations}
\label{sec:sim}

In order to establish the reliability of our fitting methodology, a number of tests were conducted. In each of these, various fits were performed for a population of simulated host galaxies. These galaxies were generated by drawing the values for $M_\star$, $Z_\star$, $t_{\textrm{age}}$, $\tau_V$, $n$, and $z$ from a series of random flat distributions. While this method does not guarantee that the simulated population will resemble an actual SN Ia host population, it is adequate for testing the performance of the fitting method across the full parameter space. 
\par
Photometry in the \textit{griz} filter bands for each of these galaxies was generated using FSPS. Gaussian noise was also applied to the photometric data, in accordance with a prescribed signal-to-noise ratio (S/N) of 10, which was chosen to roughly reflect the minimum S/N of the DES data set \citep{Kelsey_2020}. 

\subsection{Fit photometry tests}
With the first set of tests, we seek to ascertain whether \textit{griz} photometry can be used to accurately recover attenuation parameters. For this purpose, 150 galaxies were simulated employing the same CSP model as described in Section \ref{sec:methods}. We define this as the default simulation. For these fits, both the stellar mass and the redshift were fixed to their ``true'' values, following the rationale detailed in Section \ref{subsec:fits}. This represents the best case scenario for our fitting method. As a term of comparison, additional photometry in the \textit{NUV/FUV} GALEX \citep{galex} and \textit{JHKs} 2MASS \citep{2mass} filter bands was generated for half of these galaxies and used in addition to the \textit{griz} photometry in a new set of fits.
\par
Some precision and accuracy metrics for the dust parameters ($\tau_V$ and $n$) recovered in each of the fits, including the median differences between the simulated and fitted parameters (median $\Delta$) and the corresponding median relative error, are shown in Table \ref{tab:mock}. We find that the fits using additional photometry do not vary greatly from their \textit{griz} counterparts. In particular, the values for $\tau_V$ and $n$ exhibit median relative differences of $\sim 0.1\%$ and $\sim 2\%$, respectively. The full comparison between the parameters recovered for both sets of fits can be found in Fig. \ref{fig:mock_uv_nir}.
\par
Regarding the role of dust emission from far infrared (FIR) data on the determination of attenuation parameters $\tau_V$ and $n$, we find from analogous simulations and fits that its inclusion results in a similar improvement to the one discussed above.
\par

It is also crucial to establish that the fitting methodology does not impart any nonphysical correlation unto the recovered parameters. For the \textit{griz} test fit, the Spearman correlation coefficient for $\tau_V$ and $n$ is found to be $0.038$, meaning that no correlation is found between these two parameters. Given that the mock galaxy set was generated at random, without imposing any prior relation between the model parameters, the lack of correlation found indicates that no relation is being artificially introduced by the fitting process.

\subsection{Fixed mass error tests}
 With the second set of tests, we explore how a possible variation in the fixed value of stellar mass affects the recovered fit parameters. Photometry in the \textit{griz} filter bands for the same simulated galaxies used in the previous set of tests was used, with the fixed value of the mass being varied by $\pm 4\%$, according to the median difference between \cite{Kelsey_2020} and our own inferred masses, recovered in Section \ref{subsec:fits}. The median difference was chosen to vary the value of the fixed mass because, as seen from Fig. \ref{fig:mass_comp}, the bias between our fits and the mass values from \cite{Kelsey_2020} is the dominant source of the deviations between the two, even more so than the scatter of the fit.
\par

Once again, precision and accuracy metrics for the dust parameters recovered in each of the fits are shown in Table \ref{tab:mock}. These results demonstrate that the fixing of the stellar mass does not compromise the determination of the other fit parameters.

\subsection{Alternative model simulations tests}

For the third set of tests, new simulations were performed using different models from those assumed in the fitting process, with the goal of understanding how the methodology would cope with these variations. Six different simulation models were tested, with 75 galaxies being  simulated for each case. Each of the models used in these tests differs from the model presented in Section \ref{sec:methods} in one of the following ways: a \cite{Cardelli} attenuation law, with varying $R_V$ instead of $n$; a delayed $\tau$-model SFH with $\tau=0.1$; a delayed $\tau$-model SFH with $\tau=10$; a delayed $\tau$-model SFH truncated at $t_{\textrm{trunc}}=7.5$ Gyr\footnote{For the truncated models, the SFR is set to 0 at the indicated age to completely shut down star formation.}; a delayed $\tau$-model SFH truncated at $t_{\textrm{trunc}}=5$ Gyr; a delayed $\tau$-model SFH truncated at $t_{\textrm{trunc}}=3$ Gyr;

\par

The fits were once again performed using \textit{griz} photometry, with the default fitting CSP model defined in Section \ref{sec:methods}. Precision and accuracy metrics for the dust parameters recovered in each of the fits are again displayed in Table \ref{tab:mock}. This final test set shows that, for the most part, our methodology is able to cope with variations in the model used for simulation without too much impact on the dust parameters.
\par

The largest changes are found, as expected, when different attenuation laws are introduced, which cannot be avoided due to the different parameterizations used. In this case, $R_V$ is the most affected parameter, with a constant bias offset of 0.261 present in the recovered fit parameters. A more detailed look at this particular case is offered in Appendix \ref{app:B}.

\begin{table*}
\centering
    \caption{Precision and accuracy metrics for the dust parameters in the mock galaxy fits, for each of the test sets.}
    \begin{tabular}{cccccccc}
    \hline
Test & Parameter & RMS & Median $\Delta$ & Median Rel. Error (\%) & Median Fit Error & MAD & Outliers (\%) \\\hline
\multirow{2}{*}{Default (\textit{griz})}& $\tau_V$  & 1.332 & 0.263 & 10.1& 0.695 & 0.739  & 6\\
& $n$ & 0.552 & 0.006 & 10.3 & 0.262 & 0.386 & 5.33\\
\multirow{2}{*}{Default (\textit{griz}+UV+NIR)}& $\tau_V$ & 0.854 & 0.039 & 0.9 & 0.238 & 0.356 & 5.33\\
& $n$ & 0.274 & 0.004 & 0.8 & 0.072 & 0.087 & 8 \\ 
Default + Emission & $\tau_V$ & 0.629 &-0.014 & 0.3 & 0.280 & 0.390 & 7.33\\
 (\textit{griz}+FIR) & $n$ & 0.548 & -0.019 & 12.9 & 0.255 & 0.416 & 5.33\\ \hline
\multirow{2}{*}{Default (Mass+4\%)} & $\tau_V$  & 1.050 & -0.014 & 0.3 & 0.737 & 0.640 & 5.33\\
& $n$ & 0.380 & $-7.89\times10^{-5}$ & 6.4 & 0.325 & 0.231 & 6.67 \\
\multirow{2}{*}{Default (Mass-4\%)} & $\tau_V$ & 0.983 & -0.019 & 0.6 & 0.720 & 0.624 &5.33 \\ 
& $n$ & 0.350 & -0.004 & 3.7 & 0.291 & 0.229 & 6.67 \\\hline
\multirow{2}{*}{Cardelli} & $\tau_V$ & 1.221 & -0.036 & 1.0 & 0.789 & 0.786 & 8 \\
& $n$ & 0.840 & 0.261 & 10.6 & 1.761 & 0.720 & 4 \\
\multirow{2}{*}{$\tau=0.1$} & $\tau_V$ & 1.182 & 0.179 & 6.9 & 0.607 & 0.866 & 5.33 \\
& $n$ &0.532 & -0.002 & 2.6 & 0.231 & 0.286 & 2.67 \\
\multirow{2}{*}{$\tau=10$} & $\tau_V$ & 0.865 & -0.061 & 1.7 & 0.715 & 0.557 & 6.67 \\
& $n$ & 0.480 & 0.027 & 12.4 & 0.324 & 0.283 & 8 \\
\multirow{2}{*}{$t_{\textrm{trunc}}=7.5$ Gyr} & $\tau_V$ & 1.140 & 0.042 & 3.0 & 0.239 & 0.840 & 10.67 \\
& $n$ & 0.388 & -0.027 & 4.4 & 0.280 & 0.311 & 4 \\
\multirow{2}{*}{$t_{\textrm{trunc}}=5$ Gyr} & $\tau_V$ & 1.038 & 0.054 & 1.4 & 0.645 & 0.537 & 8\\
& $n$ &0.539 & 0.031 & 9.4 &0.220 & 0.224 & 4\\
\multirow{2}{*}{$t_{\textrm{trunc}}=3$ Gyr} & $\tau_V$ & 1.038 & $-2.10\times10^{-4}$ & 0.01 & 0.570 & 0.671 & 8\\
& $n$ & 0.512 & -0.083 & 4.6 & 0.310 & 0.307 & 6.67 \\\hline
    \end{tabular}
\tablefoot{The median fit error was calculated based on all the error values for each of the fit parameters. As the fit error bars are asymmetric, the largest of the median values was taken in each case. The median $\Delta$, median absolute deviation (MAD) and median relative error were calculated from the difference between the best-fit result and the ``true'' value for each parameter. The percentage of outliers was calculated by assuming a normal distribution for the data and identifying the points more than 2 standard deviations away from the mean.
}
    \label{tab:mock}
\end{table*}

Overall, although all the simulation tests show discrepancies in the recovered stellar age and metallicity parameters (see Appendix \ref{app:B}), the fits show a remarkable recovery of the original dust parameters, as seen in Table \ref{tab:mock}. Across the board, we find relatively low root mean square ($\textrm{RMS}$) values for these parameters, as well as median relative errors of $\sim10\%$ or lower when comparing to the true simulation values, suggesting a high degree of accuracy. In addition, we find that the respective median fit errors are, in general, conservative, as they are very close to the dispersion (MAD) of the fitted parameters. There are, however, some biases observed, the most significant of which is the bias of $0.263$ in $\tau_V$ observed for the default simulation test. There is also a large bias of $0.261$ in $n$ in the Cardelli test, which is most likely due to the different parameterization for different attenuation laws. Further discussion of the various tests can be found in Appendix \ref{app:B}.

\par
In general, it is thus reasonable to conclude that, although our fitting method might struggle to break the dust-metallicity-age degeneracy and accurately determine all free fit parameters, resulting in not very robust non-dust parameters, it is sufficient to satisfactorily infer and map dust properties for SN host galaxies. As such, we can be fairly confident in the results presented and discussed in Section \ref{sec:des_fits}.

\section{DES SN Ia host galaxies}
\label{sec:des_fits}

In this section, we present and discuss the results of the CSP fits to the DES SN Ia host galaxies\footnote{Cornerplots and SED fit plots for the host galaxies can be found at \url{https://github.com/SN-CRISP/DES-SN_Host-Galaxies}.}, placing particular emphasis on the recovered attenuation laws. The fits were performed following the methodology described in Section \ref{subsec:fits} and using both global and local photometric data in the \textit{griz} filter bands, originally computed by \cite{Kelsey_2020}. Additionally, we compare the recovered host attenuation laws with literature low redshift SN Ia extinction laws.
\subsection{Dust attenuation for DES galaxies}
\label{sec:des_fits_attenuation}

\par
\begin{figure*}
        \includegraphics[width=1\textwidth]{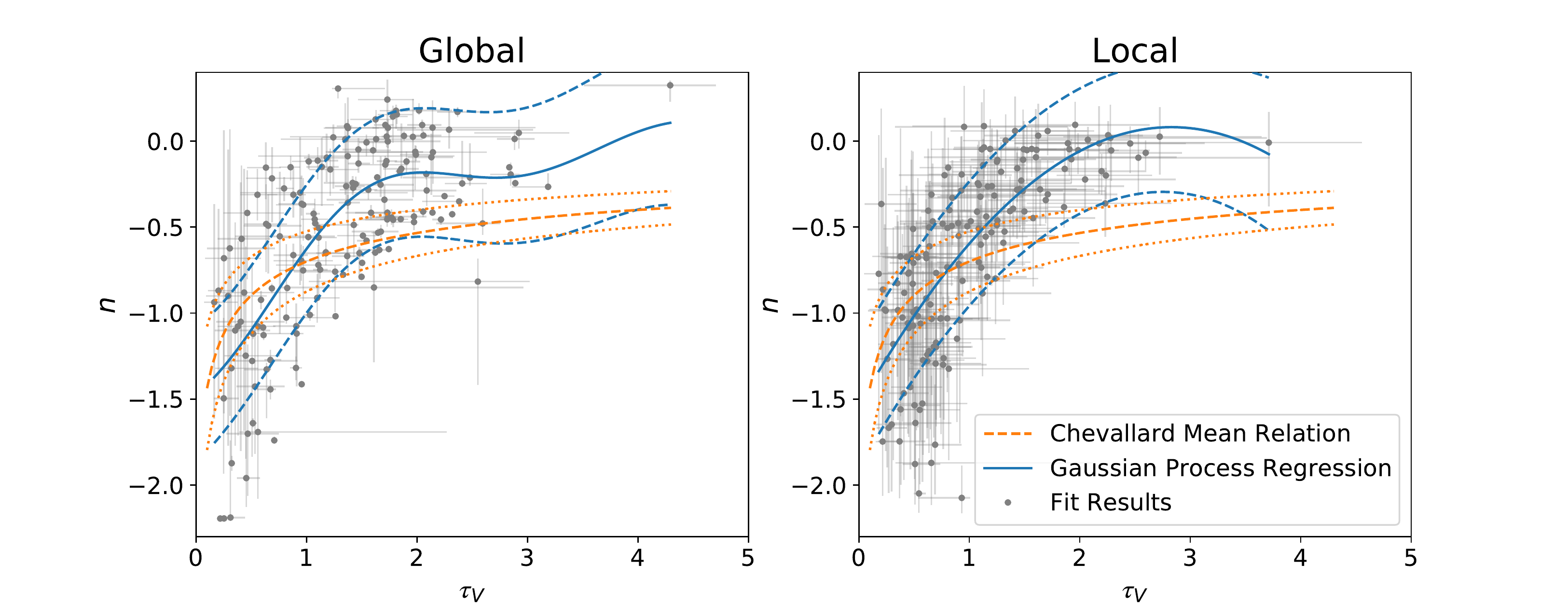}
        \caption{Best-fit values of $n$ as a function of $\tau_V$ for the fitted DES galaxies with DECam global (Left panel) and local (Right panel) \textit{griz} photometry. Results for the different galaxies are shown in gray. A Gaussian Process Regression is shown in solid blue, with dashed blue lines defining a 68\% credible interval. The mean relation found for similar simulated galaxies by \protect\cite{Chevallard_2013} is shown in dashed orange, with $\pm25\%$ error margins shown in dotted orange.}
        \label{fig:griz_av_n}
\end{figure*}

The central aspect in the characterization of the recovered attenuation laws is the correlation between the best-fit values for the $\tau_V$ and $n$ parameters, which are plotted in Fig.\ref{fig:griz_av_n}. A Gaussian Process Regression and its respective 68\% credible interval, performed using the {\sc GauPro}\footnote{\url{https://github.com/CollinErickson/GauPro}} \citep{gaupro} R package, are also plotted to better illustrate the observed correlation. \footnote{It should be noted that the regression curves are not representative of the data for $\tau_V>2.5$, due to the limited number of data points in this region.}

Despite some scatter, there is a clear correlation between the two quantities in both the global and the local cases, with larger optical depths corresponding to larger values of $n$, and thus shallower attenuation curves. The Spearman correlation coefficient between these two parameters for the DES data set is found to be 0.671 for the global measurements and 0.777 for the local measurements.
\par
\begin{figure*}
        \includegraphics[width=1\textwidth]{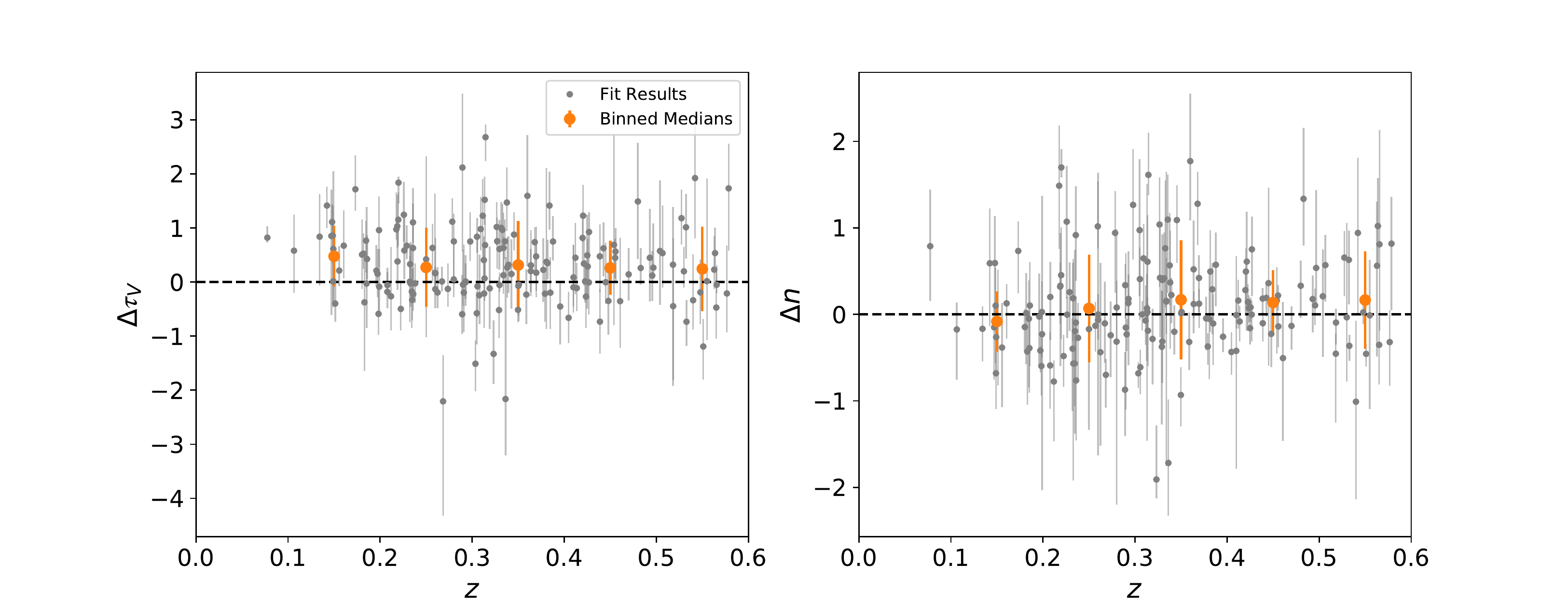}
        \caption{Differences ($\Delta=$ Global-Local) between best-fit results for the parameters obtained with global and local DECam \textit{griz} photometry for the DES galaxies, as a function of $z$. The left-hand side shows the results for $\tau_V$ and the right-hand side the ones for $n$. Results for the different galaxies are shown in gray. Binned means for each parameter are shown in orange, with error bars given by the standard deviation in each bin.}
\label{fig:griz_params_z}
\end{figure*}

Comparing the global and the local fit results offers an interesting insight. As seen in Fig. \ref{fig:griz_params_z}, where we plot the deviations $\Delta \tau_V=\tau_{V_{\textrm{Global}}}-\tau_{V_{\textrm{Local}}}$ and $\Delta n = n_{\textrm{Global}}-n_{\textrm{Local}}$ as a function of the redshift $z$, the results for both fits appear fairly consistent, with median differences of $\Delta \tau_V=0.26$ and $\Delta n=0.026$.
\par
There are some differences between the fits, as the values obtained for $\tau_V$ tend to be larger for the global photometry. There is also a large level of scatter present in the plots, with standard deviations of 0.712 for $\Delta \tau_V$ and 0.575 for $\Delta n$. Furthermore, the percentage of galaxies for which the best-fit values are not compatible with each other is found to be 39.8\% for $\tau_V$ and 36.0\% for $n$.
\par
The consistency of the observed correlation between the two result sets indicates that an aperture of 4 kpc is not sufficiently small to negate the geometrical effects associated with attenuation (see next section). Likely for the same reason, \cite{Kelsey_2020} observe a similar consistency between global and local measurements, particularly when looking at U-R steps. In fact, the global and the local mass values obtained by \cite{Kelsey_2020} indicate that, in some cases, a 4 kpc aperture is actually as large as the galaxy itself. This means one needs even smaller physical apertures to correctly estimate dust extinction at the SN location. However, this is difficult to achieve in practical terms due to instrument limitations. The fact that the results remain constant as the redshift increases indicates that the diminishing angular size of the aperture does not, in this case, play a large role in the results obtained.

 \par
\par
We used either complete or partial additional global photometry in the \textit{NUV/FUV} GALEX \citep{galex} and \textit{JHKs} 2MASS \citep{2mass} filter bands for 33 of the studied galaxies. Although this additional set of measurements is not enough to draw any conclusions by itself, it is useful to confirm our previous fit results. Combining the new photometry with the \textit{griz} global data points, a new set of fits was performed for these select galaxies. The differences between the new fit values for $\tau_V$ and $n$ and the previously discussed \textit{griz} results exhibit median relative values of $\sim 0.5\%$ and $\sim 3\%$, with a standard deviation of $1.14$ and $0.40$, respectively. Despite the somewhat large value of scatter for $\tau_V$, these values confirm the reliability of the results obtained with \textit{griz} measurements.
\subsection{Attenuation laws and galaxy orientation} 
\par
 
\begin{figure*}
        \centering
        \includegraphics[width=\textwidth]{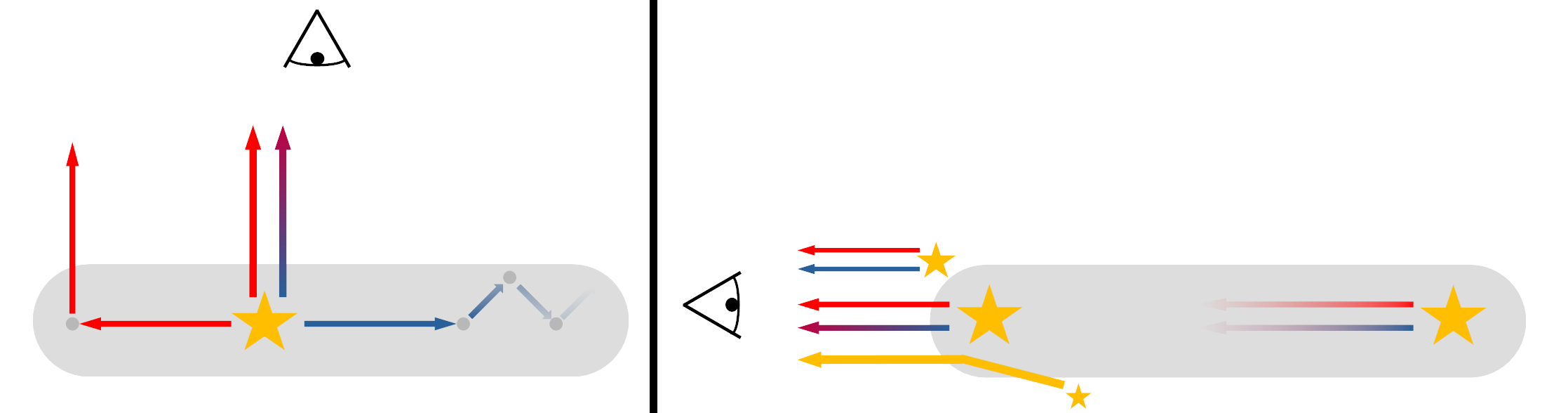}
        \caption{Schematic representation of dust attenuation for different galaxy orientations: Head-on (Left panel) and Edge-on (Right Panel).}
        \label{fig:dust_galaxy}
\end{figure*}

For comparison, the mean relation obtained by \cite{Chevallard_2013} for the overall correlation between $\tau_V$ and $n$ for a set of simulated galaxies observed at different orientation angles is also shown in Fig. \ref{fig:griz_av_n}. This relation, which results from a slightly different attenuation law than the one discussed in this paper, differs somewhat from our data, with the simulated galaxies appearing in general redder. Even so, the overall qualitative tendency is the same as the one found for DES galaxies. Additionally, the recovered tendency also matches the behavior observed by both \cite{Leja_2017} and \cite{Narayanan_2018}, which further suggests we are indeed recovering a physical relation. Given the agreement with the \cite{Chevallard_2013} mean relation, it is reasonable to conclude that the overall correlation between the recovered dust parameters can be well explained by different galaxy orientations, as also described by \cite{Viaene_2017} and \cite{Narayanan_2018} and schematically illustrated on Fig. \ref{fig:dust_galaxy}. This same trend between $\tau_V$ and $n$ is also present for elliptical galaxies, as recently shown by \cite{Sachdeva_2022}.

As stated by those works, cases with small $\tau_V$, would mostly correspond to galaxies that are being observed face-on. In this case, we must consider both the photons emitted parallel and perpendicularly to the galactic plane. Photons emitted perpendicularly to the galactic plane (along the line of sight for a face-on galaxy) suffer minimal attenuation, independent of their wavelength. Additionally, we must also take into account photons emitted along the equatorial plane of a galaxy, which might be scattered away from the plane and into the line of sight. On one hand, blue photons tend to be scattered more forwardly, meaning they tend to remain confined to the galactic plane, eventually being fully absorbed \citep{Chevallard_2013}. On the other hand, red photons tend to be scattered more isotropically and have a much lower interaction cross-section with dust, allowing them to more consistently escape the galactic plane \citep{Chevallard_2013}. This means that an excess of red radiation is added to our line of sight, leading to higher values of reddening, reflected in more negative values for $n$.
 \par

The cases with a larger $\tau_V$ statistically correspond to an edge-on view of the galaxies. In these cases, radiation emitted from the deepest layers of the galaxy is greatly absorbed, independently of wavelength. The radiation reaching an observer is thus dominated by stars located in the outermost layers of the galaxy, unobscured stars and light scattered into the line of sight \citep{Narayanan_2018}. This leads to an overall lower level of reddening and values of $n$ closer to $0$. 
\par
Galaxy orientation is not the only mechanism behind the correlation observed in Fig. \ref{fig:griz_av_n}, being mostly relevant for spiral galaxies. For elliptical galaxies, galaxy orientation is not an important factor. However, due to their general elliptical shape and overall low dust content, these particular galaxies end up exhibiting a behavior similar to the one observed for head-on spiral galaxies, leading to an overall large level of reddening. Additionally, different galaxies can have different intrinsic optical depths, which is one of the possible reasons for the scatter observed in the figure and the relations discussed in the next sections. It has also been recorded that there is some degree of degeneracy between galaxy geometry and intrinsic dust composition, which somewhat muddles the analysis \citep{Viaene_2017}.

\subsection{Attenuation laws and stellar mass}
\label{sec:fits_mass_dust}
\begin{figure*}
        \includegraphics[width=1\textwidth]{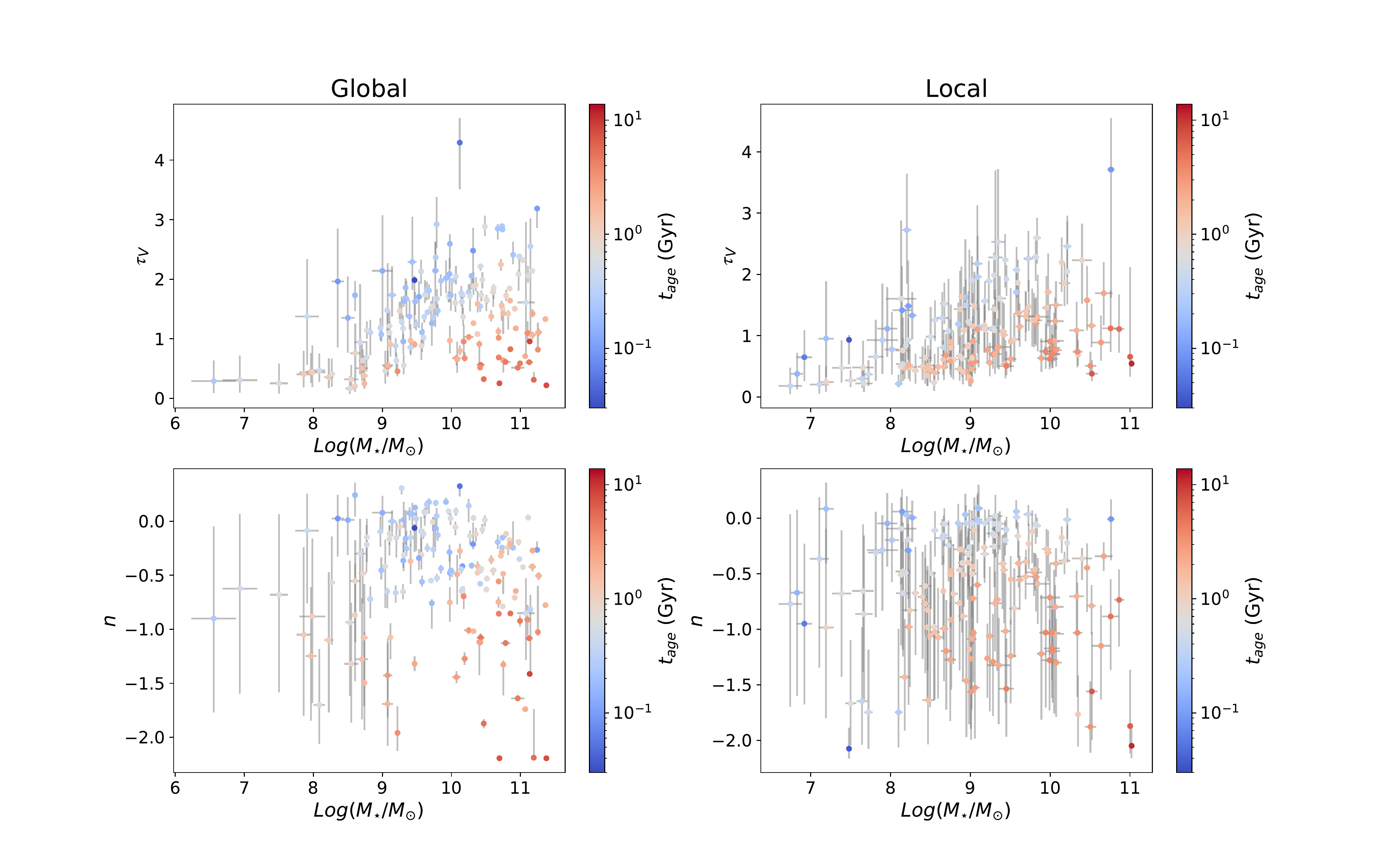}
        \caption{Best-fit results for the dust parameters as a function of $\log(M_\star/M_\odot)$ for the fitted DES galaxies with DECam global (Left panels) and local (Right panels)\textit{griz} photometry. Plots for $\tau_V$ are shown in the top row, while results for $n$ are shown in the bottom row. The best-fit for the age of each galaxy is plotted on a gradient from blue (younger) to red (older).}
\label{fig:griz_m_params}
\end{figure*}

Another important factor in the characterization of the recovered attenuation laws is the possible correlations of the dust parameters with the stellar mass and age, which are plotted in Fig. \ref{fig:griz_m_params}. Despite some scatter, the relation between $\tau_V$ and $\log(M_\star/M_\odot)$ seems to display two separate behaviors depending on the age of the corresponding stellar population. For younger galaxies, there is a steady increase of $\tau_V$ with $\log(M_\star/M_\odot)$, which is in agreement with results obtained for star-forming galaxies by \cite{Salim_2018,Garn_2010,Zahid_2013}, as more massive galaxies have more gas and dust content with more active star formation. A separate tendency is observed for older galaxies, which tend to exhibit both larger masses and smaller optical depths. This matches what is expected for quiescent galaxies \citep{Salim_2018,Zahid_2013}, which have less dust and make up a large fraction of high-mass galaxies \citep{Peng_2010}.
\par
The behavior of the two age populations is most prominent when looking at the global mass, but it is also apparent in the local case. This suggests a link not only with the total stellar mass of the galaxy, but also with the local stellar density, with older galaxies being generally more dense. It is also interesting to notice that a larger stellar density does not necessarily equate to a larger dust column density, as seen from the relatively low values of $\tau_V$ for high local masses. Although the large level of scatter complicates the analysis of the relation between $n$ and $\log(M_\star/M_\odot)$, we find that older galaxies tend to have more negative values of $n$ and thus steeper attenuation curve slopes, which once again matches previous observations for quiescent galaxies \citep{Salim_2018,wiseman_2022}. 

\par
\cite{Meldorf} examine the attenuation laws for a larger sample of DES SN Ia host galaxies and recover mostly the same trend for the relation between $n$ and $\log(M_\star/M_\odot)$. However, for the relation between $A_V$ and $\log(M_\star/M_\odot)$, they only observe the roughly linear trend associated with star-forming galaxies and do not find a decrease of $A_V$ for massive galaxies. As previously stated, this decrease is expected due to the presence of high-mass quiescent galaxies \citep[e.g.,][]{Zahid_2013,Salim_2018} and the trend we observe in Fig. \ref{fig:griz_m_params} follows exactly those expectations.
\par
For the global fit, we find a median $R_V=2.762$ (using Eq. \ref{eq:Calzetii_mod}) with a standard deviation of 1.096, while for the local case we find a median $R_V=2.612$ with a standard deviation of 0.943. In addition, in Table \ref{tab:rv_bins} we list the values of $R_V$ obtained for both low- and high-mass galaxies, when adopting a standard step of $\log(M_{\textrm{step}}/M_{\odot})=10$ for the global case and of $\log(M_{\textrm{step}}/M_{\odot})=9$ for the local case. These step locations are based on the median masses for both sets of measurements. We find that, for the global case, low-mass galaxies have a considerably larger median $R_V$, meaning that, for the same amount of dust, they appear to be systematically subject to a lower level of reddening. This once again points to the presence of massive quiescent galaxies in our sample, as for star-forming galaxies the value of $R_V$ is expected to increase with the stellar mass \citep{Salim_2018}. 
\par
Despite having standard deviations of $ R_V\sim 1$, both of the populations exhibit low standard errors, with the difference between the median values of $R_V$ for each of them being significant at $\sim 2.5 \sigma$. Additionally, although the values do not match exactly, this order relation between low- and high-mass dust properties and the observed level of scatter is the same as the one recovered by \cite{Brout_2021} and \cite{gaitan}.
\par
The results for the local case exhibit an opposite tendency, with a higher median $R_V$ for the high-mass population. This is probably due to the step location considered being too low, which does not allow the quiescent galaxies to meaningfully impact the high-mass population.

\begin{table*}
\centering
    \caption{Median global and local $R_V$ values, standard deviations ($STD$) and standard errors ($SE$) for low- and high-mass galaxies.}
    \begin{tabular}{ccccccccc}
    \hline
    
 \multirow{2}{*}{Standardization} &\multirow{2}{*}{Step}& Step Location & \multicolumn{3}{c}{Low-Mass} & \multicolumn{3}{c}{High-Mass} \\
& &$\log(M_{\textrm{step}}/M_{\odot})$ &$R_V$ & $STD$ & $SE$  & $R_V$ & $STD$ & $SE$\\\hline
\multirow{2}{*}{-} & Global & 10 & 3.092 & 1.155 & 0.128 & 2.670 & 0.966 & 0.109\\
& Local & 9 & 2.383 & 0.881 &0.104 & 2.658 & 0.978 & 0.104 \\ \hline
\multirow{2}{*}{Tripp} & Global & 9.73& 3.037 & 1.172 & 0.142 & 2.728 & 1.022 & 0.106\\
                        & Local & 9.405 & 2.625 & 0.958 & 0.094 &2.552 &0.907 & 0.121\\\hline

    \end{tabular}
\tablefoot{The thresholds for the two populations are defined as the optimal step value recovered for the Tripp standardization. Additionally, the values for the standard thresholds of $\log(M_{\textrm{step}}/M_{\odot})=10$ for the global case and of $\log(M_{\textrm{step}}/M_{\odot})=9$ for the local case are also shown.}
    \label{tab:rv_bins}
\end{table*}

\subsection{Host galaxy attenuation and SN Ia extinction}
\label{sec:ext_vs_att}
As discussed in Section \ref{sec:intro}, the phenomena governing dust extinction differ from those governing attenuation, as the geometrical effects related to the spatial distribution of dust and light sources cease to be relevant for the former, and only the amount and type of dust are important. It is thus crucial to take notice of the differences between the extinction laws affecting SNe Ia and the attenuation laws affecting their respective host galaxies.
\par
\begin{figure*}
        \includegraphics[width=1\textwidth]{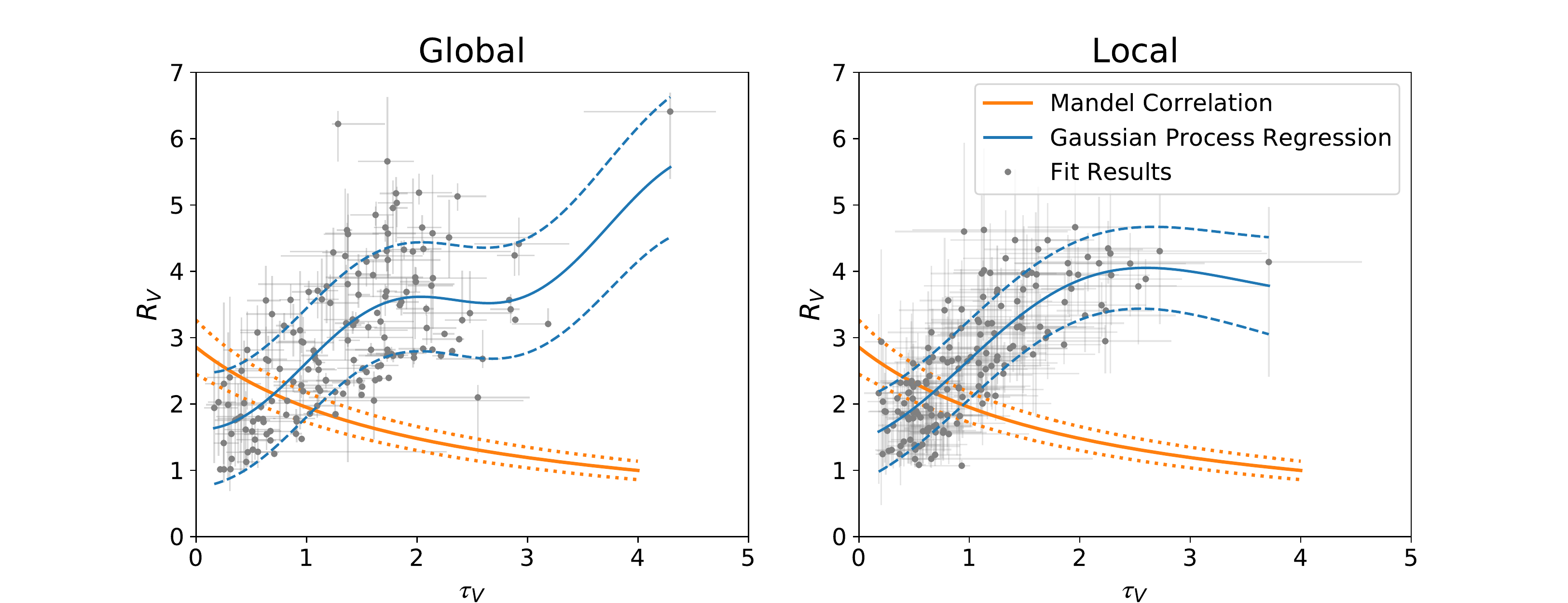}
        \caption{Best-fit values of $R_V$ as a function of $\tau_V$ for the fitted DES galaxies with DECam Global (Left panel) and Local (Right panel) \textit{griz} photometry. Results for the different galaxies are shown in gray. A Gaussian Process Regression is shown in solid blue, with dashed blue lines defining a 68\% credible interval. The correlation found by \protect\cite{Mandel_2011} between these two quantities for the extinction laws of a set of nearby SNe Ia is shown in orange.}
        \label{fig:griz_av_rv}
\end{figure*}

With this in mind, in Fig. \ref{fig:griz_av_rv} we plot the best-fit values for $R_V$ and $\tau_V$ obtained for the attenuation laws for the DES host galaxies. As a term of comparison, we look at the extinction laws obtained by \cite{Mandel_2011} for a set of SNe Ia at low redshift. We note that \cite{Burns_2014} find a similar relation to Mandel, whereas recently \cite{Rose_22} find a flat distribution of $R_V$ for various extinction amounts of SNe Ia.

\par
It is clear that the extinction for SNe Ia and the attenuation for host galaxies exhibit two different tendencies. This is due to the fact that, for a point source, the effects arising from the dust-to-star geometry and observation angle are no longer relevant. 

\par
As shown in Section \ref{sec:des_fits_attenuation}, the global and the local fit results are consistent, indicating that a 4 kpc aperture is not small enough to eliminate the dust-to-star geometry effects associated with dust attenuation.
\par
For a more direct comparison with our sample, we can look at the light-curve parameters recovered for each of the DES SN Ia by \cite{Brout_2019}. In particular, looking at the Spearman correlation coefficients between the host galaxy attenuation parameters and the respective SN Ia light-curve parameters, shown in Table \ref{tab:spearman}, it is clear that $\tau_V$ and $n$ do not meaningfully correlate with the SN properties. In particular, there is no evident correlation with light-curve color, which once again points to the difficulty of using environmental attenuation properties to infer SN dust extinction. These results show that there is not a one-to-one relation between SN extinction and host galaxy attenuation, meaning it is difficult to take one as a proxy for the other.
\begin{table}
\centering
    \caption{Spearman correlation coefficients between the global and the local host galaxy attenuation parameters and the respective SN Ia light-curve parameters.}
    \begin{tabular}{c c c c c}
    \hline
    \multirow{2}{*}{Light-Curve Parameter} & \multicolumn{2}{c}{Global} & \multicolumn{2}{c}{Local} \\
   & $\tau_V$ & $n$ & $\tau_V$ & $n$ \\ \hline
   $m_B$ & 0.082 & 0.277 &0.126 & 0.168 \\
   $c$ & 0.104 & 0.043 & 0.143 & 0.080 \\
   $x_1$ &0.041 &0.288 & 0.169 & 0.191 \\
    
    \hline

    \end{tabular}

    \label{tab:spearman}
\end{table}

\section{SN Ia cosmology}
\label{sec:hubble}
In this section, we focus on the effects of SN Ia standardization on the Hubble residuals. In particular, we examine whether there is evidence for a dust step in the data and how it compares to the more commonly used mass step. 
\subsection{SN Ia standardization}
\label{sec:Tripp_method}
For each SN Ia, the SALT2 model \citep{Guy_2007,Guy_2010} can be used to fit its light curve, returning values for the light-curve width $x_1$, observed color at maximum $c$, and observed magnitude at maximum $m_B$. Following the standardization formula introduced by \cite{Tripp98} with the addition of a $\delta_{\mu_{bias}}$ term, the corrected distance modulus $\mu$ for each SN is then given by:

\begin{equation}
    \mu=m_B-M+\alpha x_1 - \beta c +\delta_{\mu_{bias}},
    \label{eq:Tripp}
\end{equation}
\noindent
where $\alpha$, $\beta$ and $M$ are nuisance parameters obtained from a cosmological fit, describing the shape-luminosity and color-luminosity corrections, as well as the absolute magnitude of a fiducial SN Ia with $x_1 = 0$ and $c = 0$, respectively, and $\delta_{\mu_{bias}}$ is a 1D bias correction term, introduced as a function of the redshift and obtained from survey simulations. To account for the mass step, an additional correction term $\delta \mu_M$, dependent on the host galaxy mass $M_\star$, is usually added to the previous expression \citep[e.g.,][]{Betoule_2014}.

\par
For the cosmological fit, we take advantage of the fact that $\mu$ can be expressed as a function of the luminosity distance $d_L$ which, for a Friedmann-Robertson-Walker cosmology, only depends on the cosmological parameters and the redshift \citep{Riess98}. Generally, the model parameters $H_0$ and $\Omega_m$ are kept as free parameters, along with $\alpha$, $\beta$ and $M$. By minimizing the Hubble residuals $\Delta \mu=\mu-  \mu_{\textrm{model}}$ these parameters could therefore be constrained via the fit. In this work, the cosmological parameters are kept fixed to $\Omega_m=0.3$, $\Omega_\Lambda=0.7$, $w=-1$ and $H_{0}=70\SI{}{km\,s^{-1}\,Mpc^{-1}}$, as we are mainly interested in the effects of dust and the color-luminosity correction on the standardizations.

For the standardization fits we use the {\sc emcee} package to apply a Bayesian fitting procedure, with overall flat priors for the fit parameters. The likelihood function is defined as:
\begin{equation}
\label{eq:likelihood_cosmo}
    \ln{(\mathcal{L})}=-\frac{1}{2}\sum\limits_{i=0}^N \frac{\Delta \mu_i^2}{\sigma_i^2},
\end{equation}

\noindent
where $\sigma^2$ is defined as:

\begin{equation}
\begin{multlined}
    \sigma^2=\sigma_{m_B}^2+(\alpha \sigma_{x1})^2 +(\beta \sigma_c)^2+\sigma_{\textrm{int}}^2\\
    -2\beta\sigma_{m_B,c}+2\alpha\sigma_{m_B,x_1}-2\alpha\beta\sigma{x_1,c}.
    \label{eq:si_1}
\end{multlined}
\end{equation}

In the previous expression, $\sigma_{m_B}$, $\sigma_{x1}$ and $\sigma_{c}$ are the uncertainties associated with each of the light-curve fit parameters, $\sigma_{m_B,c}$, $\sigma_{m_B,x_1}$ and $\sigma_{x_1,c}$ are their covariance terms and $\sigma_{\textrm{int}}$ is a parameter that accounts for possible intrinsic variations in a SN Ia's luminosity. We fix $\sigma_{\textrm{int}}=0.107$, following the value obtained by \cite{gaitan}\footnote{A cosmological fit with a free $\sigma_{\textrm{int}}$ is explored in Appendix \ref{app:steps}.}.
\par
It is known that the likelihood defined by Eq. \ref{eq:likelihood_cosmo} can introduce biases into the recovered fit parameters \citep{Kessler_2017}. However, as our main objective is a wholesale comparison between two standardization processes using the same likelihood, these biases should not be of much relevance. This comparison is focused on the improvement of the RMS of the Hubble residuals with respect to the fiducial model.
\par
The cosmological fit parameters recovered following this method for the Tripp standardization are listed in Table. \ref{tab:cosmo_fit_1}, with the corresponding values of $\textrm{RMS}$ for the residuals also shown. The light-curve parameters recovered for each of the DES SN Ia by \cite{Brout_2019} using the SALT2 model, as well as the corresponding 1D bias corrections, were used for the fits.

\par
A separate ``Fixed-Extinction'' standardization, in which the extinction contribution to the color-luminosity correction can be fixed using our prior measurements of $R_V$ and $E(B-V)$ for the host galaxies from Section \ref{sec:des_fits}, is described in Appendix \ref{app:fixed_ext}. For this standardization, an additional free intrinsic color-luminosity parameter $\beta_\textrm{int}$ is introduced. The corresponding fit parameters for this standardization are also displayed in Table. \ref{tab:cosmo_fit_3}. It however results in much worse fits, as can be seen in the large $\textrm{RMS}$.

\renewcommand{\arraystretch}{1.75}
\begin{table}
\centering
    \caption{Fit parameters for the Tripp standardization, using a likelihood defined by Eq. \ref{eq:likelihood_cosmo}. The corresponding value of $\textrm{RMS}$ is also shown.}
    \begin{tabular}{c c c c c c}
    \hline
 $\alpha$ & $\beta$ & $\beta_{\textrm{int}}$ & $M$ & $\textrm{RMS}$  \\ \hline
        $0.157^{+0.011}_{-0.009}$ & $3.115^{+0.009}_{-0.011}$ & - & $-19.397^{+0.009}_{-0.009}$ & $0.141$ \\\hline
    \end{tabular}

    \label{tab:cosmo_fit_1}
\end{table}

\subsection{Hubble residuals steps}
\label{sec:steps}
We now examine possible ``steps'' in the Hubble residuals for the Tripp standardization, associated with both the local and global properties of the host. We investigate ``steps'' associated with both the host mass and dust properties.
\par
To look for a step, we divide the Hubble residuals into two populations, according to the properties of the SN host. By varying the threshold values at which the population division is made, we can determine the optimal location for the step. This can be done by optimizing several quantities, namely the step magnitude, the step significance and the $\Delta \textrm{RMS}$ with respect to the fiducial Tripp standardization. 
\par
For a description with two populations A and B, the step magnitude $\gamma$ is defined as the difference between the mean values of the Hubble residuals for each of the two populations, while the step significance is defined as the ratio between the step magnitude and the step error $\sigma_{\textrm{step}}$, given by:

\begin{equation}
    \sigma_{\textrm{step}}=\sqrt{\frac{\sigma_A^2}{N_A}+\frac{\sigma_B^2}{N_B}},
\end{equation}
\noindent
where $N_A$ and $N_B$ are the number of data points in each of the populations and $\sigma_A$ and $\sigma_B$ are the STDs for each of the populations.

\par
$\Delta \textrm{RMS}$ is defined as the difference between the standard $\textrm{RMS}_{\textrm{single}}$ obtained when considering the Hubble residuals as a single population and the $\textrm{RMS}_{\textrm{dual}}$ obtained when considering them as two separate populations, each with a separate mean and divided at the step in mass or dust. Thus, the larger the value of $\Delta \textrm{RMS}$, the more the two-population description is favored.

\subsubsection{The mass step}
\label{sec:mass_steps}
\begin{figure*}
        \centering
        \includegraphics[width=\textwidth]{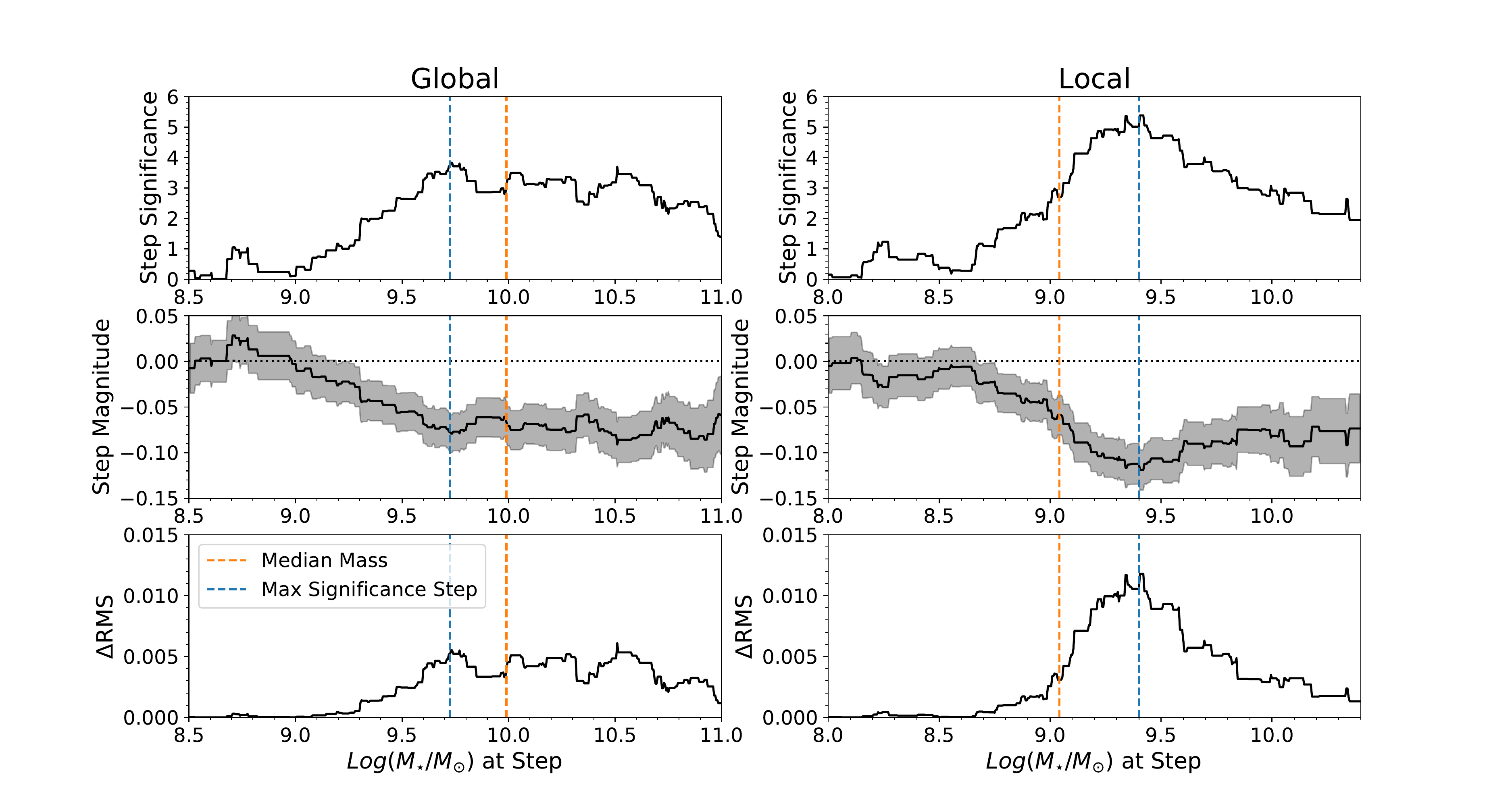}
        \caption{Variation of the mass step with step location for the Tripp standardization, for the global (Left panels) and the local (Right panels) cases. The top panel shows the significance of the step in $\sigma$, the middle panel shows the magnitude of the step in solid black, with the gray region showing the uncertainty, and the lower panel shows $\Delta \textrm{RMS}$. In these three panels, the location of the step of maximum significance is shown in blue and the median mass of the sample is shown in orange. Inspired by \protect\cite{Kelsey_2020}.}
\label{fig:hubble_res_beta_des}
\end{figure*}

\begin{figure*}
        \centering
        \includegraphics[width=\textwidth]{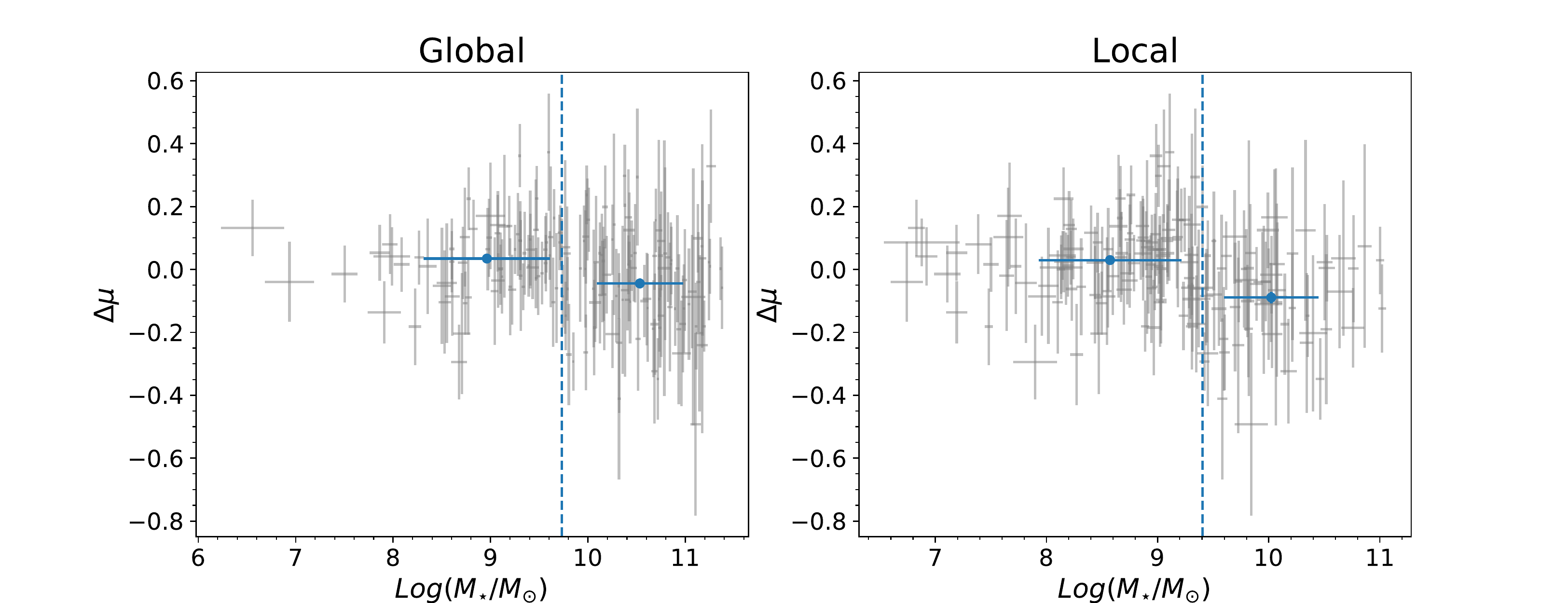}
        \caption{Hubble residuals $\Delta \mu$ for the DES SNe Ia as a function of the stellar mass for the respective host galaxies, plotted in gray. The optimal step locations, as recorded in Table. \ref{tab:mass_steps} are shown in dashed blue. The blue dots show the mean values of $\log(M_\star/M_\odot)$ and $\Delta \mu$ in each bin, with the error in the x-axis given by the standard deviation.}
\label{fig:hubble_res_plot}
\end{figure*}

When considering a mass step, the two populations are taken to be SNe Ia originating in either low- or high-mass galaxies. Following the procedure detailed in the previous section, we look for a division in mass between two populations that maximizes the step magnitude, its significance and $\Delta \textrm{RMS}$. The optimal step values obtained are shown in Table \ref{tab:mass_steps}, with the corresponding values of $\textrm{RMS}_{\textrm{dual}}$ also shown. The optimal values are also plotted in Fig. \ref{fig:hubble_res_beta_des}. In addition, in Fig. \ref{fig:hubble_res_plot} the Hubble residuals for the SNe are plotted, as well as the corresponding optimal steps and the mean values for $\log(M_\star/M_\odot)$ and $\Delta \mu$ found for each mass bin.

\begin{table*}
\centering
    \caption{Mass-step significance (in $\sigma$), magnitude ($\gamma_M$), $\Delta \textrm{RMS}$ and $\textrm{RMS}_{\textrm{dual}}$ for the Tripp and Tripp+$\delta \mu_D$ standardizations. }
    \begin{tabular}{c c c c c c c}
    \hline
          \multirow{2}{*}{Standardization} & \multirow{2}{*}{Step} & Step Location  & \multirow{2}{*}{Sig. (in $\sigma$)} & \multirow{2}{*}{$\gamma_M$} & \multirow{2}{*}{$\Delta \textrm{RMS}$} & \multirow{2}{*}{$\textrm{RMS}_{\textrm{dual}}$} \\ 
          & & $\log(M_{\textrm{step}}/M_{\odot})$ \\\hline
         \multirow{2}{*}{Tripp} & Global & 9.73 & 3.83 & $-0.079\pm0.021$ & 0.0055 & 0.1359 \\
         & Local &9.405 & 5.39 & $-0.119\pm0.022$ & 0.0118 & 0.1296\\ \hline
         \multirow{2}{*}{Tripp+$\delta \mu_D$} & Global & 9.73 & 2.99 & $-0.060\pm0.020$ & 0.0033 & 0.1310\\
         & Local & 9.405 & 4.31 & $-0.092\pm0.021$ & 0.0075 & 0.1253 \\
         \hline

    \end{tabular}
\tablefoot{For the Tripp+$\delta \mu_D$ standardization, the step location and magnitude recovered for the optimal dust step found in Section \ref{sec:dust_steps} are used. $\Delta \textrm{RMS}$ is defined as the difference between the $\textrm{RMS}_{\textrm{single}}$ obtained when considering the Hubble residuals as a single population and the $\textrm{RMS}_{\textrm{dual}}$ obtained when considering them as two separate populations, each with a separate mean and divided at the step.}
    \label{tab:mass_steps}
\end{table*}

\par
We obtain optimal steps that are not only both significant at $>3\sigma$, but also correspond to the best improvement in $\textrm{RMS}$. Additionally, both step locations and their respective significance levels roughly match the results obtained by \cite{Kelsey_2020} for this data set, pointing to the existence of a mass step for the Tripp standardization.

\par

In Table \ref{tab:rv_bins} we present the $R_V$ values for both low- and high-mass galaxy populations, as defined by the step values recovered for the Tripp standardization. We once again find that, for the global case, low-mass galaxies have a larger median $R_V$, with the difference between the two populations being significant at $\sim 2 \sigma$. We also find that, as the local step location shifts to a higher mass, the same tendency starts to be observed in this case. In the next section, we further explore the discrepancy between these populations, analyzing whether it can be described by a dust step.

\subsubsection{The dust step}
\label{sec:dust_steps}
Having established the presence of a mass step, we next look at whether evidence for a step related to the dust content of a SN host can be found in the data for the Tripp standardization. There are some differences to the approach discussed for the mass step. First, while the stellar mass of a galaxy can be described by a single parameter, $M_\star$, dust attenuation is parameterized by two parameters, $\tau_V$ and $n$. To account for this fact and ensure we are accurately accounting for a possible dust step, we adopt a two-dimensional population division. One-dimensional steps in both $\tau_V$ and $n$ were attempted, but they were found not to be very significant ($\sim 2 \sigma$). Second, we consider three different population splits to fully cover the $\tau_V-n$ parameter space. These are chosen to preserve a two-population division, thus making the mass and dust steps more comparable. 
\par
\begin{figure}
        \includegraphics[width=0.5\textwidth]{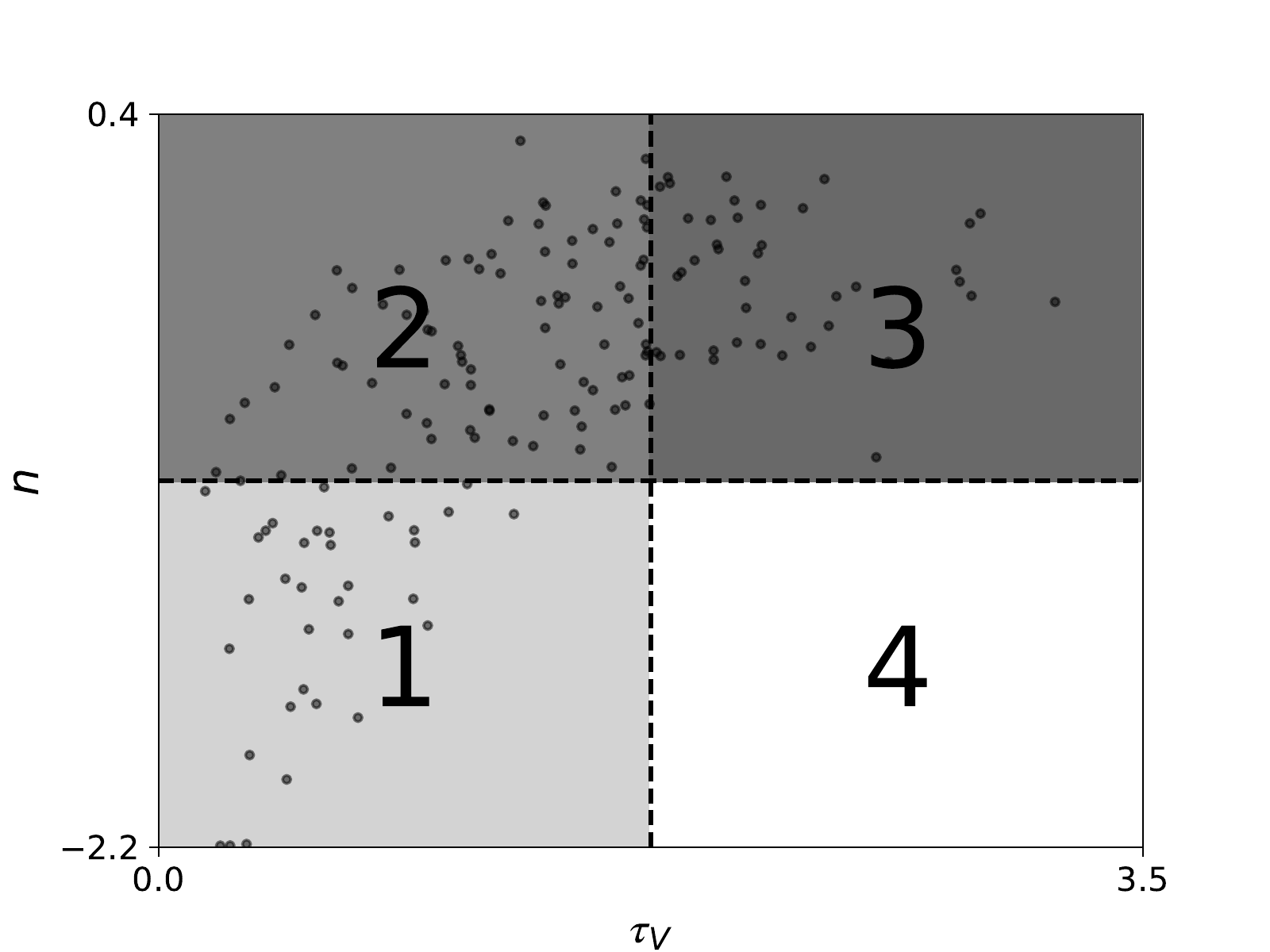}
        \caption{Schematic representations of the regions singled out by each of the proposed dust-step splits in the $\tau_V-n$ parameter space. The best-fit values of $n$ as a function of $\tau_V$ for the fitted DES galaxies with global \textit{griz} photometry are shown for reference. The dashed lines only represent possible division points for the populations, with the actual optimal values recorded in in Table. \ref{tab:dust_steps}.}
        \label{fig:regions}
\end{figure}

The three splits, schematically represented in Fig. \ref{fig:regions}, are: one population consisting of SNe Ia originating in low attenuation, high reddening galaxies (Region 1 in Fig. \ref{fig:regions}) and another consisting of the remaining SNe, defined as the First Split; one population consisting of SNe Ia originating in low attenuation, low reddening galaxies (Region 2 in Fig. \ref{fig:regions}) and another consisting of the remaining SNe, defined as the Second Split; one population consisting of SNe Ia originating in high attenuation, low reddening galaxies (Region 3 in Fig. \ref{fig:regions}) and another consisting of the remaining SNe, defined as the Third Split. There is also a fourth possible split that is not considered, as the region with high attenuation and high reddening galaxies (Region 4 in Fig. \ref{fig:regions}) is very sparsely populated.

\par

\begin{figure*}
        \centering
        \includegraphics[width=\textwidth]{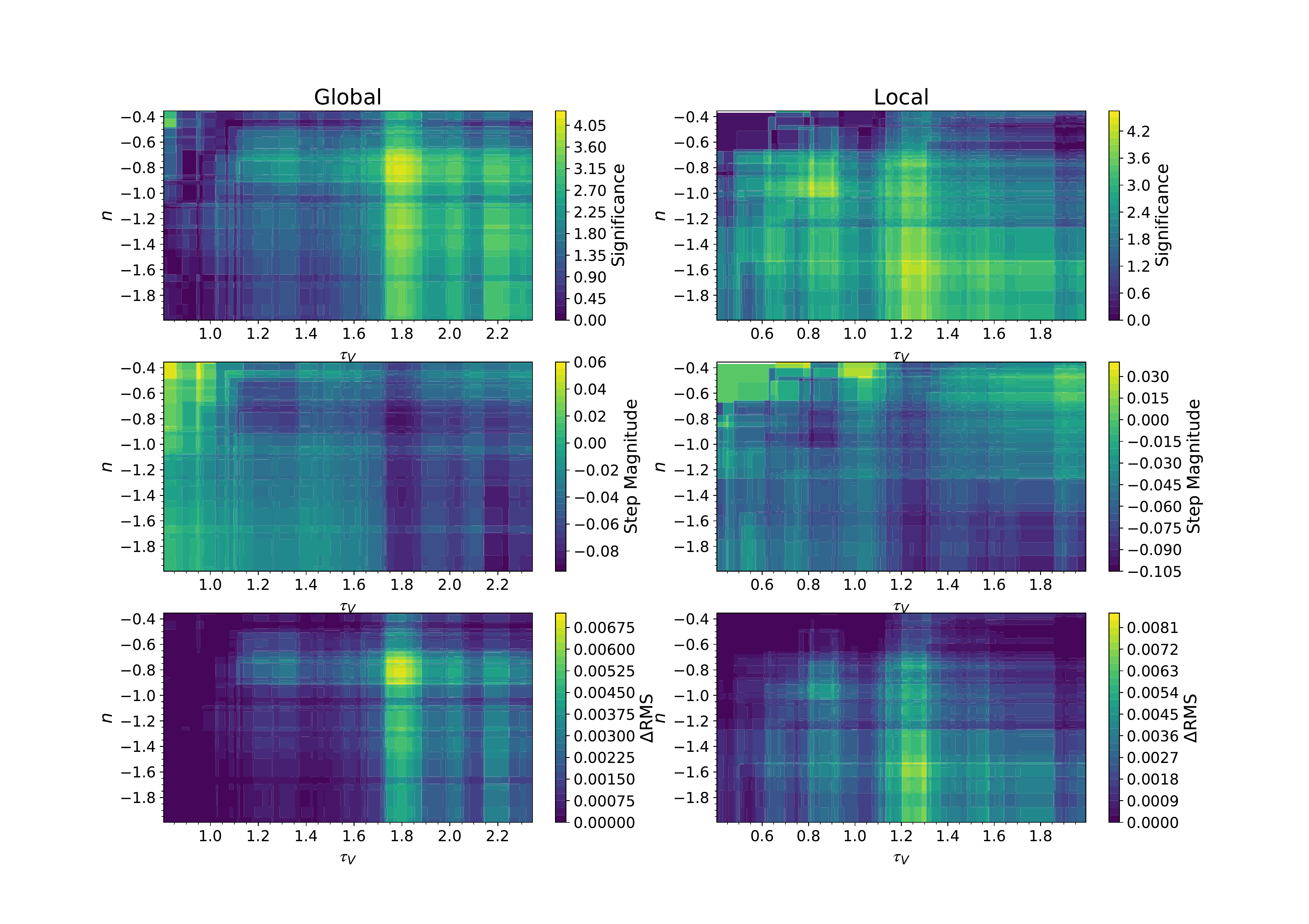}
        \caption{Variation of the second split dust step with the step location (position in $\tau_V-n$ grid) for the Tripp standardization. Results from global measurements are shown on the left, while those from local measurements are shown on the right. The top panel shows the significance of the step in $\sigma$, the middle panel shows the magnitude of the step and the lower panel shows $\Delta \textrm{RMS}$. This last quantity is defined as the difference between the $\textrm{RMS}_{\textrm{single}}$ obtained for a single population mean description and the $\textrm{RMS}_{\textrm{dual}}$ obtained for a dual population mean description with a step.}
\label{fig:dust_index_mu_2nd}
\end{figure*}

 The optimal step values for the First, Second and Third Splits are shown in Table \ref{tab:dust_steps}. The Second Split results in the best significance and $\Delta \textrm{RMS}$ values, which means it is by far the most favorable for a dust step. We can further analyze this particular case by taking a closer look at the variation of the step magnitude, significance and $\Delta \textrm{RMS}$, plotted in Fig. \ref{fig:dust_index_mu_2nd}. We can see that, not only are the optimal step locations very narrow, but they also exhibit clear delimitations for both dust parameters, suggesting we are indeed encountering a two-dimensional step. This is particularly apparent when using global data. In the local case, there appear to be two possible step locations. We take the optimal step to be the one that maximizes $\Delta \textrm{RMS}$. The significance, magnitude and $\Delta \textrm{RMS}$ values obtained for the optimal global dust step are very close to those recovered for the optimal global mass step, even though the dust step ends up being slightly stronger. In particular, the dust steps are significant at $>4\sigma$, which matches the results obtained by \cite{Meldorf} for DES-SN host galaxies.

\begin{table*}
\centering
    \caption{Dust-step significance (in $\sigma$), magnitude ($\gamma_D$), $\Delta \textrm{RMS}$ and $\textrm{RMS}_{\textrm{dual}}$ for the Tripp and Tripp+$\delta \mu_M$ standardizations. }
    \begin{tabular}{c c c c c c c c c }
    \hline
          \multirow{2}{*}{Standardization} & \multirow{2}{*}{Split} & \multirow{2}{*}{Step} & \multicolumn{2}{c}{Step Location}  & \multirow{2}{*}{Sig. (in $\sigma$)} & \multirow{2}{*}{$\gamma_D$} & \multirow{2}{*}{$\Delta \textrm{RMS}$} & \multirow{2}{*}{$\textrm{RMS}_{\textrm{dual}}$}  \\ 
          & & & $\tau_V$ & $n$\\\hline
        \multirow{6}{*}{Tripp} &  \multirow{2}{*}{First} & Global & 0.56 & -1.25 & 1.81 & $0.064\pm0.035$ & 0.0009 & 0.1405\\
        & & Local & 1.29 & -0.105 & 2.89 & $-0.073\pm0.025$ & 0.0041 & 0.1373\\
        & \multirow{2}{*}{Second} & Global & 1.785 & -0.905 & 4.20 & $-0.089\pm0.021$ & 0.0071 &0.1343\\
        & & Local & 1.29 &-1.53 & 4.55 & $-0.101\pm0.022$ & 0.0086 & 0.1328 \\
        & \multirow{2}{*}{Third} & Global & 1.785 & -0.82 & 3.45 & $0.085\pm0.025$ & 0.0047 & 0.1367\\
        & & Local & 1.29 & -0.595 & 3.76 & $0.094\pm0.025$ & 0.0065 & 0.1349 \\\hline
        \multirow{2}{*}{Tripp+$\delta \mu_M$} & \multirow{2}{*}{Second} & Global &1.785 &-0.855 & 3.52& $-0.072\pm0.021$& 0.0049 & 0.1310 \\
        & & Local &1.29 &-1.53 &3.36 &$-0.070\pm0.021$ & 0.0045 & 0.1251 \\
        
        \hline
    \end{tabular}
\tablefoot{For the Tripp+$\delta \mu_M$ standardization, the step location and magnitude recovered for the optimal mass step found in Section \ref{sec:mass_steps} are used. $\Delta \textrm{RMS}$ is defined as the difference between the $\textrm{RMS}_{\textrm{single}}$ obtained when considering the Hubble residuals as a single population and the $\textrm{RMS}_{\textrm{dual}}$ obtained when considering them as two separate populations, each with a separate mean and divided at the step. The First, Second and Third splits refer to which of the quadrants depicted in Fig. \ref{fig:regions} is isolated from the remaining SN population (1,2 and 3, respectively).}
    \label{tab:dust_steps}
\end{table*}

\begin{figure*}
        \centering
        \includegraphics[width=\textwidth]{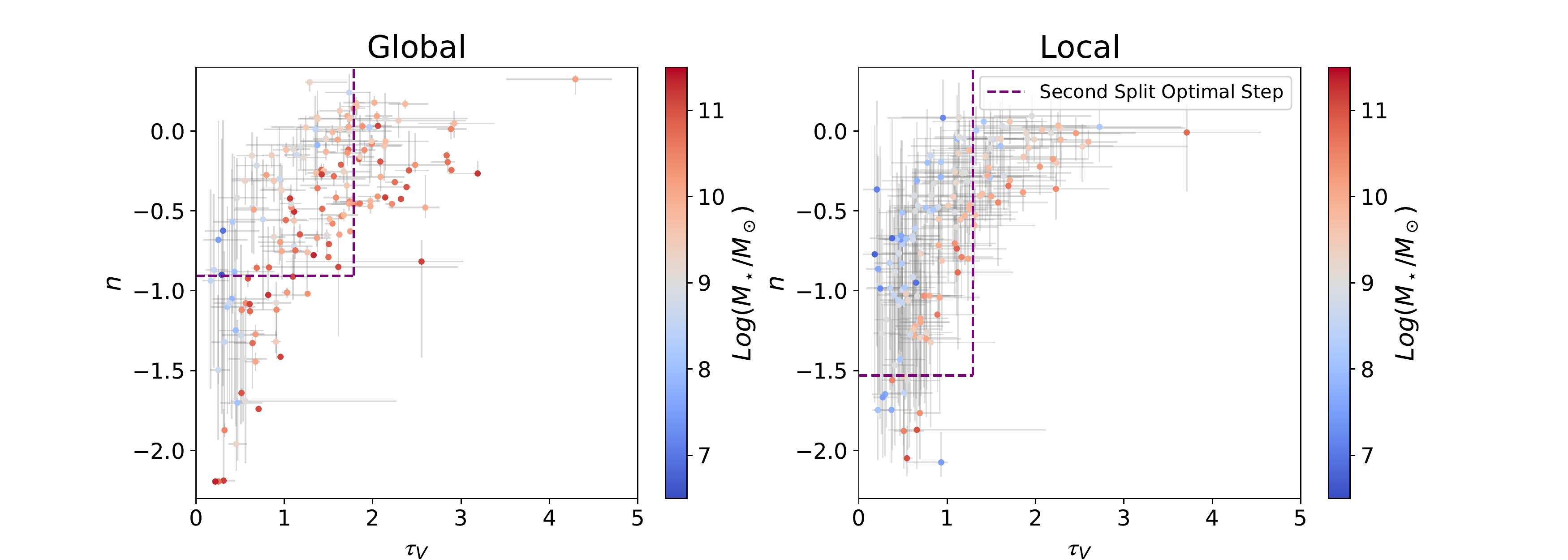}
        \caption{Best-fit values of $n$ as a function of $\tau_V$ for the fitted DES galaxies with DECam global (Left panel) and local (Right panel) \textit{griz} photometry. The corresponding value of $\log(M_\star/M_\odot)$ is plotted on a gradient from blue (lower mass) to red (higher mass). The optimal step for the Second Split is also shown in purple.}
\label{fig:dust_index_mu_mass}
\end{figure*}
\subsubsection{Mass and dust step comparison}
\label{sec:both_steps}
\par
For a more rigorous SN standardization, the steps recovered in the previous sections can be introduced into the usual Tripp standardization. In particular, we do this to ascertain whether correcting for one of the steps eliminates the other one. In such cases, the mass step takes the form:
\begin{equation}
\label{eq:mass_step}
  \delta \mu_M= 
\begin{cases}
-\frac{\gamma_{M}}{2}, &\mbox{if } \log(M_{\star}/M_{\odot}) \geq \log(M_{\textrm{step}}/M_{\odot}) \\
\frac{\gamma_{M}}{2},  &\mbox{if } \log(M_{\star}/M_{\odot})<\log(M_{\textrm{step}}/M_{\odot})
\end{cases},  
\end{equation}
\noindent
where $M_{\textrm{step}}$ is the optimal step location and $\gamma_{M}$ the corresponding step magnitude recovered in Section \ref{sec:mass_steps} and recorded in Table \ref{tab:mass_steps}. For the Second Split, the dust step takes the form:

\begin{equation}
\label{eq:dust_step}
\quad \delta \mu_D= 
\begin{cases}
\frac{\gamma_{D}}{2}, &\mbox{if } \tau_V < \tau_{V_{\textrm{step}}} \textrm{ and } n>n_{\textrm{step}}\\
-\frac{\gamma_{D}}{2},  &\mbox{otherwise}
\end{cases},
\end{equation}

\noindent
where $\tau_{V_{step}}$ and $n_{step}$ are the optimal step locations and $\gamma_{D}$ the corresponding step magnitude recovered in Section \ref{sec:dust_steps} and recorded in Table \ref{tab:dust_steps}.

 It is clear from Fig. \ref{fig:dust_index_mu_mass} that the two mass and dust populations do not exactly align. In fact, higher-mass galaxies tend to concentrate on the inside of the $\tau_V-n$ curve, while less massive galaxies concentrate on the outside. This is particularly apparent in the local case.
 This discrepancy points to the fact that the recovered dust step is not merely analogous to the mass step, as might initially be presumed, given its magnitude, significance and $\Delta \textrm{RMS}$.
 \par
For a more detailed look at this discrepancy, we can add a dust step in the form of Eq. \ref{eq:dust_step} to the Hubble residuals for the Tripp standardization. To do so, we fix the values of $\gamma_D$, $\tau_{V_{\textrm{step}}}$ and $n_{\textrm{step}}$ to those previously obtained in Section \ref{sec:dust_steps} for the optimal dust step, as recorded in Table \ref{tab:dust_steps}. Following a similar procedure as the one applied in Section \ref{sec:mass_steps}, we once again recover the optimal mass step, shown in Table \ref{tab:mass_steps}. It should be noted that, when looking for this new mass step, the baseline for comparison is not the case with no step, but rather the case already corrected for the dust step.
\par
For the local case, the optimal mass-step location remains unchanged. For the global case, however, there is a big change in the maximum significance step location, which happens because the previous optimal step significance drops slightly below the level of the new maximum. For the sake of consistency, we keep the same step location. Even so, the new maximum significance step, $\log(M_\star/M_\odot)=10.51$, has a significance of 3.20.
\par
It is clear that the introduction of the dust-step correction lowers both the significance, magnitude and $\Delta \textrm{RMS}$ of the recovered mass steps. This drop in $\Delta \textrm{RMS}$ should not be taken to mean that the inclusion of both steps results in an overall worse RMS when compared with a single step case. Rather, it signifies that the addition of a mass step to a set of Hubble residuals already corrected with a dust step results in a lower improvement in RMS than the one seen for the same mass step applied to Hubble residuals with no prior step correction. In this sense, the initial dust-step correction appears to also be partially correcting for the mass effect. However, it cannot be said that the mass step has been removed in either case. Thus, while offering some improvement, the addition of this dust-step term to the standardization is not capable of fully eliminating the mass step by itself.
\par
In the same way, we can look at the recovered dust step in the case of a Tripp+mass-step standardization, in which the mass step recovered in Section \ref{sec:mass_steps} is used. The values relating to these steps are shown in Table \ref{tab:dust_steps}. Once again, it is clear that the structure of the original steps is preserved, while the significance, magnitude and $\Delta \textrm{RMS}$ are again found to be lower. Thus, the addition of a mass-step term in the standardization is also not capable of fully eliminating the dust step by itself.
\par

\par
Thus, it seems reasonable to conclude that the mass and dust steps are two overlapping phenomena that nevertheless do not fully coincide. This points to the fact that the origin of the mass step may not be entirely driven by host dust properties. A further exploration of the relation between the two steps is offered in Appendix \ref{app:steps}, where both $\mu_M$ and $\mu_D$ are treated as free parameters in the cosmological fit. This confirms that the two steps are not analogous, as both are significantly different from zero for the best-fit cosmological model.

\par
Looking at a bigger sample of DES host galaxies that includes SN Ia photometry, \cite{Meldorf} find that the inclusion of a dust-step correction is sufficient to correct the mass step. The discrepancy between this and our results is most likely due to the different mass to dust relations recovered from the host galaxy fits, as discussed in Section \ref{sec:fits_mass_dust}.

\subsubsection{SN color and Hubble residuals steps}
\label{sec:color_steps}

Although we do not find any direct relation between SN color and the galaxy dust parameters (Section \ref{sec:ext_vs_att}), we explore the relation between SN color and the Hubble residuals, similarly to what is done in \cite[e.g.,][]{Brout_2021}. The SNe were divided into two populations: Blue SNe ($c<0$) and Red SNe ($c>0$). For an easier comparison with the previous results, the optimal locations found in Sections \ref{sec:mass_steps} and \ref{sec:dust_steps} were preserved. These results are shown in Tables \ref{tab:mass_steps_color} and \ref{tab:dust_steps_color}, for the mass and dust steps, respectively.
\par

It is clear from the significance, magnitude and $\Delta \textrm{RMS}$ values that both steps are much more pronounced for the red SNe, which matches the results obtained by \cite{Kelsey_2020} and \cite{Brout_2021}. This difference between the two populations points once again to a relation between dust reddening and the Hubble residuals steps, assuming most of the color variation for SNe stems from dust. However, it has also been suggested that there is possible indirect evidence of environmental variation of intrinsic SN color, particularly related to ejecta velocity \citep[e.g.,][]{Pang_2020,Wang_2013,Cartier_2011}, which could also be playing a partial role in this case.

\begin{table*}
\centering
    \caption{Mass-step significance (in $\sigma$), magnitude ($\gamma_M$), $\Delta \textrm{RMS}$ and $\textrm{RMS}_{\textrm{dual}}$ for different color populations.}

    \begin{tabular}{c c c c c c c c }
    \hline
          \multirow{2}{*}{Standardization} &\multirow{2}{*}{SN Color} & \multirow{2}{*}{Step} & Step Location  & \multirow{2}{*}{Sig. (in $\sigma$)} & \multirow{2}{*}{$\gamma_M$} & \multirow{2}{*}{$\Delta \textrm{RMS}$ } &  \multirow{2}{*}{$\textrm{RMS}_{\textrm{dual}}$}\\ 
         & & & $\log(M_{\textrm{step}}/M_{\odot})$ \\\hline
    \multirow{4}{*}{Tripp} & \multirow{2}{*}{Blue ($c<0$)} & Global & 9.73 & 2.12 & $-0.050\pm0.023$ & 0.0025 & 0.1241\\
       &  & Local &9.405 & 2.73 & $-0.083\pm0.031$ & 0.0059 & 0.1207\\
         & \multirow{2}{*} {Red ($c>0$)} & Global & 9.73 & 3.34 & $-0.128\pm0.038$ & 0.0125 & 0.1504\\
        & & Local & 9.405 &  5.00 & $-0.172\pm0.034$ & 0.0240 & 0.1389\\ \hline \hline
       \multirow{4}{*}{Tripp+$\delta \mu_D$} &  \multirow{2}{*}{Blue ($c<0$)} & Global & 9.73 & 1.43 & $-0.033\pm0.023$ & 0.0011 & 0.1215 \\
        & & Local &9.405 & 2.36 & $-0.071\pm0.030$ & 0.0044 & 0.1183\\
         &\multirow{2}{*} {Red ($c>0$)} & Global & 9.73 & 3.19 & $-0.110\pm0.034$ & 0.0100 &0.1393 \\
         && Local & 9.405 &  3.36 & $-0.111\pm0.033$ & 0.0110 & 0.1324\\ \hline

    \end{tabular}
    \tablefoot{The top rows show the results pertaining to the Tripp standardization, while the bottom rows show the results pertaining to the Tripp+$\delta \mu_D$. In both cases, the same step locations obtained in Section \ref{sec:mass_steps} are used. $\Delta \textrm{RMS}$ is defined as the difference between the $\textrm{RMS}_{\textrm{single}}$ obtained when considering the Hubble residuals as a single population and the $\textrm{RMS}_{\textrm{dual}}$ obtained when considering them as two separate populations, each with a separate mean and divided at the step.}
    \label{tab:mass_steps_color}
\end{table*}

\begin{table*}
\centering
    \caption{Dust-step significance (in $\sigma$), magnitude ($\gamma_D$), $\Delta \textrm{RMS}$ and $\textrm{RMS}_{\textrm{dual}}$ for different color populations. }
    \begin{tabular}{c c c c c c  c c c }
    \hline
          \multirow{2}{*}{Standardization} &\multirow{2}{*}{SN Color}  & \multirow{2}{*}{Step} & \multicolumn{2}{c}{Step Location}  & \multirow{2}{*}{Sig. (in $\sigma$)} & \multirow{2}{*}{$\gamma_D$} & \multirow{2}{*}{$\Delta \textrm{RMS}$} &\multirow{2}{*}{$\textrm{RMS}_{\textrm{dual}}$} \\ 
          && & $\tau_V$ & $n$\\\hline

        \multirow{4}{*}{Tripp} & \multirow{2}{*}{Blue ($c<0$)} &  Global & 1.785 & -0.905 & 2.53 & $-0.064\pm0.025$ & 0.0040 & 0.1226 \\
        &&  Local & 1.29 & -1.53 & 2.43 & $-0.065\pm0.027$ & 0.0039 & 0.1227 \\
        &\multirow{2}{*} {Red ($c>0$)} & Global &1.785 &-0.905 & 3.48 & $-0.131\pm0.038$& 0.0136 &0.1493 \\
        &&  Local &1.29 &-1.53 &4.36 &$-0.161\pm0.037$ & 0.0195 &0.1434\\ \hline \hline
        \multirow{4}{*}{Tripp+$\delta \mu_M$} &\multirow{2}{*}{Blue ($c<0$)} &  Global & 1.785 & -0.905 & 2.03& $-0.050\pm0.025$ & 0.0026 & 0.1215\\
        &&  Local & 1.29 & -1.53 & 1.90 & $-0.049\pm0.026$ & 0.0025 & 0.1182\\
        &\multirow{2}{*} {Red ($c>0$)} & Global &1.785 &-0.905 & 3.23 & $-0.114\pm0.035$& 0.0111 &0.1393 \\
        &&  Local &1.29 &-1.53 &4.01 &$-0.137\pm0.034$ & 0.0073 & 0.1316\\ \hline
    \end{tabular}
\tablefoot{The top rows show the results pertaining to the Tripp standardization, while the bottom rows show the results pertaining to the Tripp+$\delta \mu_M$. In both cases, the same step locations obtained in Section \ref{sec:dust_steps} are used. $\Delta \textrm{RMS}$ is defined as the difference between the $\textrm{RMS}_{\textrm{single}}$ obtained when considering the Hubble residuals as a single population and the $\textrm{RMS}_{\textrm{dual}}$ obtained when considering them as two separate populations, each with a separate mean and divided at the step.}
    \label{tab:dust_steps_color}
\end{table*}

Finally, similarly to what was done in Sec. \ref{sec:both_steps}, we examine what happens to the mass step if the dust step is corrected for and vice versa. Depending on the population in question, we use either the steps recovered for the blue or the red SNe. The results remain consistent with what we have shown before, with neither of the steps being able to fully account for the other, regardless of the color of the SNe considered.

\section{Discussion}
Several key results obtained in this work appear to support the link between environmental dust properties and the mass step.
\par
First, the mass-$\tau_V$ and mass-$n$ relations, shown in Fig. \ref{fig:griz_m_params}, make it clear that different galaxy mass populations are subject to different dust contents and dust laws. This fact is most clearly expressed in the results displayed in Table \ref{tab:rv_bins}, where it is shown that the $R_V$ values for low- and high-mass galaxies exhibit significant differences, which agrees with previous results \citep[e.g.,][]{Salim_2018,Brout_2021,gaitan}.
\par
Second, in Section \ref{sec:both_steps} we show that adding a dust step to a set of mass-step corrected Hubble residuals results in lower significances, magnitudes and $\Delta \textrm{RMS}$ than the ones obtained for the same step when no initial correction is present, and vice versa. This shows that the effects in mass and dust are at least partially related and that correcting for one also partly corrects for the other.
\par
Third, as shown in Section \ref{sec:color_steps}, both the mass and dust steps are found to have much higher values of significance, magnitude and $\Delta \textrm{RMS}$ for red SNe ($c>0$), when compared to blue SNe ($c<0$). Presuming that red SNe are also the most extinguished, we have yet another link between dust and the mass step.
\par
However, the color variations among SNe are not only a product of dust extinction, but can also originate from intrinsic differences between the SNe \citep[e.g.,][]{Conley_07}. For this reason, the fact that the mass step is found to be more prominent for red SNe might also indicate a relation between this effect and the SN intrinsic color \citep{Pang_2020,Wang_2013}.
\par
Another result that points against dust as the sole driver for the mass step is the overall lack of correlation found between SN color and the corresponding host galaxy attenuation parameters, as recorded in Table \ref{tab:spearman}. This again suggests that the differences in the mass step between different color SNe might actually be due to intrinsic factors. Another possibility is that global and local (4kpc) environmental dust properties might be very poor proxies of the real dust extinction at the SN site.
\par
This last point is further strengthened by the fact that, as shown in Fig. \ref{fig:griz_av_rv}, extinction and attenuation display very different behaviors, pointing to different physical effects. This also explains why the proposed "Fixed-Extinction" Standardization, discussed in Appendix \ref{app:fixed_ext}, gives such poor results.

\par
Furthermore, while it is true that there appears to be an overlap between the two steps, we show that accounting for one of them does not fully remove the other, suggesting that there might be other factors at play, not directly related to dust.

\par
In general, our results seem to point to the fact that the mass step might not have a fully dust-related origin, despite the evidence for at least a partial link between the two. As stated before, it is possible that a partial cause of the observed effect lies instead in intrinsic differences between SNe Ia that come about due to different progenitors. The fact that the stellar age seems to be quite important in the relations between the attenuation law and the mass
(see Fig. \ref{fig:griz_m_params}) shows that the situation is more complex and agrees with other works that show the importance of age in the mass step \citep[e.g.,][]{Rigault_2020,Rose_22,wiseman_2022}. In this sense, both the environmental mass and dust might be acting as imperfect proxies for the different types of progenitors, explaining the behavior we observe.
\par
Finally, while dust might indeed be the origin of the mass step, a two-dimensional environmental dust step might be an overly simplistic way to look at the relation between these two quantities. One might need either direct extinction data for each supernova or a better way to isolate the different mass/dust populations.

\section{Conclusions}
 In this work, we have explored ways to better probe the dust contents of SN Ia host galaxies as well as how to better make use of this information for SN standardization. Our major findings are:
 \begin{itemize}
     \item  We have shown through simulations that DECam \textit{griz} photometry is mostly sufficient to recover dust parameters for simulated SN Ia host galaxies, while discrepancies and degeneracies among other parameters do not significantly affect inferred dust properties. We found that in using both global and local DECam \textit{griz} photometry, we can recover dust properties for host galaxies that are consistent with literature predictions from both simulations and observations. We find a relation for the dust attenuation slope with the dust optical depth, both locally and globally, that is best explained with varying star-to-dust geometry with a galaxy orientation. Most importantly, we show that dust properties vary greatly across different galaxies, meaning a universal value of $R_V=3.1$ or a universal SN Ia $\beta$ correction are assumptions that are too simplistic.
     \item The relation between the attenuation parameters is found to be very different from the extinction relations obtained directly for SNe Ia, making the comparison between the two somewhat difficult. This difference is mainly due to the effects related to star-to-dust geometry that become relevant when dealing with extended objects, such as SN host galaxies.
     \item We find two different relations between a galaxy's stellar mass and its dust properties, depending on the age of the corresponding stellar population. For younger galaxies, there is a steady increase of $\tau_V$ and $n$ with $\log(M_\star/M_\odot)$, while older galaxies tend to exhibit both larger masses and smaller optical depths. A similar behavior for the two age populations can be seen when considering the local mass, pointing to an analogous relation between the results observed for the global stellar mass and the local stellar density. This population mix results in the observation that low-mass galaxies tend to have larger values of $R_V$, meaning that, for the same amount of dust, these galaxies are systematically subject to a lower level of reddening. For both mass populations, the scatter in $R_V$ is $\sim1$. All of these results match previous literature observations.
     \item We conclude that an alternative SN Ia standardization, incorporating both the $R_V$ and $E(B-V)$ values obtained for the respective host galaxies to approximate and fix SN extinction, results in a worse cosmological fit. This is very likely a consequence of the differences between host galaxy attenuation and SN Ia extinction, meaning the values for $\beta_{R_V}$ and $E(B-V)$ for the SNe Ia cannot be correctly determined. The lack of correlation between the host attenuation parameters and the respective SN light-curve parameters also points to this fact. This remains true when carrying out local studies (4 kpc), showing that even smaller apertures are needed to avoid the geometrical effects of dust attenuation. 
     \item We show there is evidence for a two-dimensional dust step, which, for the global case, results in roughly the same significance, magnitude, and residual $\textrm{RMS}$ recovered for the mass step, hinting at a relation between the two. In particular, this dust step is significant at $>4\sigma$. Furthermore, we show that these two steps are not completely analogous and that accounting for one of them in the SN standardization does not necessarily fully eliminate the other.
     \item We find that both the mass and dust steps are much more pronounced for red SNe Ia ($c>0$), which once again suggests a link between dust reddening and the Hubble residual steps.
 \end{itemize}
\par
Thus, we conclude that the need for the mass-step correction cannot be completely eliminated using only host galaxy dust data. This is true for both an alternative SN standardization, in which the color-luminosity relation is constrained with host attenuation, and for a SN standardization with a dust-step correction.

\begin{acknowledgements}
J.D.,  S.G.G., A.M. and A.P.A. acknowledge support by FCT under Project CRISP PTDC/FIS-AST-31546/2017 and Project~No.~UIDB/00099/2020. L.G. acknowledges financial support from the Spanish Ministerio de Ciencia e Innovación (MCIN), the Agencia Estatal de Investigación (AEI) 10.13039/501100011033, and the European Social Fund (ESF) “Investing in your future” under the 2019 Ramón y Cajal program RYC2019-027683-I and the PID2020-115253GA-I00 HOSTFLOWS project, from Centro Superior de Investigaciones Científicas (CSIC) under the PIE project 20215AT016, and the program Unidad de Excelencia María de Maeztu CEX2020-001058-M. L.K. thanks the UKRI Future Leaders Fellowship for support through the grant MR/T01881X/1. Computations were performed at the cluster “Baltasar-Sete-Sóis” and supported by the H2020 ERC Consolidator Grant "Matter and strong field gravity: New frontiers in Einstein’s theory" grant agreement no. MaGRaTh-646597.  P.W. acknowledges support from the Science and Technology Facilities Council (STFC) grant ST/R000506/1.
\par
\noindent
Funding for the DES Projects has been provided by the U.S. Department of Energy, the U.S. National Science Foundation, the Ministry of Science and Education of Spain, the Science and Technology Facilities Council of the United Kingdom, the Higher Education Funding Council for England, the National Center for Supercomputing Applications at the University of Illinois at Urbana-Champaign, the Kavli Institute of Cosmological Physics at the University of Chicago, the Center for Cosmology and Astro-Particle Physics at the Ohio State University, the Mitchell Institute for Fundamental Physics and Astronomy at Texas A\&M University, Financiadora de Estudos e Projetos, Fundação Carlos Chagas Filho de Amparo à Pesquisa do Estado do Rio de Janeiro, Conselho Nacional de Desenvolvimento Científico e Tecnológico and the Ministério da Ciência, Tecnologia e Inovação, the Deutsche Forschungsgemeinschaft and the Collaborating Institutions in the Dark Energy Survey.
\par
\noindent
The Collaborating Institutions are Argonne National Laboratory, the University of California at Santa Cruz, the University of Cambridge, Centro de Investigaciones Energéticas, Medioambientales y Tecnológicas-Madrid, the University of Chicago, University College London, the DES-Brazil Consortium, the University of Edinburgh, the Eidgenössische Technische Hochschule (ETH) Zürich, Fermi National Accelerator Laboratory, the University of Illinois at UrbanaChampaign, the Institut de Ciències de l’Espai (IEEC/CSIC), the Institut de Física d’Altes Energies, Lawrence Berkeley National Laboratory, the Ludwig-Maximilians Universität München and the associated Excellence Cluster Universe, the University of Michigan, the National Optical Astronomy Observatory, the University of Nottingham, The Ohio State University, the University of Pennsylvania, the University of Portsmouth, SLAC National Accelerator Laboratory, Stanford University, the University of Sussex, Texas A\&M University, and the OzDES Membership Consortium.
\par
\noindent
Based in part on observations at Cerro Tololo Inter-American Observatory, National Optical Astronomy Observatory, which is operated by the Association of Universities for Research in Astronomy (AURA) under a cooperative agreement with the National Science
Foundation.
\par
\noindent
The DES data management system is supported by the National Science Foundation under Grant Numbers AST-1138766 and AST-1536171. The DES participants from Spanish institutions are partially supported by MINECO under grants AYA2015-71825, ESP2015-66861, FPA2015-68048, SEV-2016-0588, SEV2016-0597, and MDM-2015-0509, some of which include ERDF funds from the European Union. IFAE is partially funded by the CERCA program of the Generalitat de Catalunya. Research leading to these results has received funding from the European Research Council under the European Union’s Seventh Framework Program (FP7/2007-2013) including ERC grant agreements 240672, 291329, and 306478. We acknowledge support from the Brazilian Instituto Nacional de Ciência e Tecnologia (INCT) e-Universe (CNPq grant 465376/2014-2). \\
This manuscript has been authored by Fermi Research Alliance, LLC under Contract No. DE-AC02-07CH11359 with the U.S. Department of Energy, Office of Science, Office of High Energy
Physics.

\end{acknowledgements}

%
%

\bibliographystyle{aa} 
\bibliography{apssamp.bib}

\begin{appendix}

\section{Host galaxy mass comparison}
\label{app:mass}
\begin{figure*}[t]
        \includegraphics[width=1\textwidth]{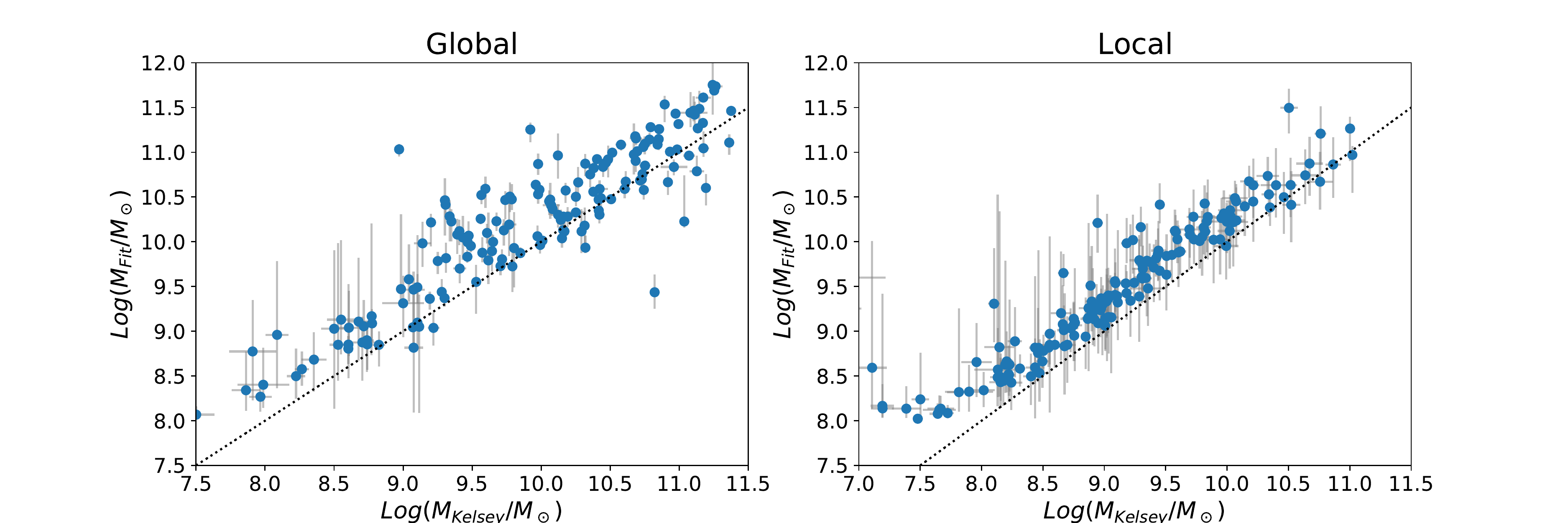}
        \caption{Comparison between the stellar mass values obtained by Kelsey et al. (2021) for the host galaxies of the DES-SN sample and the masses recovered using a five-parameter sampling with no initial minimization. In both cases, fits were performed using global (Left panel) and local (Right panel) DECam \textit{griz} photometry. The results are shown in blue. A dashed line with $y=x$ is also shown.}
        \label{fig:mass_comp}
\end{figure*}

In this appendix, we compare the stellar mass values recovered by \cite{Kelsey_2020} for the DES-SN host galaxies with the values obtained using our methodology, with five-parameter sampling and no initial minimization. The results are displayed in Fig. \ref{fig:mass_comp}. 

\par

As previously stated, the recovered values exhibit a median relative difference of $\sim 4\%$ in relation to the values computed by \cite{Kelsey_2020}, in both the global and the local cases. There is, however, a clear bias between the two fitting methods, as the fitting methodology discussed in this paper returns systematically higher-mass values. Even so, as discussed in \ref{sec:sim}, where we purposely change the mass values by $4\%$, this discrepancy does not have a noticeable impact on the recovered dust parameters.

\section{Host galaxy simulations}
\label{app:B}

\begin{figure*}
        \includegraphics[width=1\textwidth]{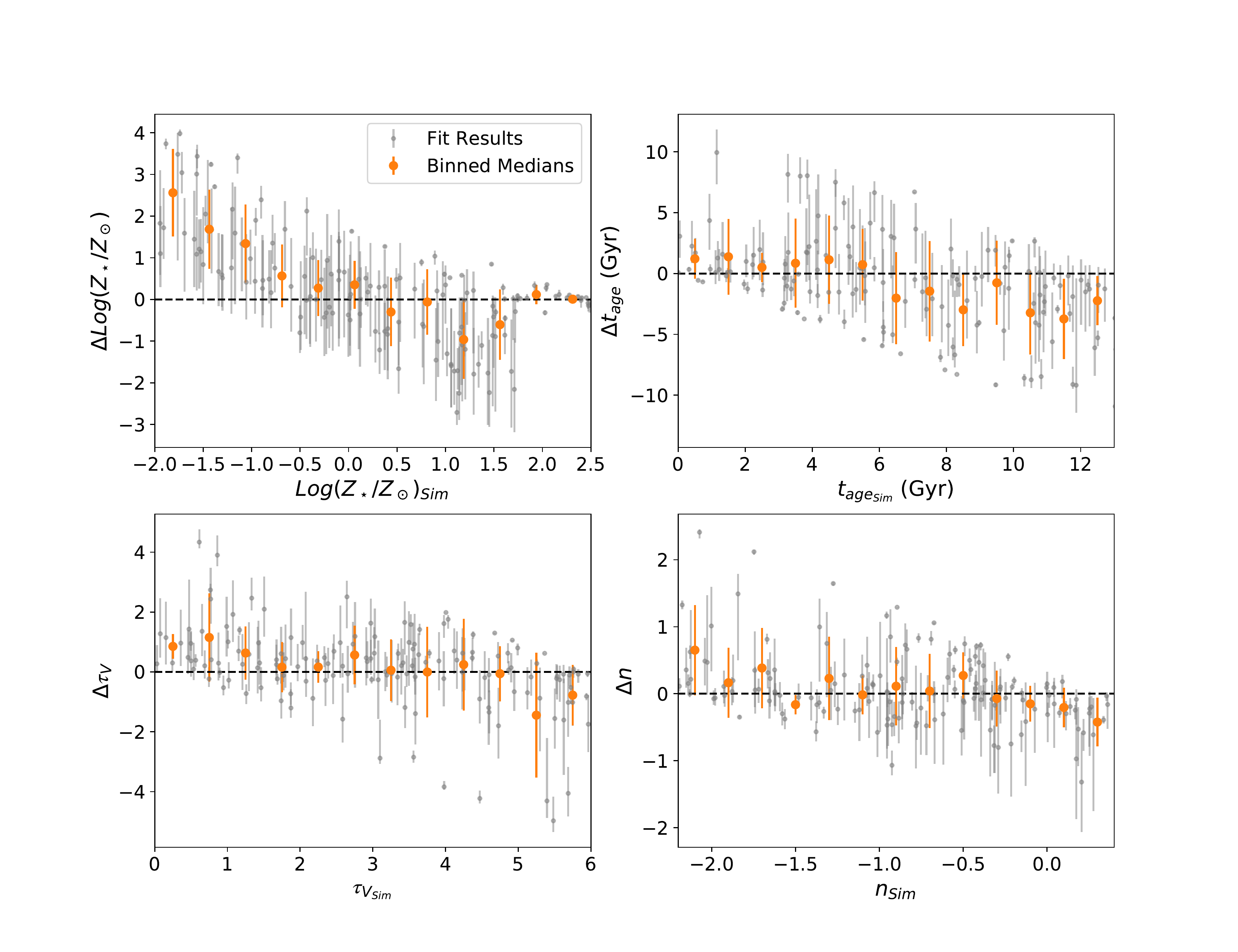}
        \caption{Residuals ($\Delta=$ Fit-Simulation) for the best-fit values for the default (\textit{griz}) test galaxy fits as a function of the original simulation parameters. From top to bottom and left to right, we have: $\log(Z_\star/Z_\odot)$, $t_{\textrm{age}}$, $\tau_V$ and $n$. Results for the different galaxies are shown in gray. Binned means for each parameter are shown in orange, with error bars given by the standard deviation in each bin.}
        \label{fig:mock_default}
\end{figure*}

In this appendix, we present a more complete view of the various simulations described in Section \ref{sec:sim}. We begin by analyzing the first test case, in which the default dust and SFH model is used for both the simulation and the fit and which relies only on \textit{griz} photometry. We can compare the recovered best-fit values and the ``true'' simulation values. The difference between these two quantities is plotted in Fig. \ref{fig:mock_default}

\par
Overall, as stated before, the fit shows remarkable quality, considering only four photometric data points were used for each galaxy. There are, however, some accuracy problems present in the fit, especially in the case of the metallicity, as there is a strong bias that skews the recovered values, resulting in a median relative error of 55\%. This inaccuracy is particularly noticeable for the extreme values of $\log(Z_\star/Z_\odot)$. It is possible that this is a byproduct of the difficulty in sourcing accurate spectra with which to build accurate CSP models for nonsolar stellar metallicities \citep{Leja_2017}. However, given that the spectra used for both the simulation and the fit are drawn from the same library, it is unlikely this is the root of the observed behavior. This also means that we do not recover the expected mass-metallicity relation in our fitted sample. Although less noticeable, there also seems to exist a slight skew for larger values of $t_{\textrm{age}}$. These skews are most likely linked to the degeneracy between $\log(Z_\star/Z_\odot)$ and $t_{\textrm{age}}$.
\par
\begin{figure*}
        \centering
        \includegraphics[width=1\textwidth]{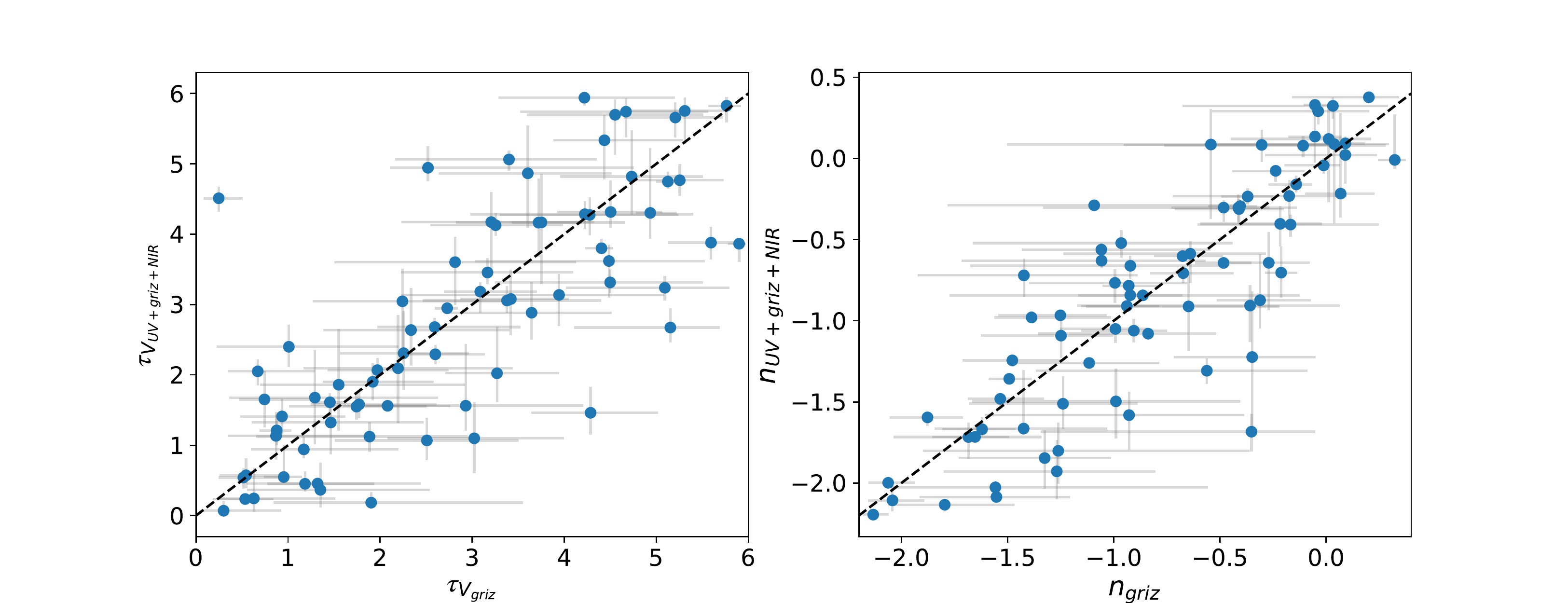}
        \caption{Comparison between the best-fit values for the simulated test galaxy fits obtained with DECam \textit{griz} photometry (x-axis) and DECam \textit{griz} + \textit{NUV/FUV} GALEX + \textit{JHKs} 2MASS photometry (y-axis). From left to right we have: $\tau_V$ and $n$.}
\label{fig:mock_uv_nir}
\end{figure*}

To better perceive the quality of these results, we look next to how they compare to the fit results obtained in the same model conditions while using a combined \textit{griz} + \textit{NUV/FUV} GALEX + \textit{JHKs} 2MASS photometry. These results are displayed in Fig. \ref{fig:mock_uv_nir}. Although some deviations are observed, on the whole the results appear mostly consistent, reinforcing the previously stated conclusion that a \textit{griz} photometry fit can adequately determine host dust properties.

\par

\begin{figure*}
        \centering
        \includegraphics[width=1\textwidth]{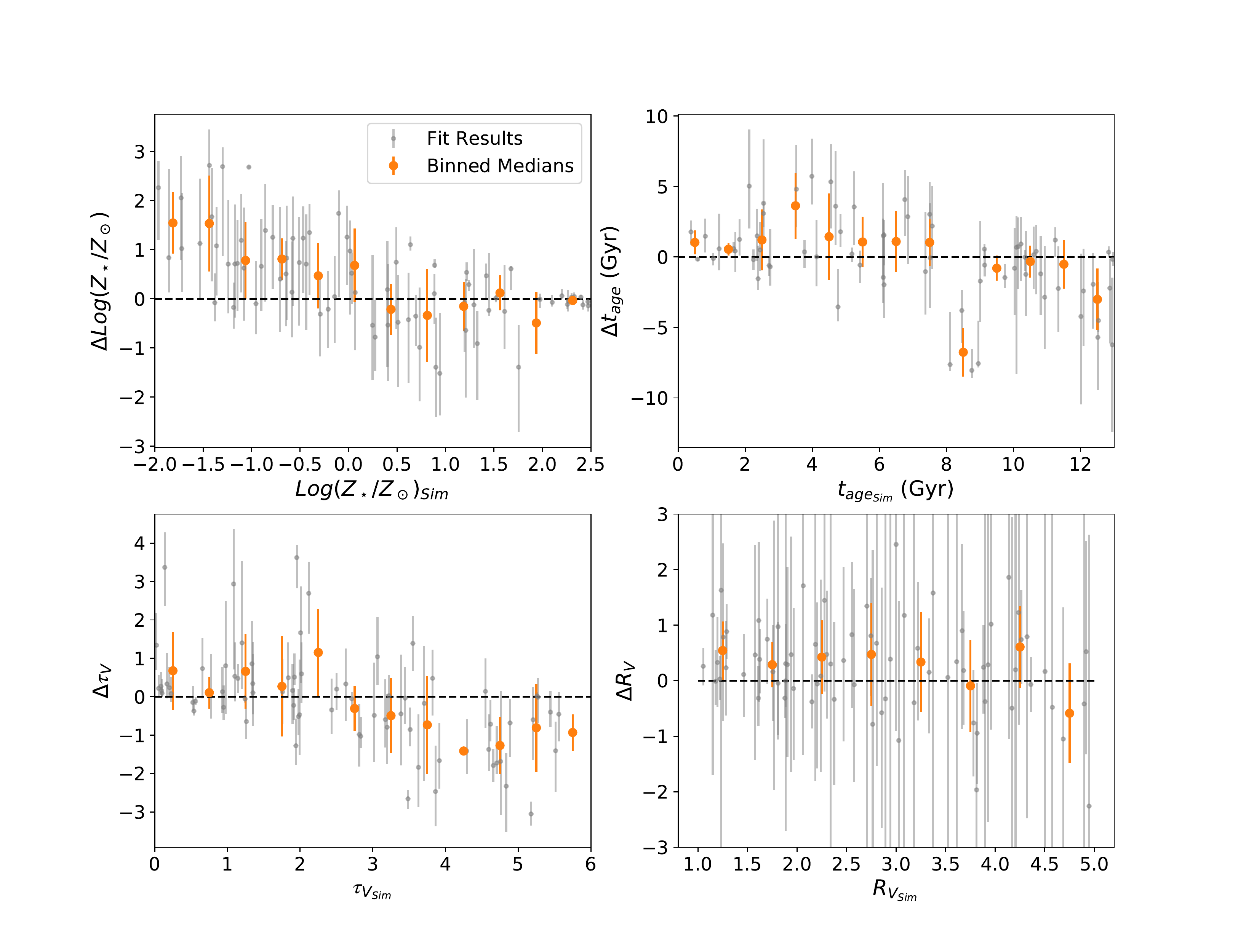}
        \caption{Residuals ($\Delta=$ Fit-Simulation) for the best-fit values for the Cardelli test galaxy fits as a function of the original simulation parameters. From top to bottom and left to right we have: $\log(Z_\star/Z_\odot)$, $t_{\textrm{age}}$, $\tau_V$ and $n$. Results for the different galaxies are shown in gray. Binned means for each parameter are shown in orange, with error bars given by the standard deviation in each bin.}
\label{fig:mock_cardelli}
\end{figure*}

As seen from Table \ref{tab:mock}, the largest changes to the recovered dust parameters occur when different attenuation models are considered for the simulation and the fits. For the test case where a Cardelli attenuation law was used for the simulations, the differences between the recovered best-fit values and the ``true'' simulation values are plotted in Fig. \ref{fig:mock_cardelli}. As previously mentioned, $R_V$ is the most affected parameter, with a clear bias observed in the recovered values. In addition, there are large errors accompanying these results, which come about due to the conversion between different $n$ and $R_V$.

\par

One final point that should be considered is that, while we have shown that, for the most part, a change in the simulation SFH does not greatly impact the determination of dust properties, there might still be an impact on the age parameter, which might influence the subsequent analysis in Section \ref{sec:des_fits}.
\par
\begin{figure*}
        \includegraphics[width=1\textwidth]{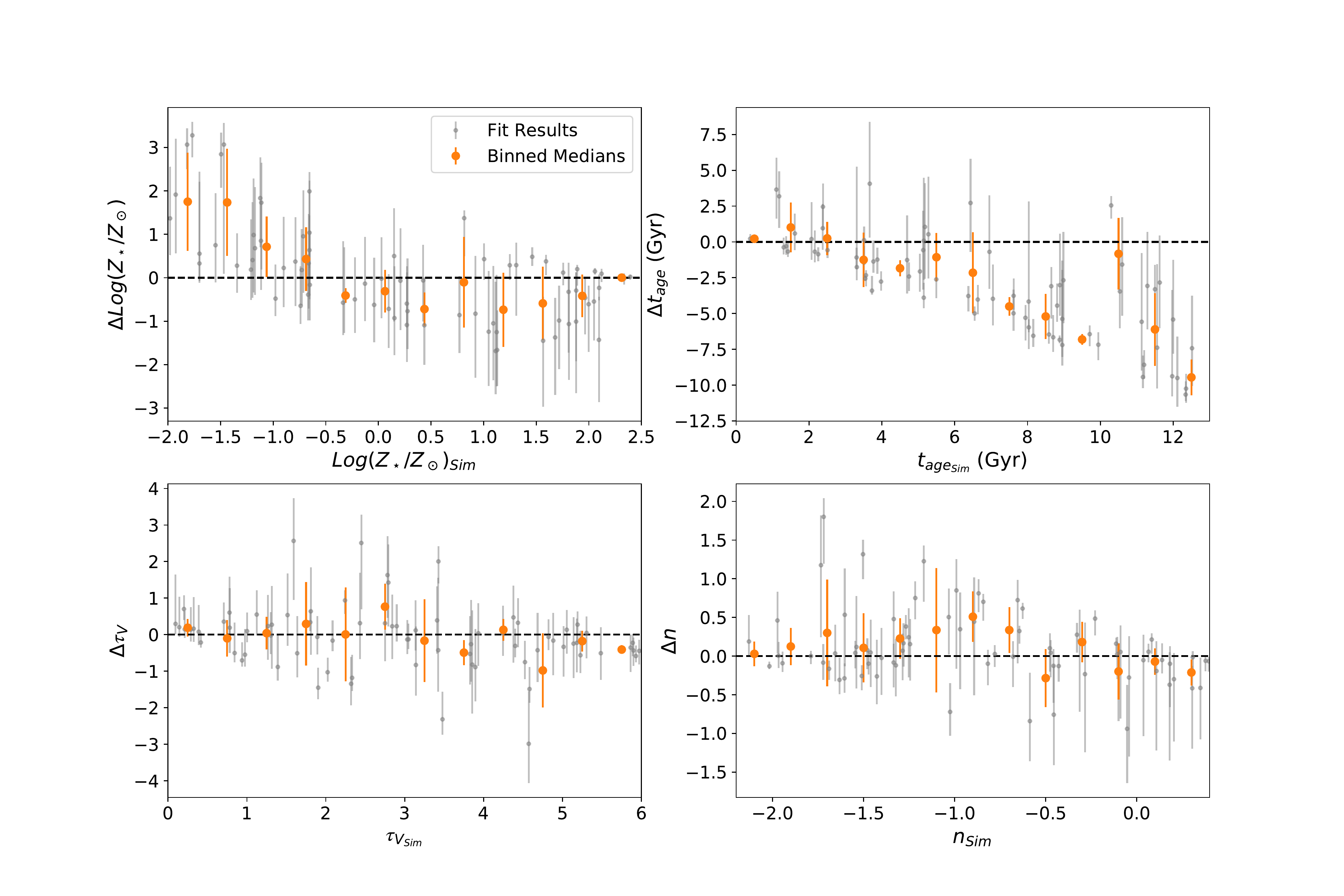}
        \caption{Residuals ($\Delta=$ Fit-Simulation) for the best-fit values for the $\tau=10$ test galaxy fits as a function of the original simulation parameters. From top to bottom and left to right we have: $\log(Z_\star/Z_\odot)$, $t_{\textrm{age}}$, $\tau_V$ and $n$. Results for the different galaxies are shown in gray. Binned means for each parameter are shown in orange, with error bars given by the standard deviation in each bin.}
        \label{fig:mock_tau_10}
\end{figure*}

To do so, we focus on the fits obtained from the simulations obtained using a delayed-$\tau$ model with $\tau=10$. The comparison between the recovered best-fit values and the ``true'' simulation values is plotted in Fig. \ref{fig:mock_tau_10}.

\par
It is clear that a discrepancy between the simulation and fit SFHs introduces a great level of bias into the recovered age results. In particular, as the SFH used for the simulation decays more slowly than the SFH presumed in the fitting process, there is a greater amount of star formation at older ages than expected. As such, the fit ends up predicting a younger age for these galaxies. If we instead consider a simulation SFH that decays faster than the one presumed for the fit, then we encounter the opposite problem, with young galaxies appearing much older in the fit. A similar effect happens for simulations with quenched SFHs.
\par
The Spearman correlation coefficients between the simulated and fitted stellar ages for the variable SFH test fits are shown in Table \ref{tab:age_corr}. While there is a degree of correlation prevalent, it starts breaking down as the simulation model moves further away from the default fit model. However, while older galaxies might appear younger and younger galaxies might appear older, an order relation appears to be maintained in the fits. As such, while the specific $t_{\textrm{age}}$ values recovered might be called into question, the broad division of the galaxies into "young" and "old" galaxies can be preserved and relied upon. 

\renewcommand{\arraystretch}{1}
\begin{table}
    \centering
    \caption{Spearman correlation coefficients between the simulated and fitted stellar ages $t_{\textrm{age}}$ for the variable SFH test fits.}
    \begin{tabular}{c c}
    \hline
    Test&Spearman Correlation Coefficient \\\hline
        $\tau=0.1$ & 0.406 \\
        $\tau=10$ & 0.432 \\
        $t_{\textrm{trunc}}=7.5$ Gyr &  0.485\\
        $t_{\textrm{trunc}}=5$ Gyr & 0.743 \\
        $t_{\textrm{trunc}}=3$ Gyr & 0.678\\\hline
    \end{tabular}

    \label{tab:age_corr}
\end{table}
\renewcommand{\arraystretch}{1.75}

\section{Fixed-extinction SN Ia standardization}
\label{app:fixed_ext}
In this appendix, we discuss the viability of an alternative SN Ia standardization. If the existence of the $\beta$ color-luminosity relation in Eq. \eqref{eq:Tripp} was entirely due to dust, then we would have \citep[e.g.,][]{gaitan}:

\begin{equation}
    \label{eq:beta}
    \beta_{R_V}=R_B=R_V+1
\end{equation}

Using this fact, we can separate the extinction $\beta_{R_V}$ and intrinsic $\beta_{\textrm{int}}$ components in the $\beta$ color-luminosity relation in Eq. \ref{eq:Tripp}, such that \citep[e.g.,][]{Brout_2021}:

\begin{equation}
\label{eq:bc_split}
\begin{aligned}
    \beta c &= \beta_{R_V}E(B-V) + \beta_{\textrm{int}}\left(c-E(B-V)\right) \\
    &=A_B + \beta_{\textrm{int}}\left(c-E(B-V)\right),
\end{aligned}
\end{equation}
\noindent
leaving out a possible additional measurement noise term \citep[e.g.,][]{popovic2021}.
\par
It is thus possible to introduce an alternative ``Fixed-Extinction'' SN Ia standardization in which the values of $R_V=\frac{\tau_V}{\tau_B-\tau_V}$ and $E(B-V)=1.086(\tau_B-\tau_V)$ can be fixed using host galaxy measurements. The new nuisance fit parameters are thus $\alpha$, $\beta_{\textrm{int}}$ and $M$. We use the values of $R_V$ and $E(B-V)$ obtained in Section \ref{sec:des_fits} from the CSP fits of the SN host galaxies to constrain the color-luminosity relation in the respective ``Fixed-Extinction'' standardization. 
\par
The $\sigma^2$ in Eq. \ref{eq:likelihood_cosmo} is thus given by:

\begin{equation}
\begin{multlined}
    \sigma^2=\sigma_{m_B}^2+(\alpha \sigma_{x1})^2+ \sigma_{A_B}^2 +\beta_{int}^2( \sigma_c^2+\sigma_{E(B-V)}^2)+\sigma_{\textrm{int}}^2\\
    -2\beta\sigma_{m_B,c}+2\alpha\sigma_{m_B,x_1}-2\alpha\beta\sigma{x_1,c},
    \label{eq:si_2}
\end{multlined}
\end{equation}
\noindent
where $\sigma_{m_B}$, $\sigma_{x1}$ and $\sigma_{c}$ are the uncertainties associated with each of the light-curve fit parameters, $\sigma_{m_B,c}$, $\sigma_{m_B,x_1}$, $\sigma_{x_1,c}$ are their covariance terms, $\sigma_{A_B}$ and $\sigma_{E(B-V)}$ are the uncertainties associated with the attenuation law parameters for the host galaxies and $\sigma_{\textrm{int}}$ is a parameter which accounts for possible intrinsic variations in a SN Ia's luminosity. We fix $\sigma_{\textrm{int}}=0.107$, following the value obtained by \cite{gaitan}.

\par
The cosmological fit parameters obtained for the "Fixed-Extinction" standardization are shown in Table. \ref{tab:cosmo_fit_3}. This standardization yields $\textrm{RMS}$ values of 0.515 and 0.575 in the global and the local cases, respectively. As previously hinted, both the global and the local measurements produce consistent results. However, judging from the $\textrm{RMS}$ values, the ``Fixed-Extinction'' standardization appears to be a much worse fit of the data, invalidating it as a replacement for the Tripp standardization, which has a $\textrm{RMS}$ value of 0.141. This is probably due to the fact that, according to Fig. \ref{fig:griz_av_rv}, host attenuation laws do not accurately describe SN extinction, meaning it is likely that the problem resides in the values used to approximate SN extinction.

\renewcommand{\arraystretch}{1.5}

\begin{table*}
\centering
    \caption{Global and Local fit parameters for the ``Fixed-Extinction'' and ``Fixed-Extinction'' + $\tau_V$ bias standardizations and respective $\textrm{RMS}$}
    \begin{tabular}{c c c c c c}
    \hline
         \multicolumn{2}{c}{Standardization} & $\alpha$ & $\beta_{\textrm{int}}$ & $M$ & $\textrm{RMS}$  \\ \hline
        \multirow{2}{*}{Fixed-Extinction} & (Global) & $0.432^{+0.010}_{-0.011}$  & $4.350^{+0.010}_{-0.010}$ & $-19.106^{+0.011}_{-0.009}$ &$0.515$ \\ 
        & (Local) & $0.425^{+0.010}_{-0.011}$  & $6.202^{+0.012}_{-0.009}$ &$-18.287^{+0.009}_{-0.009}$ &$0.575$\\
         \multirow{2}{*}{Fixed-Extinction + $\tau_V$ Bias} & (Global) & $0.351^{+0.008}_{-0.012}$  & $4.281^{+0.010}_{-0.009}$ & $-19.220^{+0.011}_{-0.011}$ &$0.420$ \\ 
        & (Local) & $0.276^{+0.009}_{-0.010}$  & $5.413^{+0.011}_{-0.010}$ &$-18.921^{+0.010}_{-0.009}$ &$0.372$\\\hline 
    \end{tabular}
    \tablefoot{A likelihood defined by Eq. 3 was used for the fits.}
    \label{tab:cosmo_fit_3}
\end{table*}

\par
A possible source of error may come from the bias in $\tau_V$ observed in Section \ref{sec:sim}, which might be skewing the standardization fit results. We can correct for this effect by subtracting the bias level from each $\tau_V$ value. For this, the largest bias value found was used, which was 0.263. For values below the bias threshold, we simply keep the original value. The results obtained from a ``Fixed-Extinction'' biased standardization are also listed in Table. \ref{tab:cosmo_fit_3}. The $\tau_V$ bias improves the standardization, with new $\textrm{RMS}$ values of 0.420 and 0.372, for the global and the local cases, respectively. Even so, the bias corrected standardization still falls short of the Tripp values.

 On the whole, the proposed ``Fixed-Extinction'' SN Ia standardization does not provide better fits than standard Tripp standardization. This is most likely a consequence of using host attenuation to approximate SN Ia extinction, as already seen in Section \ref{sec:ext_vs_att}. However, it is possible that, using extinction data directly obtained for each SN and not its environment, this method might indeed prove to significantly improve cosmological fits and possibly reduce or completely eliminate the mass step.

\section{Cosmological fits with steps}
\label{app:steps}
In this appendix, a deeper look at the relation between the mass and dust steps is presented. In particular, for a slightly more robust analysis of both these steps, we can directly introduce $\delta \mu_M$ and $\delta \mu_D$ terms in the cosmological fit. We promote the respective step magnitudes $\gamma_M$ and $\gamma_D$ to free parameters, fixing the step locations to the values previously recovered for the Tripp standardization in Section \ref{sec:steps}.

\par
As mentioned in Section \ref{sec:Tripp_method}, the likelihood defined by Eq. \ref{eq:likelihood_cosmo} can introduce biases into the recovered fit parameters. While this fact is not entirely significant when comparing two standardizations, it becomes important when dealing with a more rigorous cosmological fit. For this reason, when promoting either the mass or dust steps to free fit parameters, it is helpful to slightly modify the likelihood used \citep[e.g.,][]{gaitan}:

\begin{equation}
\label{eq:likelihood_cosmo_2}
    \ln{(\mathcal{L})}=-\frac{1}{2}\sum\limits_{i=0}^N \left(\frac{\Delta \mu_i^2}{\sigma_i^2}+ \ln{(\sigma_i^2)}\right),
\end{equation}

\par
In order to study the various steps, we look at three possible standardizations: a Tripp+mass-step standardization, in which $\gamma_M$ is promoted to a free fit parameter; a Tripp+dust-step standardization, in which $\gamma_D$ is promoted to a free fit parameter; and a Tripp+mass-step+dust-step standardization, in which both $\gamma_M$ and $\gamma_D$ are promoted to free fit parameters. It is also useful to redo the standard Tripp standardization to establish a term of comparison for the new likelihood.

\par
For an even more rigorous fit, we can also promote the $\sigma_{\textrm{int}}$ term to a free parameter. Additionally, in the cases where a step is considered, we can assign different $\sigma_{\textrm{int}}$ values to each population. Taking into account both the high and low-mass populations and the regions defined by Fig. \ref{fig:regions}, we have: Tripp (1 population): full sample ($\sigma_{\textrm{int}_1}$); Tripp + mass step (2 populations): low-mass ($\sigma_{\textrm{int}_1}$), and high-mass ($\sigma_{\textrm{int}_2}$); Tripp + dust step (2 populations): Region 2 ($\sigma_{\textrm{int}_1}$), and Regions 1,3,4 ($\sigma_{\textrm{int}_2}$); Tripp + mass step + dust step (4 populations): low-mass + region 2 ($\sigma_{\textrm{int}_1}$), low-mass + regions 1,3,4 ($\sigma_{\textrm{int}_2}$), high-mass + region 2 ($\sigma_{\textrm{int}_3}$), and high-mass + regions 1,3,4 ($\sigma_{\textrm{int}_2}$).

\par

The recovered fit parameters, as well as the corresponding $\textrm{RMS}$ values, are listed in Table \ref{tab:cosmo_fit_2}. As expected, the addition of separate mass and dust steps results in lower $\textrm{RMS}$ values, for both the global and the local cases. The dust-step value recovered for the global case is compatible with zero, which might be due to a problem with the fit convergence. The simultaneous inclusion of mass and dust steps produces the lowest values of $\textrm{RMS}$ for the respective cosmological fits. In addition, when both steps are simultaneously included, we recover both noticeable mass and dust steps, with the caveat that the global dust step is once again compatible with zero. In general, the step magnitudes in this case are overall lower than those obtained in the single step cases. As such, we once again find evidence that, although there is some level of overlap, the two steps are not completely analogous and interchangeable.

\begin{table*}

\centering
    \caption{Fit parameters for the Tripp, Tripp+$\delta \mu_M$, Tripp+$\delta \mu_D$ and Tripp+$\delta \mu_M$+$\delta \mu_D$ standardizations and respective $\textrm{RMS}$.}
    \scriptsize
    \begin{tabular}{c c c c c c c c c c c c}
    \hline
         \multicolumn{2}{c}{Standardization} & $\alpha$ & $\beta$ & $M$ & $\gamma_M/2$& $\gamma_D/2$ & $\sigma_{\textrm{
    int}_1}$& $\sigma_{\textrm{
    int}_2}$& $\sigma_{\textrm{
    int}_3}$& $\sigma_{\textrm{
    int}_4}$ &$\textrm{RMS}$  \\ \hline
        Tripp& & $0.141^{+0.012}_{-0.012}$ & $2.845^{+0.136}_{-0.137}$ & $-19.404^{+0.011}_{-0.011}$ & - & - &$0.098^{+0.010}_{-0.010}$ & - & - & -& $0.1367$\\
      \multirow{2}{*}{Tripp+$\delta \mu_M$} & (Global) & $0.155^{+0.013}_{-0.013}$ & $2.872^{+0.141}_{-0.140}$ & $-19.423^{+0.012}_{-0.012}$ & $-0.035^{+0.014}_{-0.013}$ & - & $0.097^{+0.012}_{-0.011}$& $0.071^{+0.025}_{-0.027}$ & - & - & $0.1332$\\
        & (Local)& $0.156^{+0.012}_{-0.012}$ & $2.975^{+0.138}_{-0.141}$ & $-19.422^{+0.011}_{-0.011}$ & $-0.049^{+0.012}_{-0.013}$ & - &$0.087^{+0.013}_{-0.012}$ & $0.082^{+0.020}_{-0.019}$ & - & - & $0.1341$\\
        \multirow{2}{*}{Tripp+$\delta \mu_D$} & (Global) & $0.144^{+0.012}_{-0.012}$ & $2.833^{+0.138}_{-0.133}$ & $-19.405^{+0.011}_{-0.011}$ & -& $-0.008^{+0.011}_{-0.011}$ & $0.104^{+0.014}_{-0.012}$ & $0.091^{+0.015}_{-0.014}$ & -& - & $0.1346$\\
        & (Local) & $0.145^{+0.012}_{-0.012}$ & $2.814^{+0.133}_{-0.133}$ & $-19.416^{+0.011}_{-0.011}$ & -& $-0.035^{+0.010}_{-0.011}$ & $0.101^{+0.013}_{-0.012}$ & $0.075^{+0.018}_{-0.016}$ & - & -  & $0.1310$\\
                \multirow{2}{*}{Tripp+$\delta \mu_M$+$\delta \mu_D$} & (Global) & $0.157^{+0.013}_{-0.013}$ & $2.877^{+0.145}_{-0.142}$ & $-19.422^{+0.013}_{-0.013}$ & $-0.035^{+0.014}_{-0.014}$& $-0.004^{+0.011}_{-0.011}$ & $0.0104^{+0.015}_{-0.014}$ & $0.089^{+0.019}_{-0.018}$ & $0.079^{+0.042}_{-0.039}$ & $0.069^{+0.033}_{-0.034}$  & $0.1323$\\
        & (Local) & $0.154^{+0.012}_{-0.011}$ & $2.946^{+0.012}_{-0.0122}$&$-19.426^{+0.011}_{-0.010}$  &$-0.039^{+0.011}_{-0.011}$  & $-0.026^{+0.008}_{-0.013}$ & $0.087^{+0.011}_{-0.011}$ & $0.079^{+0.009}_{-0.011}$ & $0.091^{+0.011}_{-0.011}$ & $0.069^{+0.019}_{-0.011}$  & 0.1298 \\
        \hline
    \end{tabular}
\tablefoot{A likelihood defined by Eq. D1 was used for the fits. In each one, the steps are computed based on the SN population divisions determined by the mass and dust-step locations recovered in Section 5.2. For each of these populations, we allow for a separate $\sigma_{\textrm{int}}$ free parameter. Tripp (1 population): Full sample ($\sigma_{\textrm{int}_1}$); Tripp+$\delta \mu_M$ (2 populations): Low-mass ($\sigma_{\textrm{int}_1}$), and high-mass ($\sigma_{\textrm{int}_2}$); Tripp+$\delta \mu_D$ (2 populations): Region 2 ($\sigma_{\textrm{int}_1}$), and Regions 1,3,4 ($\sigma_{\textrm{int}_2}$); Tripp+$\delta \mu_M$+$\delta \mu_D$ (4 populations): Low-mass + region 2 ($\sigma_{\textrm{int}_1}$), low-mass + regions 1,3,4 ($\sigma_{\textrm{int}_2}$), high-mass + region 2 ($\sigma_{\textrm{int}_3}$), and high-mass + regions 1,3,4 ($\sigma_{\textrm{int}_2}$).}
    \label{tab:cosmo_fit_2}
\end{table*}

\begin{table*}[t]
\section{Host galaxy fit results}
\label{app:A}
\centering
    \caption{Best-fit results for the DES host galaxy fits with DECam \textit{griz} global and local photometry.}
    \begin{tabular}{c |c c c c|c c c c }
    \hline

    \multirow{2}{*}{SNID} & \multicolumn{4}{c|}{Global} & \multicolumn{4}{c}{Local}\\
    &  $\log(Z_\star/Z_\odot)$ & $t_{\textrm{age}}$ (Gyr) & $\tau_V$ & $n$  &  $\log(Z_\star/Z_\odot)$ & $t_{\textrm{age}}$ (Gyr) & $\tau_V$ & $n$\\\hline
1248677 & $-0.008^{+0.067}_{-0.091}$ & $0.399^{+0.071}_{-0.025}$ & $1.367^{+0.047}_{-0.092} $ & $0.087^{+0.019}_{-0.058}$  & $-0.196^{+0.641}_{-0.536}$ & $0.100^{+0.172}_{-0.077}$ & $1.414^{+0.715}_{-0.584} $ & $0.059^{+0.199}_{-0.336}$ \\ 
1250017 & $-1.638^{+0.309}_{-0.176}$ & $0.359^{+0.542}_{-0.141}$ & $1.108^{+0.211}_{-0.385} $ & $-0.720^{+0.117}_{-0.178}$  & $-1.257^{+0.574}_{-0.567}$ & $0.107^{+0.737}_{-0.086}$ & $1.483^{+0.746}_{-0.886} $ & $-0.290^{+0.241}_{-0.438}$ \\ 
1253039 & $0.164^{+0.906}_{-0.658}$ & $0.209^{+0.326}_{-0.180}$ & $2.291^{+0.761}_{-0.470} $ & $0.067^{+0.105}_{-0.111}$  & $-0.585^{+0.699}_{-0.766}$ & $0.579^{+1.436}_{-0.494}$ & $1.603^{+1.275}_{-1.013} $ & $-0.094^{+0.281}_{-0.501}$ \\ 
1253101 & $-0.574^{+0.708}_{-0.629}$ & $0.933^{+0.665}_{-0.489}$ & $0.759^{+0.497}_{-0.405} $ & $-0.553^{+0.377}_{-0.568}$  & $-0.360^{+0.568}_{-0.522}$ & $0.150^{+0.188}_{-0.116}$ & $1.113^{+0.677}_{-0.461} $ & $-0.048^{+0.273}_{-0.389}$ \\ 
1253920 & $-1.976^{+0.039}_{-0.018}$ & $2.337^{+0.147}_{-0.146}$ & $0.905^{+0.056}_{-0.054} $ & $-1.319^{+0.068}_{-0.070}$  & $-1.480^{+0.551}_{-0.357}$ & $2.963^{+0.599}_{-1.049}$ & $0.695^{+0.414}_{-0.233} $ & $-1.293^{+0.431}_{-0.486}$ \\ 
1255502 & $1.287^{+0.008}_{-0.010}$ & $0.243^{+0.003}_{-0.004}$ & $2.890^{+0.022}_{-0.017} $ & $-0.244^{+0.014}_{-0.011}$  & $-0.899^{+0.678}_{-0.640}$ & $0.550^{+1.267}_{-0.526}$ & $2.277^{+1.419}_{-1.000} $ & $0.019^{+0.192}_{-0.368}$ \\ 
1257366 & $-1.665^{+0.350}_{-0.049}$ & $2.446^{+0.605}_{-0.107}$ & $0.640^{+0.037}_{-0.204} $ & $-1.327^{+0.039}_{-0.285}$  & $-1.286^{+0.805}_{-0.531}$ & $1.499^{+0.895}_{-1.241}$ & $0.903^{+0.873}_{-0.429} $ & $-0.551^{+0.209}_{-0.407}$ \\ 
1257695 & $-1.964^{+0.436}_{-0.030}$ & $0.174^{+0.281}_{-0.012}$ & $1.259^{+0.039}_{-0.375} $ & $-0.759^{+0.037}_{-0.235}$  & $-0.614^{+0.423}_{-0.426}$ & $1.096^{+0.334}_{-0.404}$ & $0.494^{+0.328}_{-0.217} $ & $-0.708^{+0.407}_{-0.560}$ \\ 
1258906 & $0.133^{+0.481}_{-0.504}$ & $0.376^{+0.225}_{-0.156}$ & $1.183^{+0.174}_{-0.206} $ & $-0.096^{+0.140}_{-0.141}$  & $-0.347^{+0.649}_{-0.502}$ & $0.530^{+0.342}_{-0.246}$ & $1.254^{+0.320}_{-0.340} $ & $-0.109^{+0.174}_{-0.238}$ \\ 
1258940 & $-0.049^{+0.218}_{-0.195}$ & $1.150^{+0.485}_{-0.148}$ & $1.418^{+0.064}_{-0.263} $ & $-0.243^{+0.040}_{-0.070}$  & $-0.852^{+0.813}_{-0.654}$ & $1.736^{+0.520}_{-0.637}$ & $0.493^{+0.299}_{-0.220} $ & $-0.994^{+0.338}_{-0.472}$ \\ 
1259412 & $-1.996^{+0.250}_{-0.003}$ & $0.304^{+0.101}_{-0.008}$ & $1.544^{+0.014}_{-0.088} $ & $-0.579^{+0.014}_{-0.008}$  & $0.195^{+0.705}_{-0.720}$ & $0.233^{+0.306}_{-0.194}$ & $1.624^{+0.950}_{-0.489} $ & $0.032^{+0.187}_{-0.306}$ \\ 
1261579 & $-1.000^{+1.250}_{-0.429}$ & $5.447^{+0.262}_{-1.446}$ & $0.312^{+0.133}_{-0.046} $ & $-2.189^{+0.450}_{-0.009}$  & $-0.543^{+0.800}_{-0.674}$ & $1.254^{+0.967}_{-0.650}$ & $1.641^{+0.488}_{-0.513} $ & $-0.280^{+0.174}_{-0.211}$ \\ 
1263369 & $-0.883^{+0.242}_{-0.151}$ & $0.563^{+0.171}_{-0.145}$ & $1.990^{+0.128}_{-0.145} $ & $-0.078^{+0.053}_{-0.037}$  & $0.259^{+0.528}_{-0.583}$ & $0.445^{+0.318}_{-0.191}$ & $1.924^{+0.225}_{-0.274} $ & $-0.105^{+0.118}_{-0.139}$ \\ 
1263715 & $0.010^{+0.559}_{-0.508}$ & $0.874^{+0.358}_{-0.303}$ & $0.635^{+0.188}_{-0.199} $ & $-0.482^{+0.231}_{-0.278}$  & $-0.505^{+0.585}_{-0.516}$ & $0.849^{+0.327}_{-0.326}$ & $0.639^{+0.310}_{-0.240} $ & $-0.611^{+0.370}_{-0.439}$ \\ 
1275946 & $-0.666^{+1.016}_{-0.910}$ & $1.597^{+0.546}_{-0.623}$ & $0.450^{+0.311}_{-0.194} $ & $-1.248^{+0.636}_{-0.601}$  & $-0.618^{+0.546}_{-0.526}$ & $0.558^{+0.258}_{-0.315}$ & $0.475^{+0.449}_{-0.240} $ & $-0.679^{+0.572}_{-0.752}$ \\ 
1279500 & $-1.696^{+0.250}_{-0.219}$ & $0.506^{+0.394}_{-0.149}$ & $1.511^{+0.121}_{-0.231} $ & $-0.550^{+0.019}_{-0.021}$  & $-0.397^{+0.983}_{-0.981}$ & $2.174^{+0.541}_{-0.701}$ & $0.407^{+0.228}_{-0.140} $ & $-1.466^{+0.544}_{-0.504}$ \\ 
1280217 & $-0.085^{+1.019}_{-0.713}$ & $0.094^{+0.543}_{-0.088}$ & $1.963^{+0.888}_{-1.112} $ & $0.025^{+0.220}_{-0.351}$  & $1.052^{+0.302}_{-1.153}$ & $0.461^{+0.695}_{-0.273}$ & $0.369^{+0.233}_{-0.194} $ & $-1.746^{+0.562}_{-0.332}$ \\ 
1281668 & $-0.092^{+0.183}_{-0.151}$ & $0.564^{+0.112}_{-0.075}$ & $1.714^{+0.077}_{-0.102} $ & $0.094^{+0.019}_{-0.037}$  & $-0.242^{+0.707}_{-0.449}$ & $0.301^{+0.323}_{-0.210}$ & $1.958^{+0.675}_{-0.431} $ & $0.095^{+0.135}_{-0.147}$ \\ 
1281886 & $-0.592^{+0.553}_{-0.320}$ & $1.556^{+0.464}_{-0.520}$ & $0.971^{+0.275}_{-0.248} $ & $-0.370^{+0.154}_{-0.210}$  & $-0.534^{+0.682}_{-0.655}$ & $1.053^{+0.735}_{-0.473}$ & $1.167^{+0.383}_{-0.427} $ & $-0.264^{+0.184}_{-0.274}$ \\ 
1282736 & $-1.947^{+0.144}_{-0.040}$ & $0.781^{+0.128}_{-0.053}$ & $1.648^{+0.032}_{-0.059} $ & $-0.533^{+0.028}_{-0.015}$  & $-0.619^{+0.640}_{-0.657}$ & $0.796^{+0.664}_{-0.403}$ & $1.864^{+0.378}_{-0.464} $ & $-0.161^{+0.124}_{-0.195}$ \\ 
1282757 & $-1.968^{+0.567}_{-0.026}$ & $0.109^{+0.206}_{-0.005}$ & $1.733^{+0.030}_{-0.379} $ & $-0.416^{+0.018}_{-0.035}$  & $-1.000^{+0.639}_{-0.623}$ & $0.559^{+0.536}_{-0.389}$ & $0.982^{+0.485}_{-0.409} $ & $-0.495^{+0.290}_{-0.325}$ \\ 
1283373 & $-0.538^{+0.067}_{-0.073}$ & $4.905^{+0.029}_{-0.031}$ & $0.563^{+0.030}_{-0.025} $ & $-1.078^{+0.040}_{-0.046}$  & $-0.160^{+0.966}_{-0.739}$ & $3.981^{+0.849}_{-1.235}$ & $0.620^{+0.213}_{-0.171} $ & $-1.278^{+0.364}_{-0.451}$ \\ 
1283878 & $1.286^{+0.005}_{-0.260}$ & $0.103^{+0.177}_{-0.002}$ & $3.187^{+0.009}_{-0.328} $ & $-0.266^{+0.080}_{-0.003}$  & $-1.021^{+1.794}_{-0.755}$ & $3.353^{+0.492}_{-1.351}$ & $0.509^{+0.227}_{-0.111} $ & $-1.878^{+0.411}_{-0.235}$ \\ 
1283936 & $-0.483^{+0.318}_{-0.483}$ & $0.521^{+0.113}_{-0.242}$ & $1.738^{+0.251}_{-0.140} $ & $0.078^{+0.048}_{-0.119}$  & $-0.318^{+0.662}_{-0.573}$ & $0.475^{+0.311}_{-0.206}$ & $1.565^{+0.326}_{-0.312} $ & $-0.045^{+0.122}_{-0.162}$ \\ 
1284587 & $-1.924^{+0.195}_{-0.074}$ & $0.700^{+0.386}_{-0.120}$ & $1.659^{+0.081}_{-0.193} $ & $-0.633^{+0.018}_{-0.023}$  & $-0.364^{+0.871}_{-0.739}$ & $2.002^{+0.849}_{-0.816}$ & $0.800^{+0.478}_{-0.320} $ & $-0.733^{+0.422}_{-0.642}$ \\ 
1285317 & $-0.061^{+0.496}_{-0.679}$ & $0.559^{+0.154}_{-0.199}$ & $0.165^{+0.212}_{-0.094} $ & $-0.937^{+0.572}_{-0.678}$  & $-1.509^{+0.564}_{-0.321}$ & $0.267^{+0.143}_{-0.079}$ & $0.215^{+0.148}_{-0.063} $ & $-1.747^{+0.751}_{-0.316}$ \\ 
1286398 & $-0.133^{+0.026}_{-0.012}$ & $3.388^{+0.015}_{-0.033}$ & $0.517^{+0.012}_{-0.006} $ & $-1.120^{+0.019}_{-0.030}$  & $0.647^{+0.909}_{-0.826}$ & $2.122^{+1.358}_{-1.042}$ & $0.914^{+0.455}_{-0.334} $ & $-1.042^{+0.578}_{-0.715}$ \\ 
1287626 & $-0.761^{+0.791}_{-0.740}$ & $1.199^{+0.438}_{-0.508}$ & $0.355^{+0.319}_{-0.185} $ & $-1.101^{+0.629}_{-0.670}$  & $2.264^{+0.169}_{-0.736}$ & $0.036^{+0.117}_{-0.005}$ & $0.932^{+0.077}_{-0.396} $ & $-2.074^{+0.189}_{-0.090}$ \\ 
1289288 & $-1.608^{+0.175}_{-0.113}$ & $0.671^{+0.344}_{-0.125}$ & $1.853^{+0.069}_{-0.185} $ & $-0.454^{+0.017}_{-0.017}$  & $-1.188^{+0.794}_{-0.509}$ & $0.558^{+0.695}_{-0.475}$ & $2.048^{+0.821}_{-0.538} $ & $-0.223^{+0.159}_{-0.177}$ \\ 
1289555 & $-0.103^{+0.415}_{-0.377}$ & $0.607^{+0.248}_{-0.290}$ & $0.464^{+0.334}_{-0.217} $ & $-0.418^{+0.248}_{-0.387}$  & $-0.707^{+0.792}_{-0.517}$ & $0.740^{+0.220}_{-0.337}$ & $0.234^{+0.293}_{-0.139} $ & $-0.979^{+0.495}_{-0.662}$ \\ 
1289600 & $-0.592^{+0.153}_{-0.049}$ & $0.185^{+0.071}_{-0.042}$ & $1.626^{+0.163}_{-0.230} $ & $0.126^{+0.021}_{-0.026}$  & $-0.416^{+0.378}_{-0.341}$ & $0.393^{+0.183}_{-0.149}$ & $0.809^{+0.285}_{-0.257} $ & $-0.154^{+0.176}_{-0.260}$ \\ 
1289656 & $-0.326^{+0.984}_{-1.044}$ & $0.949^{+0.382}_{-0.382}$ & $0.320^{+0.244}_{-0.146} $ & $-1.320^{+0.567}_{-0.542}$  & $-0.195^{+0.573}_{-0.533}$ & $0.461^{+0.387}_{-0.304}$ & $0.657^{+0.605}_{-0.397} $ & $-0.311^{+0.368}_{-0.583}$ \\ 
1289664 & $-0.093^{+0.883}_{-0.896}$ & $0.477^{+0.941}_{-0.371}$ & $1.374^{+0.967}_{-0.810} $ & $-0.087^{+0.343}_{-0.672}$  & $-0.542^{+0.891}_{-0.494}$ & $0.196^{+0.380}_{-0.172}$ & $0.953^{+0.936}_{-0.627} $ & $0.083^{+0.237}_{-0.490}$ \\ 
 \hline

    \end{tabular}

\end{table*}
\setcounter{table}{0}
\begin{table*}[t]

\centering
    \caption{Continued from above}
    \begin{tabular}{c |c c c c|c c c c }
    \hline
    \multirow{2}{*}{SNID} & \multicolumn{4}{c|}{Global} & \multicolumn{4}{c}{Local}\\
    &  $\log(Z_\star/Z_\odot)$ & $t_{\textrm{age}}$ (Gyr) & $\tau_V$ & $n$  &  $\log(Z_\star/Z_\odot)$ & $t_{\textrm{age}}$ (Gyr) & $\tau_V$ & $n$\\\hline
1290816 & $0.327^{+0.360}_{-0.379}$ & $0.582^{+0.184}_{-0.137}$ & $1.140^{+0.152}_{-0.142} $ & $-0.149^{+0.172}_{-0.228}$  & $-0.574^{+0.633}_{-0.575}$ & $0.848^{+0.552}_{-0.474}$ & $0.759^{+0.542}_{-0.388} $ & $-0.480^{+0.439}_{-0.703}$ \\ 
1291080 & $-0.347^{+1.106}_{-1.142}$ & $1.571^{+0.659}_{-0.741}$ & $0.508^{+0.411}_{-0.212} $ & $-1.278^{+0.567}_{-0.556}$  & $0.634^{+0.694}_{-1.479}$ & $1.474^{+1.007}_{-0.804}$ & $0.512^{+0.326}_{-0.192} $ & $-1.639^{+0.539}_{-0.398}$ \\ 
1291090 & $-0.112^{+0.212}_{-0.390}$ & $0.387^{+0.084}_{-0.065}$ & $0.856^{+0.087}_{-0.101} $ & $-0.152^{+0.066}_{-0.078}$  & $-0.105^{+0.576}_{-0.583}$ & $0.562^{+0.259}_{-0.223}$ & $0.847^{+0.245}_{-0.218} $ & $-0.330^{+0.219}_{-0.274}$ \\ 
1291794 & $-1.693^{+0.233}_{-0.210}$ & $0.838^{+0.563}_{-0.141}$ & $1.785^{+0.053}_{-0.252} $ & $-0.457^{+0.029}_{-0.044}$  & $-1.244^{+0.760}_{-0.492}$ & $0.484^{+0.607}_{-0.303}$ & $2.237^{+0.419}_{-0.448} $ & $-0.199^{+0.180}_{-0.137}$ \\ 
1292145 & $-0.535^{+0.369}_{-0.315}$ & $0.256^{+0.204}_{-0.144}$ & $1.736^{+0.478}_{-0.367} $ & $-0.002^{+0.101}_{-0.141}$  & $-0.330^{+0.588}_{-0.585}$ & $0.416^{+0.390}_{-0.233}$ & $1.288^{+0.425}_{-0.441} $ & $-0.178^{+0.171}_{-0.245}$ \\ 
1292332 & $-0.853^{+0.300}_{-0.374}$ & $3.376^{+0.096}_{-0.196}$ & $0.674^{+0.098}_{-0.055} $ & $-1.272^{+0.058}_{-0.058}$  & $-0.637^{+1.126}_{-0.852}$ & $3.396^{+0.957}_{-1.208}$ & $0.800^{+0.372}_{-0.273} $ & $-1.031^{+0.330}_{-0.429}$ \\ 
1292336 & $-1.887^{+0.130}_{-0.081}$ & $1.356^{+0.510}_{-0.350}$ & $0.908^{+0.168}_{-0.209} $ & $-1.076^{+0.133}_{-0.204}$  & $-1.293^{+0.739}_{-0.536}$ & $1.088^{+0.896}_{-0.863}$ & $1.085^{+0.714}_{-0.464} $ & $-0.508^{+0.221}_{-0.289}$ \\ 
1292560 & $0.278^{+0.345}_{-0.383}$ & $0.176^{+0.085}_{-0.068}$ & $1.732^{+0.241}_{-0.268} $ & $0.242^{+0.116}_{-0.198}$  & $-0.655^{+0.571}_{-0.594}$ & $1.067^{+0.404}_{-0.468}$ & $0.488^{+0.367}_{-0.232} $ & $-0.830^{+0.527}_{-0.691}$ \\ 
1293319 & $-0.744^{+0.380}_{-0.375}$ & $0.654^{+0.203}_{-0.284}$ & $1.215^{+0.255}_{-0.189} $ & $-0.165^{+0.123}_{-0.154}$  & $-1.167^{+1.050}_{-0.564}$ & $1.632^{+0.490}_{-0.590}$ & $0.466^{+0.289}_{-0.173} $ & $-1.430^{+0.522}_{-0.480}$ \\ 
1293758 & $-2.000^{+0.0005}_{-0.0001}$ & $0.771^{+0.014}_{-0.007}$ & $1.499^{+0.005}_{-0.009} $ & $-0.789^{+0.003}_{-0.006}$  & $-0.059^{+0.749}_{-0.633}$ & $1.051^{+0.863}_{-0.453}$ & $1.487^{+0.553}_{-0.472} $ & $-0.108^{+0.292}_{-0.407}$ \\ 
1294014 & $-0.082^{+0.346}_{-0.306}$ & $0.530^{+0.183}_{-0.142}$ & $0.633^{+0.133}_{-0.173} $ & $-0.154^{+0.148}_{-0.184}$  & $-0.418^{+0.588}_{-0.593}$ & $0.932^{+0.363}_{-0.351}$ & $0.431^{+0.273}_{-0.217} $ & $-0.674^{+0.416}_{-0.558}$ \\ 
1294743 & $-1.993^{+0.295}_{-0.006}$ & $0.191^{+0.124}_{-0.013}$ & $2.056^{+0.039}_{-0.222} $ & $-0.410^{+0.020}_{-0.031}$  & $-0.191^{+0.920}_{-0.974}$ & $2.004^{+0.782}_{-0.875}$ & $0.535^{+0.393}_{-0.236} $ & $-1.018^{+0.518}_{-0.647}$ \\ 
1295027 & $-0.763^{+0.298}_{-0.244}$ & $0.728^{+0.186}_{-0.191}$ & $1.603^{+0.161}_{-0.153} $ & $-0.054^{+0.051}_{-0.078}$  & $-0.519^{+0.609}_{-0.631}$ & $0.578^{+0.513}_{-0.299}$ & $1.609^{+0.366}_{-0.440} $ & $-0.051^{+0.112}_{-0.232}$ \\ 
1296321 & $-1.617^{+0.196}_{-0.170}$ & $0.494^{+0.336}_{-0.209}$ & $1.176^{+0.222}_{-0.218} $ & $-0.652^{+0.070}_{-0.070}$  & $-0.982^{+0.598}_{-0.399}$ & $0.513^{+0.734}_{-0.461}$ & $1.204^{+1.199}_{-0.606} $ & $-0.262^{+0.337}_{-0.443}$ \\ 
1296657 & $-1.279^{+0.219}_{-0.344}$ & $0.030^{+0.178}_{-0.014}$ & $1.985^{+0.070}_{-0.811} $ & $-0.062^{+0.038}_{-0.222}$  & $-0.671^{+0.451}_{-0.298}$ & $0.205^{+0.295}_{-0.178}$ & $1.132^{+0.791}_{-0.649} $ & $0.088^{+0.214}_{-0.381}$ \\ 
1297026 & $-1.004^{+0.003}_{-0.007}$ & $5.109^{+0.024}_{-0.024}$ & $0.826^{+0.007}_{-0.007} $ & $-0.854^{+0.011}_{-0.010}$  & $0.159^{+1.749}_{-1.012}$ & $6.508^{+4.506}_{-6.180}$ & $0.656^{+1.462}_{-0.326} $ & $-1.871^{+0.508}_{-0.245}$ \\ 
1297465 & $0.127^{+0.198}_{-0.161}$ & $0.564^{+0.056}_{-0.051}$ & $1.726^{+0.046}_{-0.044} $ & $0.027^{+0.048}_{-0.062}$  & $-0.988^{+0.601}_{-0.606}$ & $1.813^{+0.641}_{-0.734}$ & $0.610^{+0.387}_{-0.292} $ & $-0.914^{+0.438}_{-0.627}$ \\ 
1298281 & $-1.407^{+0.539}_{-0.353}$ & $3.489^{+0.253}_{-0.433}$ & $0.455^{+0.126}_{-0.087} $ & $-1.959^{+0.245}_{-0.167}$  & $-0.524^{+1.067}_{-0.889}$ & $2.839^{+0.763}_{-1.050}$ & $0.678^{+0.353}_{-0.223} $ & $-1.196^{+0.459}_{-0.526}$ \\ 
1298893 & $-0.860^{+0.061}_{-0.067}$ & $4.534^{+0.046}_{-0.053}$ & $0.611^{+0.027}_{-0.025} $ & $-1.128^{+0.030}_{-0.030}$  & $-0.452^{+0.984}_{-0.790}$ & $2.705^{+1.050}_{-0.974}$ & $1.200^{+0.429}_{-0.365} $ & $-0.530^{+0.203}_{-0.311}$ \\ 
1299643 & $-0.957^{+0.687}_{-0.692}$ & $0.454^{+0.316}_{-0.223}$ & $2.131^{+0.195}_{-0.214} $ & $-0.093^{+0.137}_{-0.218}$  & $-0.267^{+0.723}_{-0.537}$ & $0.494^{+0.396}_{-0.242}$ & $1.906^{+0.392}_{-0.382} $ & $-0.047^{+0.116}_{-0.152}$ \\ 
1299775 & $-1.724^{+0.019}_{-0.018}$ & $1.372^{+0.245}_{-0.190}$ & $2.250^{+0.089}_{-0.108} $ & $-0.320^{+0.005}_{-0.006}$  & $-0.007^{+1.043}_{-0.934}$ & $2.616^{+1.336}_{-1.101}$ & $1.577^{+0.562}_{-0.494} $ & $-0.447^{+0.219}_{-0.400}$ \\ 
1299785 & $-0.444^{+0.195}_{-0.238}$ & $1.704^{+0.142}_{-0.180}$ & $0.797^{+0.090}_{-0.072} $ & $-0.275^{+0.066}_{-0.075}$  & $-0.491^{+0.695}_{-0.612}$ & $1.381^{+0.469}_{-0.554}$ & $0.448^{+0.346}_{-0.228} $ & $-0.770^{+0.411}_{-0.625}$ \\ 
1300516 & $1.386^{+0.045}_{-0.373}$ & $0.165^{+0.535}_{-0.045}$ & $2.594^{+0.162}_{-0.419} $ & $-0.479^{+0.203}_{-0.066}$  & $-0.838^{+0.720}_{-0.671}$ & $1.850^{+0.985}_{-0.912}$ & $1.369^{+0.520}_{-0.450} $ & $-0.407^{+0.201}_{-0.218}$ \\ 
1300912 & $0.026^{+0.672}_{-0.334}$ & $0.793^{+0.370}_{-0.351}$ & $0.882^{+0.114}_{-0.237} $ & $-0.312^{+0.157}_{-0.134}$  & $0.187^{+0.749}_{-0.848}$ & $0.998^{+0.751}_{-0.560}$ & $0.612^{+0.449}_{-0.325} $ & $-0.680^{+0.445}_{-0.635}$ \\ 
1301933 & $-0.576^{+0.234}_{-0.324}$ & $0.707^{+0.068}_{-0.186}$ & $1.544^{+0.169}_{-0.069} $ & $-0.007^{+0.045}_{-0.080}$  & - & - & - & - \\ 
1302058 & $-1.871^{+0.209}_{-0.104}$ & $0.249^{+0.212}_{-0.052}$ & $1.973^{+0.112}_{-0.255} $ & $-0.438^{+0.034}_{-0.045}$  & $-0.525^{+0.765}_{-0.709}$ & $1.400^{+0.823}_{-0.680}$ & $1.013^{+0.513}_{-0.431} $ & $-0.466^{+0.323}_{-0.481}$ \\ 
1302187 & $-1.743^{+0.300}_{-0.067}$ & $0.412^{+0.514}_{-0.048}$ & $2.382^{+0.047}_{-0.337} $ & $-0.350^{+0.009}_{-0.003}$  & $-1.815^{+0.322}_{-0.134}$ & $11.206^{+0.701}_{-0.694}$ & $0.544^{+0.063}_{-0.049} $ & $-2.049^{+0.204}_{-0.112}$ \\ 
1302523 & $-1.702^{+0.890}_{-0.218}$ & $0.093^{+0.489}_{-0.036}$ & $2.480^{+0.384}_{-0.609} $ & $-0.212^{+0.198}_{-0.046}$  & $-0.532^{+0.566}_{-0.483}$ & $0.542^{+0.373}_{-0.248}$ & $1.985^{+0.370}_{-0.378} $ & $-0.053^{+0.098}_{-0.140}$ \\ 
1302648 & $-1.943^{+0.086}_{-0.042}$ & $2.940^{+0.188}_{-0.123}$ & $1.031^{+0.041}_{-0.060} $ & $-1.010^{+0.031}_{-0.034}$  & $-0.168^{+0.994}_{-1.085}$ & $3.360^{+0.957}_{-1.181}$ & $0.699^{+0.318}_{-0.227} $ & $-1.198^{+0.376}_{-0.462}$ \\ 
1303279 & $1.138^{+0.005}_{-0.005}$ & $0.223^{+0.002}_{-0.002}$ & $2.836^{+0.008}_{-0.008} $ & $-0.152^{+0.005}_{-0.005}$  & $-1.457^{+0.533}_{-0.398}$ & $5.104^{+1.496}_{-1.902}$ & $1.120^{+0.622}_{-0.387} $ & $-0.885^{+0.336}_{-0.483}$ \\ 
1303496 & $-1.999^{+0.003}_{-0.001}$ & $0.464^{+0.075}_{-0.019}$ & $2.217^{+0.022}_{-0.078} $ & $-0.456^{+0.009}_{-0.030}$  & $-0.629^{+0.702}_{-0.582}$ & $1.558^{+1.107}_{-0.851}$ & $1.710^{+0.630}_{-0.561} $ & $-0.309^{+0.199}_{-0.259}$ \\ 
1303883 & $0.030^{+0.081}_{-0.033}$ & $3.056^{+0.074}_{-0.131}$ & $1.017^{+0.032}_{-0.025} $ & $-0.557^{+0.018}_{-0.021}$  & $0.076^{+0.825}_{-0.834}$ & $1.854^{+0.962}_{-0.792}$ & $1.148^{+0.379}_{-0.331} $ & $-0.556^{+0.215}_{-0.270}$ \\ 
1303952 & $-0.251^{+0.005}_{-0.007}$ & $2.009^{+0.104}_{-0.109}$ & $1.110^{+0.061}_{-0.056} $ & $-0.505^{+0.033}_{-0.036}$  & $-0.955^{+1.469}_{-0.758}$ & $2.745^{+0.774}_{-1.169}$ & $0.575^{+0.409}_{-0.191} $ & $-1.526^{+0.520}_{-0.456}$ \\ 
1304442 & $-0.141^{+0.078}_{-0.607}$ & $0.435^{+0.031}_{-0.189}$ & $1.286^{+0.426}_{-0.054} $ & $0.306^{+0.020}_{-0.056}$  & $-0.101^{+0.730}_{-0.647}$ & $1.340^{+0.393}_{-0.518}$ & $0.311^{+0.257}_{-0.135} $ & $-1.181^{+0.680}_{-0.675}$ \\

 \hline

    \end{tabular}

\end{table*}

\setcounter{table}{0}
\begin{table*}[t]

\centering
    \caption{Continued from above}
    \begin{tabular}{c |c c c c|c c c c }
    \hline
    \multirow{2}{*}{SNID} & \multicolumn{4}{c|}{Global} & \multicolumn{4}{c}{Local}\\
    &  $\log(Z_\star/Z_\odot)$ & $t_{\textrm{age}}$ (Gyr) & $\tau_V$ & $n$  &  $\log(Z_\star/Z_\odot)$ & $t_{\textrm{age}}$ (Gyr) & $\tau_V$ & $n$\\\hline

1304678 & $1.256^{+0.009}_{-0.010}$ & $0.696^{+0.009}_{-0.008}$ & $2.140^{+0.012}_{-0.012} $ & $-0.415^{+0.008}_{-0.008}$  & $-1.266^{+0.716}_{-0.536}$ & $5.240^{+1.701}_{-1.847}$ & $1.108^{+0.566}_{-0.386} $ & $-0.736^{+0.273}_{-0.421}$ \\ 
1305504 & $0.248^{+0.004}_{-0.009}$ & $3.520^{+0.058}_{-0.692}$ & $0.817^{+0.453}_{-0.027} $ & $-1.026^{+0.427}_{-0.035}$  & $-0.563^{+1.344}_{-1.037}$ & $2.884^{+0.896}_{-1.185}$ & $0.551^{+0.366}_{-0.181} $ & $-1.563^{+0.474}_{-0.433}$ \\ 
1305626 & $-0.284^{+0.469}_{-0.383}$ & $0.837^{+0.182}_{-0.257}$ & $1.470^{+0.153}_{-0.143} $ & $-0.049^{+0.067}_{-0.091}$  & $-0.466^{+0.912}_{-0.883}$ & $2.113^{+0.602}_{-0.738}$ & $0.452^{+0.304}_{-0.199} $ & $-1.087^{+0.473}_{-0.632}$ \\ 
1306073 & $-0.994^{+0.020}_{-0.011}$ & $5.212^{+0.027}_{-0.027}$ & $0.323^{+0.009}_{-0.009} $ & $-1.873^{+0.042}_{-0.044}$  & $-1.073^{+0.844}_{-0.679}$ & $6.839^{+0.977}_{-1.403}$ & $0.378^{+0.225}_{-0.120} $ & $-1.560^{+0.537}_{-0.437}$ \\ 
1306141 & $0.702^{+0.247}_{-0.694}$ & $0.311^{+0.214}_{-0.146}$ & $1.360^{+0.167}_{-0.101} $ & $-0.262^{+0.299}_{-0.076}$  & $-0.606^{+0.731}_{-0.774}$ & $1.575^{+0.439}_{-0.562}$ & $0.394^{+0.313}_{-0.185} $ & $-1.026^{+0.485}_{-0.606}$ \\ 
1306360 & $-0.750^{+0.012}_{-0.013}$ & $0.893^{+0.041}_{-0.039}$ & $1.845^{+0.035}_{-0.036} $ & $-0.175^{+0.011}_{-0.012}$  & $-0.568^{+0.582}_{-0.571}$ & $1.317^{+0.632}_{-0.508}$ & $1.396^{+0.317}_{-0.343} $ & $-0.393^{+0.110}_{-0.161}$ \\ 
1306390 & $-1.248^{+0.261}_{-0.204}$ & $1.592^{+0.340}_{-0.510}$ & $1.585^{+0.220}_{-0.145} $ & $-0.417^{+0.045}_{-0.036}$  & $-0.866^{+0.674}_{-0.595}$ & $1.881^{+0.851}_{-0.775}$ & $1.499^{+0.389}_{-0.394} $ & $-0.409^{+0.137}_{-0.119}$ \\ 
1306537 & $-0.092^{+0.591}_{-0.625}$ & $0.291^{+0.299}_{-0.186}$ & $2.922^{+0.458}_{-0.403} $ & $0.048^{+0.076}_{-0.090}$  & $-0.412^{+0.619}_{-0.759}$ & $1.164^{+1.217}_{-0.680}$ & $1.432^{+0.625}_{-0.657} $ & $-0.283^{+0.217}_{-0.284}$ \\ 
1306626 & $-0.442^{+0.447}_{-0.392}$ & $0.481^{+0.203}_{-0.200}$ & $0.688^{+0.274}_{-0.234} $ & $-0.216^{+0.181}_{-0.256}$  & $-0.422^{+0.609}_{-0.547}$ & $1.070^{+0.313}_{-0.360}$ & $0.348^{+0.227}_{-0.172} $ & $-0.826^{+0.396}_{-0.544}$ \\ 
1306785 & $-0.301^{+0.349}_{-0.258}$ & $3.466^{+0.461}_{-0.684}$ & $1.097^{+0.202}_{-0.158} $ & $-0.912^{+0.138}_{-0.144}$  & $-0.431^{+0.812}_{-0.795}$ & $1.685^{+1.275}_{-1.002}$ & $1.308^{+0.687}_{-0.564} $ & $-0.591^{+0.231}_{-0.361}$ \\ 
1306980 & $-0.479^{+0.237}_{-0.157}$ & $0.559^{+0.116}_{-0.160}$ & $1.882^{+0.230}_{-0.141} $ & $0.031^{+0.028}_{-0.030}$  & $-0.258^{+0.534}_{-0.379}$ & $0.418^{+0.398}_{-0.214}$ & $2.073^{+0.374}_{-0.451} $ & $0.008^{+0.075}_{-0.108}$ \\ 
1306991 & $-0.744^{+0.037}_{-0.022}$ & $2.058^{+0.060}_{-0.061}$ & $1.175^{+0.033}_{-0.033} $ & $-0.647^{+0.022}_{-0.023}$  & $-0.669^{+0.753}_{-0.703}$ & $1.996^{+0.827}_{-0.764}$ & $1.163^{+0.359}_{-0.353} $ & $-0.789^{+0.188}_{-0.255}$ \\ 
1307277 & $-0.823^{+0.365}_{-0.246}$ & $0.890^{+0.150}_{-0.210}$ & $1.906^{+0.123}_{-0.109} $ & $-0.118^{+0.057}_{-0.070}$  & $-0.531^{+0.806}_{-0.707}$ & $1.219^{+0.997}_{-0.638}$ & $1.456^{+0.518}_{-0.562} $ & $-0.278^{+0.182}_{-0.254}$ \\ 
1307830 & $-0.752^{+0.003}_{-0.006}$ & $7.099^{+0.013}_{-0.013}$ & $0.254^{+0.002}_{-0.002} $ & $-2.194^{+0.009}_{-0.005}$  & $0.008^{+0.951}_{-0.916}$ & $2.480^{+0.868}_{-1.072}$ & $0.767^{+0.293}_{-0.224} $ & $-1.262^{+0.314}_{-0.357}$ \\ 
1308326 & $1.064^{+0.085}_{-0.370}$ & $0.206^{+0.351}_{-0.067}$ & $1.110^{+0.166}_{-0.334} $ & $-0.561^{+0.056}_{-0.046}$  & $-0.560^{+0.660}_{-0.700}$ & $1.205^{+0.357}_{-0.459}$ & $0.354^{+0.273}_{-0.177} $ & $-0.983^{+0.516}_{-0.640}$ \\ 
1308582 & $-0.002^{+0.015}_{-0.747}$ & $1.706^{+0.952}_{-0.071}$ & $1.428^{+0.042}_{-0.347} $ & $-0.487^{+0.020}_{-0.187}$  & $1.400^{+0.194}_{-0.944}$ & $1.242^{+1.265}_{-0.386}$ & $0.691^{+0.238}_{-0.181} $ & $-1.765^{+0.350}_{-0.289}$ \\ 
1308884 & $-1.250^{+0.00005}_{-0.0001}$ & $1.133^{+0.005}_{-0.005}$ & $1.079^{+0.003}_{-0.003} $ & $-0.478^{+0.002}_{-0.002}$  & $-0.618^{+0.419}_{-0.513}$ & $2.182^{+0.206}_{-0.456}$ & $0.258^{+0.209}_{-0.092} $ & $-1.266^{+0.652}_{-0.631}$ \\ 
1309288 & $-1.999^{+0.002}_{-0.001}$ & $1.113^{+0.032}_{-0.026}$ & $1.504^{+0.014}_{-0.016} $ & $-0.708^{+0.006}_{-0.007}$  & $-0.595^{+0.414}_{-0.502}$ & $0.092^{+1.961}_{-0.070}$ & $3.711^{+0.842}_{-2.096} $ & $-0.009^{+0.179}_{-0.372}$ \\ 
1309492 & $0.852^{+0.153}_{-0.845}$ & $0.197^{+0.261}_{-0.075}$ & $1.859^{+0.156}_{-0.165} $ & $-0.160^{+0.263}_{-0.058}$  & $0.441^{+0.543}_{-0.660}$ & $0.738^{+0.755}_{-0.424}$ & $1.226^{+0.495}_{-0.489} $ & $-0.316^{+0.229}_{-0.321}$ \\ 
1312274 & $-0.700^{+0.346}_{-0.310}$ & $0.677^{+0.166}_{-0.207}$ & $2.884^{+0.182}_{-0.160} $ & $0.013^{+0.042}_{-0.063}$  & $-0.347^{+0.677}_{-0.618}$ & $0.674^{+0.536}_{-0.295}$ & $2.597^{+0.329}_{-0.419} $ & $-0.068^{+0.073}_{-0.103}$ \\ 
1313594 & $-0.272^{+0.142}_{-0.217}$ & $0.053^{+0.116}_{-0.036}$ & $4.293^{+0.414}_{-0.784} $ & $0.325^{+0.027}_{-0.096}$  & $-0.546^{+0.684}_{-0.628}$ & $0.286^{+0.571}_{-0.222}$ & $2.174^{+0.956}_{-0.762} $ & $-0.013^{+0.217}_{-0.293}$ \\ 
1314897 & $0.966^{+0.142}_{-0.964}$ & $0.228^{+0.419}_{-0.098}$ & $2.410^{+0.223}_{-0.159} $ & $-0.246^{+0.247}_{-0.021}$  & $0.878^{+0.433}_{-1.026}$ & $0.984^{+1.025}_{-0.509}$ & $0.939^{+0.430}_{-0.373} $ & $-0.812^{+0.359}_{-0.440}$ \\ 
1315192 & $-1.536^{+0.082}_{-0.080}$ & $0.711^{+0.193}_{-0.116}$ & $1.738^{+0.072}_{-0.111} $ & $-0.441^{+0.010}_{-0.010}$  & $-0.965^{+0.579}_{-0.588}$ & $1.212^{+0.775}_{-0.791}$ & $1.153^{+0.543}_{-0.428} $ & $-0.438^{+0.196}_{-0.259}$ \\ 
1315259 & $-2.000^{+0.0003}_{-0.0001}$ & $1.884^{+0.022}_{-0.023}$ & $1.332^{+0.009}_{-0.009} $ & $-0.777^{+0.005}_{-0.005}$  & $-0.911^{+0.869}_{-0.717}$ & $2.828^{+0.658}_{-0.934}$ & $0.659^{+0.371}_{-0.234} $ & $-1.033^{+0.337}_{-0.444}$ \\ 
1315296 & $-0.662^{+0.363}_{-0.093}$ & $0.464^{+0.107}_{-0.090}$ & $1.021^{+0.142}_{-0.152} $ & $-0.118^{+0.045}_{-0.047}$  & $-0.385^{+0.461}_{-0.373}$ & $0.312^{+0.181}_{-0.135}$ & $1.133^{+0.323}_{-0.272} $ & $-0.036^{+0.133}_{-0.184}$ \\ 
1316385 & $1.379^{+0.007}_{-0.008}$ & $0.554^{+0.006}_{-0.006}$ & $1.745^{+0.011}_{-0.011} $ & $-0.628^{+0.013}_{-0.012}$  & $-0.152^{+0.918}_{-0.846}$ & $2.506^{+0.859}_{-0.900}$ & $0.635^{+0.318}_{-0.224} $ & $-1.221^{+0.531}_{-0.592}$ \\ 
1316431 & $-0.560^{+0.884}_{-0.620}$ & $0.254^{+0.150}_{-0.162}$ & $0.291^{+0.349}_{-0.199} $ & $-0.900^{+0.853}_{-0.871}$  & $1.096^{+0.312}_{-0.339}$ & $0.151^{+0.184}_{-0.082}$ & $0.376^{+0.347}_{-0.254} $ & $-0.671^{+0.746}_{-0.931}$ \\ 
1316437 & $-0.326^{+0.400}_{-0.319}$ & $0.575^{+0.145}_{-0.176}$ & $0.556^{+0.180}_{-0.154} $ & $-0.312^{+0.151}_{-0.150}$  & $1.103^{+0.332}_{-1.768}$ & $0.356^{+0.496}_{-0.192}$ & $0.297^{+0.194}_{-0.149} $ & $-1.648^{+0.667}_{-0.390}$ \\ 
1316465 & $-0.589^{+0.285}_{-0.333}$ & $0.676^{+0.111}_{-0.199}$ & $1.102^{+0.187}_{-0.114} $ & $-0.113^{+0.079}_{-0.109}$  & $-0.493^{+0.539}_{-0.554}$ & $0.763^{+0.369}_{-0.322}$ & $0.491^{+0.348}_{-0.270} $ & $-0.511^{+0.450}_{-0.635}$ \\ 
1317164 & $-1.998^{+0.003}_{-0.001}$ & $8.560^{+0.041}_{-0.042}$ & $0.957^{+0.006}_{-0.005} $ & $-1.414^{+0.014}_{-0.014}$  & $-0.135^{+1.068}_{-1.024}$ & $3.941^{+1.488}_{-1.462}$ & $0.744^{+0.452}_{-0.282} $ & $-1.030^{+0.503}_{-0.675}$ \\ 
1317277 & $0.216^{+0.353}_{-0.529}$ & $2.692^{+0.745}_{-0.418}$ & $0.674^{+0.052}_{-0.125} $ & $-1.443^{+0.057}_{-0.058}$  & $-0.399^{+1.017}_{-0.909}$ & $3.758^{+0.936}_{-1.185}$ & $0.699^{+0.317}_{-0.235} $ & $-1.172^{+0.348}_{-0.446}$ \\ 
1317286 & $-0.024^{+0.809}_{-0.916}$ & $0.743^{+0.335}_{-0.403}$ & $0.253^{+0.334}_{-0.173} $ & $-0.681^{+0.745}_{-0.903}$  & $0.233^{+0.556}_{-1.020}$ & $0.396^{+0.213}_{-0.215}$ & $0.204^{+0.320}_{-0.151} $ & $-0.366^{+0.555}_{-0.982}$ \\ 
1317666 & $-1.995^{+0.227}_{-0.004}$ & $0.300^{+0.177}_{-0.012}$ & $1.973^{+0.022}_{-0.193} $ & $-0.471^{+0.010}_{-0.047}$  & $0.016^{+0.914}_{-0.848}$ & $2.014^{+0.837}_{-0.814}$ & $0.559^{+0.327}_{-0.223} $ & $-1.061^{+0.623}_{-0.679}$ \\ 
1319366 & $-1.601^{+0.400}_{-0.301}$ & $0.133^{+0.366}_{-0.059}$ & $1.706^{+0.260}_{-0.525} $ & $-0.341^{+0.083}_{-0.111}$  & $-0.860^{+0.613}_{-0.573}$ & $0.472^{+0.436}_{-0.404}$ & $1.077^{+0.850}_{-0.422} $ & $-0.245^{+0.224}_{-0.325}$ \\ 

 \hline

    \end{tabular}

\end{table*}

\setcounter{table}{0}
\begin{table*}[t]

\centering
    \caption{Continued from above}
    \begin{tabular}{c |c c c c|c c c c }
    \hline
    \multirow{2}{*}{SNID} & \multicolumn{4}{c|}{Global} & \multicolumn{4}{c}{Local}\\
    &  $\log(Z_\star/Z_\odot)$ & $t_{\textrm{age}}$ (Gyr) & $\tau_V$ & $n$  &  $\log(Z_\star/Z_\odot)$ & $t_{\textrm{age}}$ (Gyr) & $\tau_V$ & $n$\\\hline

1319821 & $0.131^{+0.402}_{-0.365}$ & $0.291^{+0.055}_{-0.057}$ & $1.783^{+0.133}_{-0.036} $ & $0.143^{+0.064}_{-0.153}$  & $-0.054^{+0.800}_{-0.725}$ & $1.276^{+0.816}_{-0.618}$ & $0.803^{+0.461}_{-0.354} $ & $-0.504^{+0.335}_{-0.478}$ \\ 
1322229 & $-0.114^{+0.317}_{-0.226}$ & $0.312^{+0.120}_{-0.079}$ & $1.818^{+0.144}_{-0.171} $ & $0.155^{+0.036}_{-0.055}$  & $-0.465^{+0.717}_{-0.630}$ & $0.896^{+0.475}_{-0.421}$ & $0.636^{+0.381}_{-0.312} $ & $-0.501^{+0.269}_{-0.392}$ \\ 
1322979 & $1.006^{+0.508}_{-2.104}$ & $0.496^{+0.627}_{-0.321}$ & $0.469^{+0.283}_{-0.194} $ & $-1.701^{+0.518}_{-0.362}$  & $-1.146^{+1.792}_{-0.609}$ & $0.679^{+0.340}_{-0.280}$ & $0.271^{+0.166}_{-0.109} $ & $-1.667^{+0.571}_{-0.379}$ \\ 
1324542 & $-0.247^{+0.785}_{-0.643}$ & $0.136^{+0.545}_{-0.129}$ & $2.141^{+0.932}_{-0.971} $ & $0.078^{+0.158}_{-0.163}$  & $-0.359^{+0.815}_{-0.738}$ & $0.552^{+0.229}_{-0.279}$ & $0.218^{+0.300}_{-0.139} $ & $-0.863^{+0.708}_{-0.780}$ \\ 
1327978 & $1.487^{+0.012}_{-0.019}$ & $2.081^{+0.037}_{-0.025}$ & $0.709^{+0.005}_{-0.006} $ & $-1.740^{+0.021}_{-0.015}$  & $0.861^{+0.775}_{-1.070}$ & $2.797^{+1.702}_{-1.551}$ & $0.889^{+0.440}_{-0.295} $ & $-1.149^{+0.366}_{-0.485}$ \\ 
1328066 & $-0.106^{+0.478}_{-0.338}$ & $1.697^{+0.511}_{-0.588}$ & $0.970^{+0.252}_{-0.215} $ & $-0.752^{+0.141}_{-0.179}$  & $-0.733^{+0.749}_{-0.677}$ & $1.450^{+0.836}_{-0.679}$ & $1.319^{+0.408}_{-0.412} $ & $-0.526^{+0.175}_{-0.206}$ \\ 
1328105 & $-0.358^{+0.230}_{-0.375}$ & $0.534^{+0.110}_{-0.106}$ & $2.059^{+0.161}_{-0.117} $ & $0.033^{+0.028}_{-0.017}$  & $-0.565^{+0.567}_{-0.590}$ & $0.815^{+0.585}_{-0.362}$ & $1.434^{+0.316}_{-0.403} $ & $-0.157^{+0.126}_{-0.197}$ \\ 
1329312 & $-0.742^{+0.022}_{-0.009}$ & $1.763^{+0.036}_{-0.039}$ & $1.641^{+0.022}_{-0.020} $ & $-0.210^{+0.008}_{-0.007}$  & $-1.014^{+1.055}_{-0.695}$ & $2.800^{+0.784}_{-0.966}$ & $0.763^{+0.392}_{-0.270} $ & $-1.300^{+0.396}_{-0.489}$ \\ 
1329615 & $-0.392^{+0.115}_{-0.095}$ & $4.416^{+0.132}_{-0.139}$ & $0.588^{+0.031}_{-0.030} $ & $-0.923^{+0.051}_{-0.054}$  & $-0.672^{+0.858}_{-0.726}$ & $2.082^{+1.032}_{-0.888}$ & $1.249^{+0.447}_{-0.435} $ & $-0.488^{+0.178}_{-0.212}$ \\ 
1330031 & $-1.989^{+0.306}_{-0.009}$ & $0.386^{+0.555}_{-0.036}$ & $1.480^{+0.052}_{-0.394} $ & $-0.651^{+0.036}_{-0.156}$  & $-0.955^{+0.446}_{-0.436}$ & $1.616^{+0.729}_{-0.986}$ & $0.900^{+0.611}_{-0.383} $ & $-0.477^{+0.278}_{-0.424}$ \\ 
1330426 & $-0.135^{+0.998}_{-0.962}$ & $1.212^{+0.600}_{-0.650}$ & $0.405^{+0.396}_{-0.223} $ & $-1.050^{+0.811}_{-0.752}$  & $-0.415^{+0.998}_{-0.903}$ & $0.955^{+0.379}_{-0.449}$ & $0.244^{+0.264}_{-0.162} $ & $-0.986^{+0.890}_{-0.815}$ \\ 
1331123 & $0.010^{+0.023}_{-0.012}$ & $4.561^{+0.031}_{-0.048}$ & $0.607^{+0.008}_{-0.007} $ & $-1.083^{+0.016}_{-0.015}$  & $-0.683^{+0.868}_{-0.819}$ & $2.609^{+1.045}_{-1.106}$ & $1.104^{+0.468}_{-0.396} $ & $-0.602^{+0.230}_{-0.344}$ \\ 
1333246 & $-0.680^{+0.181}_{-0.202}$ & $0.965^{+0.105}_{-0.141}$ & $1.724^{+0.097}_{-0.089} $ & $-0.116^{+0.021}_{-0.036}$  & $-0.574^{+0.636}_{-0.681}$ & $0.386^{+0.461}_{-0.239}$ & $2.456^{+0.503}_{-0.498} $ & $-0.014^{+0.100}_{-0.172}$ \\ 
1334084 & $-1.602^{+0.199}_{-0.141}$ & $0.185^{+0.360}_{-0.085}$ & $2.089^{+0.269}_{-0.473} $ & $-0.287^{+0.069}_{-0.089}$  & $-0.633^{+0.491}_{-0.422}$ & $1.036^{+0.558}_{-0.533}$ & $1.250^{+0.520}_{-0.428} $ & $-0.120^{+0.188}_{-0.286}$ \\ 
1334087 & $-1.145^{+0.145}_{-0.147}$ & $4.462^{+0.055}_{-0.081}$ & $0.515^{+0.034}_{-0.029} $ & $-1.640^{+0.029}_{-0.026}$  & $-0.643^{+1.080}_{-0.931}$ & $4.253^{+0.749}_{-1.165}$ & $0.504^{+0.237}_{-0.151} $ & $-1.536^{+0.454}_{-0.430}$ \\ 
1334302 & $-1.385^{+0.216}_{-0.322}$ & $0.227^{+0.133}_{-0.061}$ & $2.848^{+0.082}_{-0.185} $ & $-0.194^{+0.069}_{-0.064}$  & $-0.561^{+0.537}_{-0.725}$ & $0.502^{+0.422}_{-0.248}$ & $2.285^{+0.348}_{-0.389} $ & $-0.054^{+0.092}_{-0.177}$ \\ 
1334423 & $0.756^{+0.120}_{-0.723}$ & $0.169^{+0.242}_{-0.046}$ & $0.956^{+0.122}_{-0.212} $ & $-0.364^{+0.165}_{-0.069}$  & $-0.537^{+0.557}_{-0.556}$ & $0.292^{+0.228}_{-0.156}$ & $1.189^{+0.347}_{-0.335} $ & $-0.046^{+0.173}_{-0.254}$ \\ 
1334448 & $1.516^{+0.082}_{-0.202}$ & $0.332^{+1.399}_{-0.179}$ & $2.551^{+0.470}_{-1.531} $ & $-0.817^{+0.135}_{-0.602}$  & $0.026^{+0.672}_{-0.706}$ & $0.891^{+1.120}_{-0.641}$ & $2.231^{+0.600}_{-0.712} $ & $-0.362^{+0.095}_{-0.193}$ \\ 
1334597 & $-1.316^{+0.700}_{-0.494}$ & $2.955^{+0.453}_{-0.769}$ & $0.534^{+0.270}_{-0.168} $ & $-1.427^{+0.339}_{-0.393}$  & $-0.179^{+0.973}_{-0.961}$ & $2.400^{+0.820}_{-1.046}$ & $0.581^{+0.376}_{-0.225} $ & $-1.274^{+0.522}_{-0.552}$ \\ 
1334620 & $-1.998^{+0.006}_{-0.001}$ & $0.451^{+0.033}_{-0.017}$ & $1.623^{+0.021}_{-0.037} $ & $-0.648^{+0.012}_{-0.022}$  & $-0.550^{+0.612}_{-0.523}$ & $1.046^{+0.625}_{-0.451}$ & $1.470^{+0.442}_{-0.392} $ & $-0.229^{+0.179}_{-0.230}$ \\ 
1334644 & $-1.000^{+0.006}_{-0.006}$ & $7.900^{+0.012}_{-0.012}$ & $0.219^{+0.002}_{-0.002} $ & $-2.194^{+0.009}_{-0.004}$  & $0.958^{+0.642}_{-1.763}$ & $2.261^{+2.550}_{-1.503}$ & $0.814^{+0.726}_{-0.364} $ & $-1.324^{+0.424}_{-0.533}$ \\ 
1334645 & $-1.958^{+0.812}_{-0.037}$ & $2.501^{+1.265}_{-0.169}$ & $0.911^{+0.062}_{-0.413} $ & $-1.119^{+0.058}_{-0.307}$  & $-0.646^{+0.913}_{-0.789}$ & $2.385^{+0.878}_{-0.921}$ & $0.905^{+0.414}_{-0.340} $ & $-0.722^{+0.294}_{-0.415}$ \\ 
1335717 & $0.769^{+0.069}_{-0.118}$ & $0.324^{+0.063}_{-0.038}$ & $1.446^{+0.039}_{-0.043} $ & $-0.250^{+0.044}_{-0.036}$  & $0.030^{+0.680}_{-0.614}$ & $1.410^{+0.676}_{-0.558}$ & $0.608^{+0.310}_{-0.252} $ & $-0.657^{+0.321}_{-0.434}$ \\ 
1336008 & $-0.526^{+0.236}_{-0.144}$ & $0.291^{+0.088}_{-0.096}$ & $2.367^{+0.261}_{-0.183} $ & $0.170^{+0.028}_{-0.031}$  & $-0.429^{+0.572}_{-0.473}$ & $0.474^{+0.461}_{-0.216}$ & $1.892^{+0.329}_{-0.485} $ & $-0.012^{+0.092}_{-0.166}$ \\ 
1336009 & $0.884^{+0.208}_{-1.025}$ & $0.244^{+0.457}_{-0.129}$ & $1.671^{+0.304}_{-0.312} $ & $-0.253^{+0.164}_{-0.071}$  & $-0.322^{+0.677}_{-0.579}$ & $0.520^{+0.337}_{-0.249}$ & $1.522^{+0.342}_{-0.328} $ & $-0.052^{+0.141}_{-0.193}$ \\ 
1336453 & $-1.372^{+0.134}_{-0.078}$ & $0.371^{+0.200}_{-0.106}$ & $0.881^{+0.130}_{-0.180} $ & $-0.663^{+0.068}_{-0.093}$  & $-0.595^{+0.576}_{-0.464}$ & $1.181^{+0.356}_{-0.410}$ & $0.455^{+0.310}_{-0.195} $ & $-0.764^{+0.434}_{-0.605}$ \\ 
1336480 & $-1.975^{+0.341}_{-0.020}$ & $0.208^{+0.201}_{-0.020}$ & $1.731^{+0.053}_{-0.260} $ & $-0.455^{+0.025}_{-0.041}$  & $-0.870^{+0.494}_{-0.522}$ & $0.763^{+0.526}_{-0.440}$ & $1.100^{+0.389}_{-0.387} $ & $-0.323^{+0.253}_{-0.260}$ \\ 
1336687 & $-0.369^{+0.440}_{-0.406}$ & $1.133^{+0.162}_{-0.276}$ & $0.204^{+0.156}_{-0.096} $ & $-0.869^{+0.476}_{-0.612}$  & $-0.818^{+0.436}_{-0.437}$ & $0.487^{+0.277}_{-0.309}$ & $0.535^{+0.426}_{-0.279} $ & $-0.675^{+0.462}_{-0.571}$ \\ 
1337117 & $-0.696^{+0.744}_{-0.665}$ & $0.484^{+0.596}_{-0.399}$ & $0.943^{+0.975}_{-0.540} $ & $-0.299^{+0.322}_{-0.538}$  & $-0.298^{+0.524}_{-0.764}$ & $0.306^{+0.438}_{-0.266}$ & $0.926^{+0.921}_{-0.552} $ & $-0.288^{+0.315}_{-0.545}$ \\ 
1337228 & $-0.905^{+0.353}_{-0.391}$ & $0.175^{+0.736}_{-0.109}$ & $2.144^{+0.486}_{-0.950} $ & $-0.065^{+0.133}_{-0.183}$  & $-0.568^{+0.807}_{-0.658}$ & $1.210^{+0.409}_{-0.520}$ & $0.411^{+0.341}_{-0.205} $ & $-0.882^{+0.406}_{-0.584}$ \\ 
1337272 & $-0.251^{+0.556}_{-0.404}$ & $0.136^{+0.237}_{-0.100}$ & $1.351^{+0.699}_{-0.583} $ & $0.011^{+0.210}_{-0.324}$  & $-0.433^{+0.522}_{-0.764}$ & $0.256^{+0.268}_{-0.190}$ & $0.776^{+0.606}_{-0.413} $ & $-0.198^{+0.333}_{-0.380}$ \\ 
1337649 & $-0.032^{+0.598}_{-0.785}$ & $1.759^{+0.358}_{-0.542}$ & $0.251^{+0.138}_{-0.091} $ & $-1.496^{+0.467}_{-0.438}$  & $-0.092^{+0.939}_{-0.991}$ & $1.474^{+0.591}_{-0.697}$ & $0.444^{+0.328}_{-0.202} $ & $-1.058^{+0.568}_{-0.642}$ \\ 
1337687 & $-0.094^{+0.624}_{-0.562}$ & $0.695^{+0.311}_{-0.293}$ & $0.413^{+0.257}_{-0.219} $ & $-0.568^{+0.426}_{-0.574}$  & $-0.507^{+0.552}_{-0.520}$ & $0.537^{+0.317}_{-0.257}$ & $0.931^{+0.354}_{-0.298} $ & $-0.194^{+0.224}_{-0.327}$ \\ 
1337703 & $-0.252^{+0.022}_{-0.027}$ & $1.879^{+0.115}_{-0.120}$ & $1.124^{+0.068}_{-0.064} $ & $-0.747^{+0.049}_{-0.052}$  & $-0.268^{+0.485}_{-0.386}$ & $0.952^{+0.570}_{-0.404}$ & $1.858^{+0.332}_{-0.379} $ & $-0.384^{+0.081}_{-0.121}$ \\ 

 \hline

    \end{tabular}

\end{table*}

\setcounter{table}{0}
\begin{table*}[t]

\centering
    \caption{Continued from above}
    \begin{tabular}{c |c c c c|c c c c }
    \hline
    \multirow{2}{*}{SNID} & \multicolumn{4}{c|}{Global} & \multicolumn{4}{c}{Local}\\
    &  $\log(Z_\star/Z_\odot)$ & $t_{\textrm{age}}$ (Gyr) & $\tau_V$ & $n$  &  $\log(Z_\star/Z_\odot)$ & $t_{\textrm{age}}$ (Gyr) & $\tau_V$ & $n$\\\hline
1337838 & $-0.405^{+0.873}_{-0.874}$ & $1.142^{+0.575}_{-0.589}$ & $0.436^{+0.454}_{-0.247} $ & $-0.880^{+0.721}_{-0.803}$  & $0.619^{+1.675}_{-0.936}$ & $0.060^{+0.153}_{-0.046}$ & $0.648^{+0.443}_{-0.389} $ & $-0.950^{+0.718}_{-0.727}$ \\ 
1338128 & $0.834^{+0.141}_{-0.250}$ & $0.218^{+0.114}_{-0.064}$ & $1.075^{+0.119}_{-0.122} $ & $-0.455^{+0.120}_{-0.086}$  & $-0.433^{+0.632}_{-0.503}$ & $0.867^{+0.437}_{-0.347}$ & $0.669^{+0.316}_{-0.299} $ & $-0.467^{+0.336}_{-0.464}$ \\ 
1338170 & $1.004^{+0.050}_{-0.071}$ & $0.696^{+0.183}_{-0.166}$ & $1.369^{+0.154}_{-0.140} $ & $-0.668^{+0.041}_{-0.054}$  & $-0.159^{+0.885}_{-1.054}$ & $2.286^{+0.747}_{-0.910}$ & $0.620^{+0.319}_{-0.211} $ & $-1.241^{+0.331}_{-0.426}$ \\ 
1338278 & $-0.275^{+0.275}_{-0.223}$ & $1.977^{+0.448}_{-0.601}$ & $0.654^{+0.301}_{-0.224} $ & $-0.492^{+0.181}_{-0.279}$  & $-0.326^{+0.499}_{-0.408}$ & $0.470^{+0.316}_{-0.222}$ & $1.126^{+0.384}_{-0.351} $ & $-0.139^{+0.151}_{-0.182}$ \\ 
1338430 & $0.241^{+0.051}_{-0.411}$ & $0.726^{+0.485}_{-0.067}$ & $2.085^{+0.067}_{-0.273} $ & $-0.191^{+0.022}_{-0.032}$  & $0.146^{+1.067}_{-0.853}$ & $0.518^{+1.465}_{-0.481}$ & $2.530^{+1.188}_{-1.096} $ & $-0.096^{+0.140}_{-0.222}$ \\ 
1338471 & $-0.859^{+0.244}_{-0.283}$ & $0.669^{+0.306}_{-0.259}$ & $1.714^{+0.209}_{-0.257} $ & $-0.137^{+0.050}_{-0.081}$  & $-0.897^{+0.786}_{-0.654}$ & $1.914^{+0.777}_{-0.787}$ & $0.700^{+0.407}_{-0.325} $ & $-0.767^{+0.288}_{-0.432}$ \\ 
1338675 & $0.422^{+0.212}_{-0.433}$ & $0.182^{+0.051}_{-0.050}$ & $1.631^{+0.091}_{-0.097} $ & $0.012^{+0.168}_{-0.075}$  & $-0.022^{+0.658}_{-0.762}$ & $0.747^{+0.424}_{-0.325}$ & $0.628^{+0.385}_{-0.303} $ & $-0.404^{+0.361}_{-0.625}$ \\ 
1339002 & $-1.000^{+0.028}_{-0.026}$ & $4.212^{+0.057}_{-0.057}$ & $0.687^{+0.021}_{-0.021} $ & $-0.856^{+0.028}_{-0.028}$  & $-0.173^{+0.761}_{-0.695}$ & $1.044^{+0.835}_{-0.498}$ & $2.198^{+0.404}_{-0.484} $ & $-0.175^{+0.102}_{-0.139}$ \\ 
1339149 & $0.319^{+0.223}_{-0.217}$ & $0.258^{+0.063}_{-0.077}$ & $1.471^{+0.106}_{-0.098} $ & $-0.130^{+0.073}_{-0.053}$  & $-0.227^{+0.504}_{-0.345}$ & $0.144^{+0.141}_{-0.074}$ & $1.329^{+0.381}_{-0.387} $ & $0.004^{+0.152}_{-0.221}$ \\ 
1339392 & $-1.997^{+0.005}_{-0.002}$ & $1.708^{+0.051}_{-0.049}$ & $1.263^{+0.020}_{-0.021} $ & $-1.018^{+0.014}_{-0.014}$  & $0.102^{+0.895}_{-0.853}$ & $3.421^{+1.144}_{-1.243}$ & $0.739^{+0.357}_{-0.253} $ & $-1.032^{+0.451}_{-0.577}$ \\ 
1339450 & $-0.251^{+0.004}_{-0.006}$ & $1.812^{+0.125}_{-0.097}$ & $1.064^{+0.057}_{-0.072} $ & $-0.422^{+0.031}_{-0.043}$  & $-0.260^{+0.511}_{-0.423}$ & $0.393^{+0.486}_{-0.203}$ & $2.257^{+0.454}_{-0.540} $ & $0.035^{+0.082}_{-0.104}$ \\ 
1340454 & $-0.367^{+0.330}_{-0.308}$ & $0.498^{+0.106}_{-0.148}$ & $2.047^{+0.174}_{-0.134} $ & $0.094^{+0.032}_{-0.040}$  & $-0.464^{+0.697}_{-0.611}$ & $1.159^{+0.646}_{-0.554}$ & $0.820^{+0.397}_{-0.353} $ & $-0.401^{+0.233}_{-0.333}$ \\ 
1341370 & $-0.033^{+0.808}_{-0.893}$ & $1.352^{+0.509}_{-0.634}$ & $0.380^{+0.338}_{-0.185} $ & $-1.077^{+0.495}_{-0.636}$  & $-0.607^{+0.758}_{-0.800}$ & $0.856^{+0.490}_{-0.478}$ & $0.478^{+0.444}_{-0.284} $ & $-0.655^{+0.600}_{-0.727}$ \\ 
1341894 & $-1.654^{+0.283}_{-0.167}$ & $0.407^{+0.349}_{-0.104}$ & $1.770^{+0.118}_{-0.222} $ & $-0.444^{+0.047}_{-0.042}$  & $-0.484^{+0.634}_{-0.626}$ & $0.658^{+0.498}_{-0.309}$ & $1.087^{+0.385}_{-0.426} $ & $-0.255^{+0.255}_{-0.365}$ \\ 
1342255 & $-1.792^{+0.290}_{-0.170}$ & $0.385^{+0.379}_{-0.119}$ & $1.673^{+0.160}_{-0.265} $ & $-0.526^{+0.039}_{-0.036}$  & $-0.252^{+0.956}_{-0.898}$ & $1.915^{+0.668}_{-0.775}$ & $0.521^{+0.344}_{-0.223} $ & $-0.979^{+0.518}_{-0.702}$ \\ 
1343208 & $-0.254^{+0.021}_{-0.024}$ & $1.605^{+0.138}_{-0.172}$ & $1.374^{+0.109}_{-0.083} $ & $-0.358^{+0.045}_{-0.039}$  & $-0.557^{+0.518}_{-0.625}$ & $1.455^{+0.727}_{-0.568}$ & $1.254^{+0.352}_{-0.387} $ & $-0.462^{+0.148}_{-0.189}$ \\ 
1343337 & $-0.039^{+0.183}_{-0.338}$ & $0.253^{+0.098}_{-0.041}$ & $1.811^{+0.103}_{-0.146} $ & $0.177^{+0.035}_{-0.104}$  & $-0.307^{+0.654}_{-0.520}$ & $0.372^{+0.294}_{-0.189}$ & $1.493^{+0.423}_{-0.371} $ & $-0.047^{+0.164}_{-0.220}$ \\ 
1343401 & $-1.995^{+0.008}_{-0.004}$ & $0.736^{+0.046}_{-0.043}$ & $2.319^{+0.031}_{-0.031} $ & $-0.425^{+0.008}_{-0.008}$  & $-0.332^{+0.999}_{-1.018}$ & $3.238^{+1.488}_{-1.369}$ & $0.905^{+0.599}_{-0.414} $ & $-0.714^{+0.299}_{-0.514}$ \\ 
1343533 & $-0.044^{+0.150}_{-0.187}$ & $1.518^{+0.524}_{-0.192}$ & $1.560^{+0.082}_{-0.258} $ & $-0.283^{+0.030}_{-0.058}$  & $-0.221^{+0.917}_{-0.997}$ & $2.018^{+0.972}_{-0.956}$ & $1.086^{+0.467}_{-0.380} $ & $-0.703^{+0.233}_{-0.339}$ \\ 
1343759 & $-0.520^{+0.104}_{-0.066}$ & $0.211^{+0.077}_{-0.076}$ & $1.376^{+0.277}_{-0.222} $ & $0.076^{+0.053}_{-0.070}$  & $-0.893^{+0.827}_{-0.760}$ & $0.524^{+0.507}_{-0.385}$ & $0.841^{+0.601}_{-0.416} $ & $-0.493^{+0.294}_{-0.350}$ \\ 
1344692 & $1.174^{+0.729}_{-0.922}$ & $0.515^{+1.896}_{-0.480}$ & $1.611^{+1.353}_{-0.866} $ & $-0.851^{+0.325}_{-0.434}$  & $-0.822^{+0.754}_{-0.692}$ & $2.197^{+0.912}_{-0.829}$ & $1.237^{+0.387}_{-0.377} $ & $-0.799^{+0.196}_{-0.264}$ \\ 
1345553 & $-1.554^{+0.564}_{-0.363}$ & $3.637^{+0.704}_{-0.869}$ & $0.953^{+0.285}_{-0.250} $ & $-0.695^{+0.094}_{-0.117}$  & $-0.584^{+0.764}_{-0.666}$ & $1.270^{+0.985}_{-0.643}$ & $1.070^{+0.487}_{-0.505} $ & $-0.411^{+0.258}_{-0.408}$ \\ 
1345594 & $-0.274^{+0.034}_{-0.067}$ & $2.206^{+0.127}_{-0.153}$ & $1.422^{+0.076}_{-0.062} $ & $-0.271^{+0.025}_{-0.028}$  & $-0.786^{+0.772}_{-0.688}$ & $2.481^{+1.296}_{-1.077}$ & $1.692^{+0.510}_{-0.522} $ & $-0.343^{+0.125}_{-0.135}$ \\ 
1346137 & $-0.506^{+0.416}_{-0.152}$ & $0.194^{+0.092}_{-0.069}$ & $2.019^{+0.298}_{-0.250} $ & $0.178^{+0.041}_{-0.026}$  & $-0.287^{+0.505}_{-0.379}$ & $0.271^{+0.189}_{-0.129}$ & $1.709^{+0.356}_{-0.321} $ & $0.059^{+0.104}_{-0.142}$ \\ 
1346387 & $0.629^{+1.584}_{-1.302}$ & $1.805^{+0.919}_{-1.775}$ & $0.561^{+1.707}_{-0.168} $ & $-1.691^{+0.563}_{-0.387}$  & $-0.879^{+0.586}_{-0.586}$ & $0.254^{+0.691}_{-0.224}$ & $2.724^{+0.921}_{-0.875} $ & $0.026^{+0.173}_{-0.221}$ \\ 
1346956 & $-0.355^{+0.861}_{-0.756}$ & $0.463^{+0.263}_{-0.261}$ & $0.306^{+0.411}_{-0.208} $ & $-0.623^{+0.691}_{-0.974}$  & $-0.186^{+1.333}_{-1.269}$ & $0.362^{+0.215}_{-0.227}$ & $0.180^{+0.297}_{-0.135} $ & $-0.771^{+0.807}_{-0.929}$ \\ 
1346966 & $0.068^{+0.697}_{-0.107}$ & $0.331^{+0.058}_{-0.163}$ & $1.243^{+0.085}_{-0.101} $ & $0.022^{+0.076}_{-0.299}$  & $-0.963^{+0.895}_{-0.673}$ & $1.655^{+0.545}_{-0.641}$ & $0.492^{+0.323}_{-0.222} $ & $-1.073^{+0.446}_{-0.549}$ \\

 \hline

    \end{tabular}

\end{table*}

\end{appendix}

%
\clearpage
\onecolumn
\noindent
$^{1}$ CENTRA, Instituto Superior T\'ecnico, Universidade de Lisboa, Av. Rovisco Pais 1, 1049-001 Lisboa, Portugal \\
$^{2}$ CENTRA, Faculdade de Ci\^encias, Universidade de Lisboa, Ed. C8, Campo Grande, 1749-016 Lisboa, Portugal \\
$^{3}$ Institut d'Estudis Espacials de Catalunya (IEEC), 08034 Barcelona, Spain \\
$^{4}$ Institute of Space Sciences (ICE, CSIC),  Campus UAB, Carrer de Can Magrans, s/n,  08193 Barcelona, Spain \\
$^{5}$ Institute of Cosmology and Gravitation, University of Portsmouth, Portsmouth, PO1 3FX, UK\\ 
$^{6}$ School of Physics and Astronomy, University of Southampton,  Southampton, SO17 1BJ, UK \\
$^{7}$ Department of Physics, Duke University Durham, NC 27708, USA \\
$^{8}$ Center for Astrophysics $\vert$ Harvard \& Smithsonian, 60 Garden Street, Cambridge, MA 02138, USA \\
$^{9}$ Department of Astronomy, University of California, Berkeley,  501 Campbell Hall, Berkeley, CA 94720, USA \\
$^{10}$ Laborat\'orio Interinstitucional de e-Astronomia - LIneA, Rua Gal. Jos\'e Cristino 77, Rio de Janeiro, RJ - 20921-400, Brazil \\
$^{11}$ Department of Physics, University of Michigan, Ann Arbor, MI 48109, USA \\
$^{12}$ CNRS, UMR 7095, Institut d'Astrophysique de Paris, F-75014, Paris, France \\
$^{13}$ Sorbonne Universit\'es, UPMC Univ Paris 06, UMR 7095, Institut d'Astrophysique de Paris, F-75014, Paris, France \\
$^{14}$ University Observatory, Faculty of Physics, Ludwig-Maximilians-Universit\"at, Scheinerstr. 1, 81679 Munich, Germany \\
$^{15}$ Department of Physics \& Astronomy, University College London, Gower Street, London, WC1E 6BT, UK \\
$^{16}$ Kavli Institute for Particle Astrophysics \& Cosmology, P. O. Box 2450, Stanford University, Stanford, CA 94305, USA \\
$^{17}$ SLAC National Accelerator Laboratory, Menlo Park, CA 94025, USA \\
$^{18}$ Instituto de Astrofisica de Canarias,E-38205 La Laguna, Tenerife, Spain
$^{19}$ Universidad de La Laguna, Dpto.\\ Astrofísica, E-38206 La Laguna, Tenerife, Spain\\ 
$^{20}$ Center for Astrophysical Surveys, National Center for Supercomputing Applications, 1205 West Clark St., Urbana, IL 61801, USA \\
$^{21}$ Department of Astronomy, University of Illinois at Urbana-Champaign, 1002 W. Green Street, Urbana, IL 61801, USA \\
$^{22}$ Institut de F\'{\i}sica d'Altes Energies (IFAE), The Barcelona Institute of Science and Technology, Campus UAB, 08193 Bellaterra (Barcelona) Spain \\
$^{23}$ Astronomy Unit, Department of Physics, University of Trieste, via Tiepolo 11, I-34131 Trieste, Italy \\
$^{24}$ INAF-Osservatorio Astronomico di Trieste, via G. B. Tiepolo 11, I-34143 Trieste, Italy \\
$^{25}$ Institute for Fundamental Physics of the Universe, Via Beirut 2, 34014 Trieste, Italy \\
$^{26}$ Hamburger Sternwarte, Universit\"{a}t Hamburg, Gojenbergsweg 112, 21029 Hamburg, Germany\\
$^{27}$ School of Mathematics and Physics, University of Queensland,  Brisbane, QLD 4072, Australia \\
$^{28}$ Centro de Investigaciones Energ\'eticas, Medioambientales y Tecnol\'ogicas (CIEMAT), Madrid, Spain \\
$^{29}$ Department of Physics, IIT Hyderabad, Kandi, Telangana 502285, India \\
$^{30}$ Fermi National Accelerator Laboratory, P. O. Box 500, Batavia, IL 60510, USA \\
$^{31}$ Jet Propulsion Laboratory, California Institute of Technology, 4800 Oak Grove Dr., Pasadena, CA 91109, USA \\
$^{32}$ Institute of Theoretical Astrophysics, University of Oslo. P.O. Box 1029 Blindern, NO-0315 Oslo, Norway \\
$^{33}$ Kavli Institute for Cosmological Physics, University of Chicago, Chicago, IL 60637, USA \\
$^{34}$ Instituto de Fisica Teorica UAM/CSIC, Universidad Autonoma de Madrid, 28049 Madrid, Spain \\
$^{35}$ Department of Physics and Astronomy, University of Pennsylvania, Philadelphia, PA 19104, USA \\
$^{36}$ Department of Astronomy, University of Michigan, Ann Arbor, MI 48109, USA \\
$^{37}$ Santa Cruz Institute for Particle Physics, Santa Cruz, CA 95064, USA \\
$^{38}$ Center for Cosmology and Astro-Particle Physics, The Ohio State University, Columbus, OH 43210, USA \\
$^{39}$ Department of Physics, The Ohio State University, Columbus, OH 43210, USA \\
$^{40}$ Australian Astronomical Optics, Macquarie University, North Ryde, NSW 2113, Australia \\
$^{41}$ Lowell Observatory, 1400 Mars Hill Rd, Flagstaff, AZ 86001, USA \\
$^{42}$ Department of Astrophysical Sciences, Princeton University, Peyton Hall, Princeton, NJ 08544, USA \\
$^{43}$ Instituci\'o Catalana de Recerca i Estudis Avan\c{c}ats, E-08010 Barcelona, Spain \\
$^{44}$ Institute of Astronomy, University of Cambridge, Madingley Road, Cambridge CB3 0HA, UK \\
$^{45}$ Observat\'orio Nacional, Rua Gal. Jos\'e Cristino 77, Rio de Janeiro, RJ - 20921-400, Brazil \\
$^{46}$ Computer Science and Mathematics Division, Oak Ridge National Laboratory, Oak Ridge, TN 37831\\ 
$^{47}$ Lawrence Berkeley National Laboratory, 1 Cyclotron Road, Berkeley, CA 94720, USA \\
\end{document}